\documentclass[a4paper,12pt]{article}
\usepackage{graphics}
\usepackage{ifthen}
\usepackage{cite}

\author{The OPAL Collaboration}
\title{Search for Chargino and Neutralino Production
at $\sqrt{s}=192-209$~GeV at LEP}
\newcommand{\Title}     {Search for Chargino and Neutralino Production
at $\sqrt{s}=192-209$~GeV at LEP}
\newcommand{\EPnum}     {CERN-EP/2003-090}

\newcommand{\Date}      {8 December 2003}
\newcommand{\Author}    {D.~Toya, I.~Trigger, T.~Marchant, S.~Asai, K.~Nagai}
\newcommand{\MailAddr}  {Isabel.Trigger@cern.ch}
\newcommand{\EdBoard}   {G.~Azuelos, T.~Junk, H.~Neal, P.~Tran} 
\newcommand{\DraftVer}  {Version 2.0}
\newcommand{\DraftDate} {\today}
\newcommand{\TimeLimit} {Thursday, November 27}
\newboolean{Draft}
\setboolean{Draft}{false}
\newcommand{\W}{\mbox{$\mathrm{W}$}}
\newcommand{\Z}{\mbox{$\mathrm{Z}$}}
\newcommand{\e}{\mbox{$\mathrm{e}$}}
\newcommand{\q}{\mbox{$\mathrm{q}$}}

\newcommand{\se}{\mbox{$\tilde\mathrm{e}$}}

\newcommand{\snue}{\mbox{$\tilde\nu_{\mathrm{e}}$}}

\begin{document}
\begin{titlepage}
%
%
\begin{center}
    \large
     EUROPEAN ORGANIZATION FOR NUCLEAR RESEARCH
\end{center}\bigskip
\begin{flushright}
    \large
   \EPnum\\
   \Date
\end{flushright}
%
%
%
%
\bigskip\bigskip
\begin{center}
\Large\bf\boldmath
    \Title
\end{center}
\bigskip
\bigskip
\begin{center}{{\Large The OPAL Collaboration}\\
  \ifthenelse{\boolean{Draft}} {
    Authors: \Author \\
    Editorial Board: \EdBoard
  }{}   
}\end{center}\bigskip\bigskip
\bigskip
%
%
\begin{abstract}
Approximately 438~pb$^{-1}$ of ${\e}^+{\e}^-$ data from the OPAL
detector, taken 
with the LEP collider running at centre-of-mass energies of
192-209~GeV, are analyzed to search for evidence 
of chargino pair production, ${\e}^+{\e}^-\to\tilde\chi^+_1\tilde\chi^-_1$,
or neutralino associated production,
${\e}^+{\e}^-\to\tilde\chi^0_2\tilde\chi^0_1$. 
Limits are set at the 95\% confidence level on
the product of the cross-section for the process
${\e}^+{\e}^-\to\tilde\chi^+_1\tilde\chi^-_1$ and its branching
ratios to topologies containing jets and missing energy, or jets with a
lepton and missing energy, 
and on the product of the cross-section for 
${\e}^+{\e}^-\to\tilde\chi^0_2\tilde\chi^0_1$ and its branching ratio
to jets.
R-parity conservation is assumed throughout this paper.
When these results are interpreted in the context of the Constrained
Minimal Supersymmetric Standard Model, limits are also set on the
masses of the $\tilde\chi^{\pm}_1, \tilde\chi^0_1$ and
$\tilde\chi^0_2$, and regions of the parameter space of the model are
ruled out.
Nearly model-independent limits are also set at the 95\% confidence level on
$\sigma({\e}^+{\e}^-\to\tilde\chi^+_1\tilde\chi^-_1)$
with the assumption that each chargino decays via a $\W$ boson, and on
$\sigma({\e}^+{\e}^-\to\tilde\chi^0_2\tilde\chi^0_1)$
with the $\tilde\chi^0_2$ assumed to decay via a $\Z^0$.
\end{abstract}

\bigskip
\bigskip
 \begin{center}
 {\Large To be submitted to Euro. Phys. J. C}\\
 \ifthenelse{\boolean{Draft}} {
   {\large \bf Please send comments to {\MailAddr}\\  by {\TimeLimit}}\\
   {\large \bf Final Draft}
   {\large \bf \DraftVer{} (\DraftDate)}\\
  }{}
 \end{center}
\bigskip

\end{titlepage}
%
%
\begin{center}{
G.\thinspace Abbiendi$^{  2}$,
C.\thinspace Ainsley$^{  5}$,
P.F.\thinspace {\AA}kesson$^{  3,  y}$,
G.\thinspace Alexander$^{ 22}$,
J.\thinspace Allison$^{ 16}$,
P.\thinspace Amaral$^{  9}$, 
G.\thinspace Anagnostou$^{  1}$,
K.J.\thinspace Anderson$^{  9}$,
S.\thinspace Arcelli$^{  2}$,
S.\thinspace Asai$^{ 23}$
% ...
}\end{center}\bigskip
\bigskip
$^{  1}$School of Physics and Astronomy, University of Birmingham,
Birmingham B15 2TT, UK
\newline
$^{  2}$Dipartimento di Fisica dell' Universit\`a di Bologna and INFN,
I-40126 Bologna, Italy
\newline
$^{  3}$Physikalisches Institut, Universit\"at Bonn,
D-53115 Bonn, Germany
\newline
% ...
$^{  5}$Cavendish Laboratory, Cambridge CB3 0HE, UK
\newline
% ...
$^{  9}$Enrico Fermi Institute and Department of Physics,
University of Chicago, Chicago IL 60637, USA
\newline
% ...
$^{ 16}$Department of Physics, Schuster Laboratory, The University,
Manchester M13 9PL, UK
\newline
% ...
$^{ 22}$Department of Physics and Astronomy, Tel Aviv University,
Tel Aviv 69978, Israel
\newline
$^{ 23}$International Centre for Elementary Particle Physics and
Department of Physics, University of Tokyo, Tokyo 113-0033, and
Kobe University, Kobe 657-8501, Japan
\newline
% ...
$^{  y}$ now at CERN
\newline
% ...
%
\clearpage
\section{Introduction and Theory}
Supersymmetric (SUSY) extensions of the Standard Model postulate the
existence of fermionic partners for all Standard Model
bosons, scalar bosonic partners for all Standard Model fermions, and
at least two
complex Higgs doublet fields, each containing a charged and a 
neutral component~\cite{susy}. 
The partners of the $U(1)_Y$ weak hypercharge field $\mathrm{B}^0$ and the
neutral component of the $SU(2)$ weak isospin field ${\W}^0$ (gauginos),
and of the two neutral components of the Higgs fields (higgsinos),
mix to form the four neutralinos $\tilde\chi^0_{1,2,3,4}$, where the
index increases with the neutralino mass.
The higgsino partners of the charged components of the Higgs fields
and gaugino partners of the charged components of the weak isospin
field mix to form the charginos $\tilde\chi^\pm_{1,2}$, 
where the index 1 indicates the lighter chargino.

SUSY particles couple to the same Standard Model particles
as their Standard Model partners. The lightest charginos may thus be
pair-produced at LEP in $s$-channel $\gamma$ or ${\Z}^0$
exchange. Their gaugino components can also be pair-produced in $t$-channel
scalar electron neutrino ($\snue$) exchange if the $\snue$ is
sufficiently light. 
The chargino pair production cross-section is expected to be
large~\cite{feng}.  It is typically a few picobarns (pb) and depends on
the mass and mixing parameters of the SUSY model; however, the interference
between the $s$-channel and $t$-channel processes is destructive, so
if there is a light $\snue$ the cross-section can be small. 
If R-parity conservation is assumed, as it is throughout this paper,
the lightest SUSY particle (LSP), which is expected to be either the
$\tilde\chi^0_1$ or a $\tilde\nu$, is stable and undetectable.
If the LSP is the $\tilde\chi^0_1$, neutralinos are most likely
to be detected in associated production of
$\tilde\chi^0_2\tilde\chi^0_1$,
either through $s$-channel $\gamma$ or ${\Z}^0$ exchange or by
$t$-channel $\se$ exchange if there is a light scalar electron.
Neutralino cross-sections are comparable to those for chargino pair
production only if the higgsino component is predominant; however, if
charginos are too massive to be 
pair-produced, neutralino associated production may be the only
SUSY process with a visible signature at LEP~\cite{ambrosanio}.
In such cases, the neutralino search can be used to rule out 
regions of SUSY parameter space which are not accessible to the
chargino search. 
A list of the chargino and neutralino production and decay modes
considered in this paper, along with a few final states which were not
considered, is given in Table~\ref{tab:processes}.
\begin{table}[hbtp]
  \begin{center}
  \caption{Principal production and decay modes for the lightest
  charginos and neutralinos in ${\e}^+{\e}^-$ collisions.  All
  detectable final states consist of missing energy and jets or
  charged leptons or both. The final states above the horizontal line
  in each category are considered in this paper. The chargino
  three-body leptonic decays are analyzed in~\cite{acoplanar}, and the
  efficiencies, background estimates, candidates and likelihood
  distributions from that analysis are used in this paper to set
  cross-section limits on chargino production.  Chargino two-body
  decays and neutralino invisible and leptonic decays listed below the
  lines are not considered here when setting limits, although the chargino
  two-body leptonic decays with the stable $\tilde\nu$ in the final
  state are analyzed in~\cite{acoplanar}.
    \label{tab:processes}}  
\begin{tabular}{|l|l@{$\to$}l@{$\to$}l@{$\to$}l|} 
\hline
Chargino pair production & ${\e}^+{\e}^-$ & $\tilde\chi^+_1\tilde\chi^-_1$ &
 $({\W}^{+\ast}\tilde\chi^0_1)
({\W}^{-\ast}\tilde\chi^0_1)$ &
$\ell^+\nu_\ell\tilde\chi^0_1\ell^{\prime -}{\bar{\nu_{\ell^\prime}}}\tilde\chi^0_1$
\\
& \multicolumn{2}{c}{} & &
${\q}{\bar{\q}^\prime}\tilde\chi^0_1\ell\nu_\ell\tilde\chi^0_1 $\\
& \multicolumn{2}{c}{} & &
${\q}{\bar{\q}^\prime}\tilde\chi^0_1{\q^{\prime\prime}}
  {\bar{\q}^{\prime\prime\prime}}\tilde\chi^0_1 $\\
& \multicolumn{1}{c}{} & & $(\tilde{\nu}\ell^+)(\tilde{\bar{\nu^\prime}}\ell^{\prime-})$ &
  $(\tilde\chi^0_1\nu\ell^+)(\tilde\chi^0_1\bar{\nu^\prime}\ell^{\prime-})$\\
  \cline{2-5}
& \multicolumn{1}{c}{} & & $(\tilde{\nu}\ell^+)(\tilde{\bar{\nu^\prime}}\ell^{\prime-})$ &
  stable ($\tilde\nu$ LSP)\\
\hline
Neutralino associated production & ${\e}^+{\e}^-$ &
$\tilde\chi^0_2\tilde\chi^0_1$ & $({\Z}^{0\ast}\tilde\chi^0_1)
(\tilde\chi^0_1)$ & ${\q}{\bar{\q}}\tilde\chi^0_1\tilde\chi^0_1 $\\
  \cline{2-5}
& \multicolumn{2}{c}{} & & $\ell^+\ell^-\tilde\chi^0_1\tilde\chi^0_1 $\\
& \multicolumn{2}{c}{} & & $\nu\bar{\nu}\tilde\chi^0_1\tilde\chi^0_1 $\\
& \multicolumn{1}{c}{} & & $\tilde{\ell}\ell \tilde\chi^0_1$ &
$\tilde\chi^0_1\ell\ell \tilde\chi^0_1$\\
& \multicolumn{1}{c}{} & & $\tilde{\nu}{\bar{\nu}} \tilde\chi^0_1$ &
$\tilde\chi^0_1\nu{\bar{\nu}} \tilde\chi^0_1$\\
\hline
\end{tabular}
\end{center}
\end{table}

In R-parity--conserving SUSY models where the $\tilde\chi^0_1$ is the
LSP, the $\tilde\chi^\pm_1$ decays to a $\tilde\chi^0_1$ and a
${\W}^\pm$, which is off-shell if the mass difference $\Delta M_\pm$
between the $\tilde\chi^\pm_1$ and the $\tilde\chi^0_1$ is less than
the ${\W}^\pm$ mass.  The ${\W}^\pm$ in turn decays either to
${\q}{\bar{\q}}$ or $\ell\nu$, the latter being increasingly favoured
if  $\Delta M_\pm$ is very small. 
If there is a light $\tilde\nu$, an additional decay channel
$\tilde\chi^\pm_1\to\tilde\nu\ell^\pm$ opens up, and 
will predominate if the $\tilde\nu$ is the LSP.  If the $\tilde\nu$ is
not the LSP, it will decay to $\tilde\chi^0_1 \nu$.
Neither of these decay channels leads to a new final-state topology;
they simply increase the 
fraction of leptonic final states.  The existence of light scalar
leptons would lead to similar leptonic final states.
The event topologies for chargino pair production contain
missing energy from undetected $\tilde\chi^0_1$ or $\tilde\nu$ and
neutrinos, and either four jets, two jets and a lepton, or two
leptons.
The analysis of the fully leptonic topology has been published
separately~\cite{acoplanar}. 
The kinematics of the event depend mainly on $\Delta M_\pm$, since
that determines the energy remaining for the Standard Model particles
in the final state.  There is also some dependence on the mass of the
$\tilde\chi^\pm_1$, which determines the boost of the final state
particles. 

Similarly, a $\tilde\chi^0_2$ is expected to decay to a
$\tilde\chi^0_1$ and a virtual ${\Z}^0$, which will decay to ${\q}{\bar{\q}}$,
$\ell^+\ell^-$ or $\nu{\bar \nu}$.  In the presence of light scalar
leptons or scalar neutrinos, additional decays become possible which
increase the $\ell^+\ell^-$ or the invisible $\nu{\bar \nu}$ branching
fractions. 
Detectable event topologies consist of missing energy from
the two undetected $\tilde\chi^0_1$ particles
and a pair of jets from ${\q}{\bar{\q}}$ or a
pair of leptons.  The kinematics depend on $\Delta M_0$, the mass
difference between the $\tilde\chi^0_2$ and $\tilde\chi^0_1$, in the
same way as for charginos and on the mass of the $\tilde\chi^0_2$.
If $\Delta M_0$ is small, the experimental signature becomes
monojet-like because of the small angular separation between the
${\q}{\bar{\q}}$ jets. 

The analysis for chargino and neutralino decays into hadronic and
semileptonic final states described in this paper is a new one,
different from the one used to analyze the OPAL data at centre-of-mass
energies below 192~GeV~\cite{189opal,182opal,172opal}.  The results
for charginos decaying to leptonic final states are taken from the
analysis described in~\cite{acoplanar}. 
\section{Detector and Data Samples}
A complete description of the OPAL detector can be found
in~\cite{detector} and only a brief overview is given here.  
The central detector consisted of a system of tracking chambers
covering the angular\footnote{The OPAL coordinate system is
  defined so that the $z$ axis is in the direction of the electron
  beam, the $x$ axis points toward the centre of the LEP ring, and
  $\theta$ and $\phi$ are the polar and azimuthal angles, defined
  relative to the $+z$ and $+x$ axes respectively.}
region $|\cos\theta|<0.97$,
inside a 0.435~T uniform magnetic field parallel to the beam
axis.  It was composed of a
large-volume jet chamber surrounded by a set of $z$ chambers measuring the
track coordinates along the beam direction and containing a high
precision drift chamber and a silicon microstrip vertex
detector~\cite{microvertex}.  Scintillation counters 
surrounded the solenoid and scintillating tiles~\cite{te} were
also present in the endcaps just outside the pressure bell for the
central tracking chambers; these provided time of flight
information. 
A lead-glass electromagnetic calorimeter, the barrel section of which
was located outside the magnet coil, covered
the full azimuthal range with excellent hermeticity in the polar
angle range of $|\cos\theta |<0.82$ for the barrel region and
$0.81<|\cos\theta |<0.984$ for the endcap region.  The magnet
return yoke was instrumented for hadron calorimetry and consisted of
barrel and endcap sections along with pole tip detectors that
together covered the region $|\cos\theta |<0.99$.  Four layers of
muon chambers covered the outside of the hadron calorimeter.
In the region $|\cos\theta|>0.99$,
electromagnetic calorimeters close to the beam axis completed the
geometrical acceptance down to about 33~mrad.
These included the forward detectors,
which were lead-scintillator sandwich calorimeters and, at smaller
angles, silicon tungsten calorimeters~\cite{sw} located on both
sides of the interaction point and used to monitor the luminosity,
and the forward scintillating tile counters. 
The gap between
the endcap electromagnetic calorimeter and the forward detector was
instrumented with an additional lead-scintillator electromagnetic
calorimeter, called the ``gamma catcher''. 

The data taken with the OPAL detector in 1999 and 2000 at
$\sqrt{s}=192-209$~GeV were studied for the hadronic and semileptonic
channel analyses described in this paper
(see Table~\ref{tab:lumi} for details).
Data were used for this search only if the central jet chamber, barrel and
endcap electromagnetic calorimeters, hadron calorimeters and forward
detectors, including the silicon
tungsten calorimeters, were all functioning. 
For convenience, Table~\ref{tab:lumi} also shows which data were studied
in the leptonic channel analysis~\cite{acoplanar}, which is combined
with the other two channels to produce limits on chargino pair production.
\begin{table}[hbtp]
  \begin{center}
  \caption{Data samples used in the analyses described in this paper
    (${\q}{\bar{\q}}$ for $\tilde\chi^0_2\tilde\chi^0_1$ associated
    production and ${\q}{\bar{\q}}\ell, {\q}{\bar{\q}}{\q}{\bar{\q}}$
    for chargino pair production) 
    as well as in the lepton pair analysis ($\ell\ell$)~\cite{acoplanar}.
    Energy bins are defined so that 
    $\sqrt{s}_{\mathrm{min}}<\sqrt{s}\leq\sqrt{s}_{\mathrm{max}}$.
    The total luminosity studied is 437.6~pb$^{-1}$ in the semileptonic and
    hadronic channels and 440.4~pb$^{-1}$ in the leptonic channel.
    \label{tab:lumi}}  
\begin{tabular}{|l|l|l|l|l|l|}
\hline
\multicolumn{3}{|c}{${\q}{\bar{\q}}\ell, {\q}{\bar{\q}}{\q}{\bar{\q}}$} &
\multicolumn{3}{|c|}{$\ell\ell$} \\ 
$\sqrt{s}_{\mathrm{min}}-
\sqrt{s}_{\mathrm{max}}$ &
$\sqrt{s}_{\mathrm{avg}}$ &
$\int{\cal L}$ & $\sqrt{s}_{\mathrm{min}}-
\sqrt{s}_{\mathrm{max}}$ &
$\sqrt{s}_{\mathrm{avg}}$  & $\int{\cal L}$ \\
(GeV) & (GeV) & (pb$^{-1}$) & (GeV) & (GeV) & (pb$^{-1}$)\\
\hline
190.0 -- 194.0& 191.6 & 29.14 &190.0 -- 194.0&  191.6 & 29.3  \\
194.0 -- 198.0& 195.5 & 73.96 &194.0 -- 198.0&  195.5 & 76.4  \\
198.0 -- 201.0& 199.5 & 75.40 &198.0 -- 201.0&  199.5 & 76.6  \\
201.0 -- 202.5& 201.7 & 38.27 &201.0 -- 204.0&  202.0 & 45.5  \\
202.5 -- 204.5& 203.7 & 10.08 & --- &  --- & --- \\
204.5 -- 205.5& 205.1 & 71.94 &204.0 -- 206.0&  205.1 & 79.0  \\
205.5 -- 206.5& 206.3 & 65.56 & --- &  --- & --- \\
206.5 -- 207.5& 206.6 & 64.98 &206.0 -- 208.0&  206.5 & 124.6 \\
207.5 -- 210.0& 208.0 & 8.25  &208.0 -- 210.0&  207.9 &   9.0 \\
\hline
\end{tabular}
\end{center}
\end{table}

Chargino and neutralino signal events are generated with the DFGT
generator~\cite{dfgt}, which includes spin correlations and allows
for a proper treatment of the ${\W}^\pm$ and ${\Z}^0$ boson width
effects in the chargino and neutralino decays.  The
generator includes initial state radiation (ISR) and uses
JETSET~7.4~\cite{jetset} for the hadronization of the
quark--anti-quark system in the hadronic decays of charginos and
neutralinos.  Signal events are generated at centre-of-mass
energies of 196, 202 and 207~GeV.
For chargino pair production, 
5000 events are generated at each point on a grid with 5~GeV
intervals in chargino mass starting at 75~GeV, with the last point at
0.5~GeV below half the beam energy.
For example, for $\sqrt{s}=196$~GeV, the
generated chargino masses are 75, 80, 85, 90, 95 and 97.5~GeV.
For each chargino mass, 22 equally spaced lightest neutralino masses are
generated, from zero to 3~GeV below the chargino mass.  For
$\tilde\chi^0_2\tilde\chi^0_1$ production, 5000 events are
generated at each point on a grid with 2, 5, 10 or 20~GeV spacing
in $\Delta M_0$ from 3~GeV to $M(\tilde\chi^0_1)$, with closer spacing
for smaller $\Delta M_0$, and 5, 10 or 20~GeV spacing in
$M(\tilde\chi^0_2)+M(\tilde\chi^0_1)$ from 100~GeV up to 1~GeV below
the centre-of-mass energy, with closer spacing nearer to the
centre-of-mass energy.

\begin{table}[tbp]
  \caption{Standard Model background cross-sections at
    $\sqrt{s}=208$~GeV. With the exception of the PHOJET and
    Vermaseren samples, all background samples are generated at all of
    the centre-of-mass energies in Table~\ref{tab:lumi}.  For
    processes where a sample was not generated 
    at 208~GeV, the cross-section shown here was re-scaled from the
    207~GeV sample. 
    The last column shows the approximate cross-section
    surviving the preselection cuts described in
    Section~\ref{sec:presel}. \label{tab:mc}}
  \begin{minipage}{\textwidth}
  \begin{tabular}{|l|l|r|r|}
    \hline
    Background Process & Monte Carlo &
    \multicolumn{2}{c|}{Cross-section (pb)} \\ 
    & Generator & Generator & Preselection \\
    \hline \hline
    $ \left.
    \begin{array}{l}
    \nu{\bar \nu}\gamma     \end{array} \right. $ 
    & NUNUGPV~\cite{nunugpv} & \begin{tabular}{r}7.90\end{tabular} &
    \begin{tabular}{r}--\end{tabular} \\ 
    \hline
    $ \left.
    \begin{array}{l}{\e}^+{\e}^-     \end{array} \right. $
    \parbox{5.5cm}{\sloppy \raggedright ($\e^+$ and 
    $\e^-$ scatter $>12.5^\circ$)} & 
    BHWIDE~\cite{bhwide} & \begin{tabular}{r}493\end{tabular} &
    \begin{tabular}{r}15\end{tabular} \\  
    \hline
    $ \left.
    \begin{array}{l}{\e}^+{\e}^- \end{array} \right. $ (near beam
    axis) & TEEGG~\cite{teegg} & \begin{tabular}{r}637.6\end{tabular}
    & \begin{tabular}{r}20\end{tabular}\\  
    \hline
    $ \left.
    \begin{array}{l}
      \mu^+\mu^-\\\tau^+\tau^-\\{\q}{\bar{\q}}(\gamma)
    \end{array}\right\} $
    &KK2f~\cite{kk2f} &
    \begin{tabular}{r}
      6.815\\ 6.62\\78.44
    \end{tabular} &
    \begin{tabular}{r}
      0.2 \\ 0.2 \\ 3
    \end{tabular} \\
    \hline
    $ \left.
    \begin{array}{l}\gamma\gamma(\gamma...) \end{array} \right. $ &
      RADCOR~\cite{radcor} & \begin{tabular}{r}19.818\end{tabular} & --  \\ 
    \hline
    $ \left.
    \begin{array}{l}
      \ell\ell\ell\ell\\
      \ell\ell {\q}{\bar{\q}}\\
      {\q}{\bar{\q}}{\q}{\bar{\q}}
    \end{array} \right\} $
      \parbox{5cm}{\sloppy \raggedright (four-fermion final states
    without ${\e}^+{\e}^-$)} & 
    KoralW\footnote{Four-fermion events without an ${\e}^+{\e}^-$ pair
    in the final state are simulated with KoralW using matrix elements
    from grc4f to ensure a correct treatment of interference among the
    diagrams, as the treatment of ISR in KoralW is superior to that in
    grc4f.}~\cite{koralw} & 
    \begin{tabular}{r}
      3.416\\
      9.426\\
      8.966
    \end{tabular}
    & \begin{tabular}{r}10\end{tabular}\\
    \hline
    $ \left.
    \begin{array}{l}
      {\e}^+{\e}^-{\e}^+{\e}^-\\
      {\e}^+{\e}^-\mu^+\mu^-\\
      {\e}^+{\e}^-\tau^+\tau^-\\
      {\e}^+{\e}^-{\q}{\bar{\q}}
    \end{array} \right\} $
      \parbox{4cm}{\sloppy \raggedright (excluding multiperipheral
      two-photon process)} & 
      grc4f~\cite{grc4f} &
    \begin{tabular}{r}
      16.749\\
      11.575\\
      1.726\\
      37.788
    \end{tabular} &
    \begin{tabular}{r}2\end{tabular} \\
    \hline
    $ \left.
    \begin{array}{l}{\e}^+{\e}^-{\e}^+{\e}^-\end{array} \right.$ (two-photon) &
    Vermaseren\footnote{No Vermaseren samples were generated at
    192~GeV, 204~GeV, 205~GeV or 207~GeV.}~\cite{vermaseren} & 
    \begin{tabular}{r}876\end{tabular}
    & \begin{tabular}{r}0.3\end{tabular}\\  
    \hline
    $ \left.
    \begin{array}{l}
      {\e}^+{\e}^-\mu^+\mu^-\\
      {\e}^+{\e}^-\tau^+\tau^-
    \end{array} \right\} $ \parbox{3cm}{\sloppy \raggedright
      (two-photon)} & BDK\footnote{Includes effects of additional ISR
      compared with Vermaseren.}~\cite{bdk}&
      \begin{tabular}{r}619.3\\469.0\end{tabular} & 
      \begin{tabular}{r}0.2\\0.9\end{tabular} \\  
    \hline
    $ \left.
    \begin{array}{l}
      {\e}^+{\e}^-{\q}{\bar{\q}}
    \end{array} \right.$
    \parbox{5cm}{\sloppy \raggedright(untagged two-photon)} &
    \parbox{3.5cm}{\sloppy \raggedright PHOJET\footnote{No PHOJET
    samples were generated at 192~GeV,
    204~GeV, 205~GeV, 207~GeV or 208~GeV.}~\cite{phojet}  
    (PYTHIA~\cite{pythia} for hadronic channel analyses)} &
    \begin{tabular}{r}11166\end{tabular} &
    \begin{tabular}{r}6\end{tabular} \\ 
    \hline
    $ \left.
    \begin{array}{l}{\e}^+{\e}^-{\q}{\bar{\q}}\end{array} \right.$
    \parbox{5cm}{\sloppy \raggedright(single-tagged two-photon)} &
    HERWIG~\cite{herwig} &  \begin{tabular}{r}327.8\end{tabular} &
    \begin{tabular}{r}0.7\end{tabular} \\ 
    \hline
    $ \left.
    \begin{array}{l}{\e}^+{\e}^-{\q}{\bar{\q}}\end{array} \right.$
    \parbox{5cm}{\sloppy \raggedright(double-tagged two-photon)} &
    PHOJET &
    \begin{tabular}{r}3.34\end{tabular} & 
    \begin{tabular}{r}--\end{tabular}\\ 
    \hline
  \end{tabular}
  \end{minipage}
\end{table}
The main sources of background (see Table~\ref{tab:mc}) are events
with genuine missing energy either from neutrinos, as in the case of 
four-fermion final states including a ${\W}\to\ell\nu$ or
${\Z}^0\to\nu{\bar\nu}$ decay or any events with $\tau$ decays, or
from particles escaping down the beampipe as in events with 
ISR and ``two-photon'' interactions, where the
interaction is between initial state photons radiated by the
${\e}^+$ and ${\e}^-$
($\e^+\e^-\to\e^+\e^-\gamma\gamma\to\e^+\e^-f\bar{f}$) rather than
directly between the electron and positron. 
In referring to two-photon events,
``untagged'' events are those where the photons are 
both real and the electron and positron are lost in the beampipe,
``singly-tagged'' means that one photon is
sufficiently off-shell to kick an electron into the detector, and
in ``doubly tagged'' events, both photons are virtual and
are seen in the detector.
The generated luminosities are generally several hundred times the
data luminosity, although for the two-photon processes they were only
a few times to a few tens of times the data luminosity, and the
available Bhabha luminosity at 205-207~GeV was only a few times the
data luminosity.

Signal and background events are processed through a detailed simulation of
the OPAL detector~\cite{gopal} and analyzed in the same way as the
OPAL data.
\section{Analysis \label{sec:analysis}}
The main goal of this analysis is to have a selection which is
optimized at every kinematically accessible value of
$(M_{\tilde\chi^\pm_1},\Delta M_\pm)$ or 
$(M_{\tilde\chi^0_2},\Delta M_0)$, henceforth shortened to
$(M,\Delta M)$ when referring generically to the chargino and
neutralino analyses.
This ensures that there are no ``efficiency valleys'', values of
$(M,\Delta M)$ where a signal might be missed because the analysis was
optimized for other regions.
In order to make this task feasible, the same variables must be used
for all $(M,\Delta M)$ so that the
optimization can be done systematically.  Since many of the
most useful selection variables are properties of leptons or jets,
a single set of variables cannot describe all of the
possible signal topologies.
The chargino pair production ($\tilde\chi^+_1\tilde\chi^-_1$) analysis is
therefore split into three channels: fully hadronic, 
semileptonic (one lepton and some jets) and fully leptonic.

The analysis of fully leptonic final states has been published
separately~\cite{acoplanar}.
This paper describes the analyses of the hadronic and semileptonic
final states, which arise from chargino decays
via a ${\W}$. 
If all charginos decay to a lightest neutralino and a ${\W}$,
which decays to all leptons and to the four lightest quarks when
kinematically accessible, 
then the fractions of events falling 
into each channel will be the same as for ${\W}$-pair
decays:  46\% $\q{\bar{\q}}\q{\bar{\q}}$, 44\%
$\q{\bar{\q}}\ell\nu$, 10\% $\ell^+\ell^{'-}\nu{\bar \nu}$.
The branching ratios are very sensitive to $\Delta M$ when $\Delta
M$ is below about 4~GeV.

The neutralino associated production ($\tilde\chi^0_2\tilde\chi^0_1$)
search is performed only in the
fully hadronic channel, as this final state is expected to account for
70\% of the signal rate if ${\Z}^0$ branching ratios are assumed.  This
assumption is valid for $\Delta M$ above about 10~GeV; for smaller
$\Delta M$, the leptonic and invisible branching fractions become
relatively more important as mass thresholds are crossed.

In order to combine the results of the three chargino channels, they
are designed to select non-overlapping subsets of the data.
There is, by construction, no overlap between the hadronic and
semileptonic channels.
There remains a small overlap between
the leptonic channel and the others.  The overlap has  
been checked for candidates in the signal 
Monte Carlo passing stringent signal-selection requirements, 
which dominate the limit-setting,
and found to be negligible for the hadronic
channel and less than 1\% for the semileptonic channel.

\subsection{Construction of the Likelihood Function \label{subsec:lh}}
The description that follows is for the hadronic and semileptonic
channels only, although the analysis of the leptonic channel is very
similar~\cite{acoplanar}.

After the common preselection described in Section~\ref{sec:presel}, a set
of additional cuts, described in Sections~\ref{sec:hadrsel} --
\ref{sec:nnsel}, is applied in each channel.
After these cuts, a set of discriminant variables is
computed.  The distributions of these variables are formed for each of
the Standard Model background Monte Carlo samples used, as well as for
the signal Monte Carlo samples for all the generated
$(M,\Delta M)$ points.
Much of the background can be rejected by rejecting
events which do not fall approximately within the same range in each
discriminant variable as the signal. 
The ranges of the signal sample variable distributions are checked at each
generated value of $(M,\Delta M)$, as in general they vary with
$(M,\Delta M)$.  
A set of fairly loose cuts which reject values of
discriminant variables outside the signal range is stored for each
$(M,\Delta M)$.

The variables which have the greatest discriminating power
against the backgrounds in trial likelihoods evaluated at several different
$(M,\Delta M)$ values are then used to construct a likelihood
discriminant.  Reference histograms are made for each variable $x_i$
for the signal and the sum of all major backgrounds after the cuts for each
$(M,\Delta M)$ point.  All histograms are normalized to unity,
after which the histogram contents are denoted by $P_S(x_i)$ and
$P_B(x_i)$ for signal and background reference histograms.
These histograms are given as inputs to 
a calculation~\cite{pc} which constructs the signal and background
likelihoods $L_S\approx\prod_i P_S(x_i)$ and $L_B\approx \prod_i
P_B(x_i)$; the likelihoods are not exact products of the
probabilities because correlations between the variables are projected
out~\cite{pc}. 
A likelihood discriminant $L_R={L_S}/(L_S+L_B)$ is then
constructed at each $(M,\Delta M)$ for every event in the data
sample as well as for the signal and background Monte Carlo events.

The same likelihood function is used for the data
collected at all centre-of-mass energies.
Variables are scaled to the centre-of-mass energy wherever
possible in order to minimize energy dependences, except in the case
of four-fermion backgrounds where features of some variable
distributions depend on absolute thresholds.   The signal
samples used to construct the reference histograms are generated at
$\sqrt{s}=207$~GeV, with a cross-check set made at 196~GeV to test the
validity of the assumption of energy independence. 

Likelihood reference histograms are made for every $(M,\Delta M)$
point at  which signal was generated.
The generated 5~GeV spacing in $M_{\pm}$ is adequate, since
the kinematics do not change rapidly with $M_{\pm}$; 
however, for $\Delta M$, a spacing of 2~GeV is required.  Since the
chargino signal samples are generated with $\Delta M$ spacings ranging
from about 
3.4~GeV for $M=75$~GeV to 4.8~GeV for $M=103.5$~GeV, this involves
interpolating signal reference histograms for the likelihood.  This is
done using the linear interpolation technique described
in~\cite{futyan,marchant} and also used in~\cite{acoplanar}.
Similarly, in the case of the neutralino analysis, reference
histograms are made for every $(M,\Delta M)$ point at which signal was
generated, and are interpolated to obtain a grid with a 5~GeV
spacing in $M_{\tilde\chi^0_2}$ and a 2~GeV spacing in $\Delta M$,
which is again found to be adequate.
\subsection{Common Preselection (Charginos and Neutralinos) \label{sec:presel}}
The common preselection used for the hadronic and semileptonic
channels of the chargino and neutralino searches is designed to select
well-measured events with missing 
transverse energy and to eliminate events which are due
to cosmic rays passing through the detector or events where energy is
missing along the axis of the beam due to ISR or
escaping initial state electrons in two-photon interactions.
Events with jets or leptons detected very close to the beam
axis are rejected since they tend to be poorly measured.  This cut
also rejects a fraction of the events from LEP machine backgrounds,
such as those from beam-gas or beam-wall interactions.
Energy and momentum information from tracks in the inner detector and
energy deposit clusters in the calorimeters are combined using an
algorithm which matches tracks and clusters to avoid
double-counting~\cite{172opal}. 
This algorithm is used in the construction
of all event quantities such as the visible energy,
$E_{\mathrm{vis}}$, which is the sum of the energies associated with
all tracks and clusters in the event, corrected for double counting,
the missing transverse momentum, 
$p_{\mathrm{T}}^{\mathrm{miss}}$ (which is the component of momentum in the $x-y$ plane required
to balance the vector sum of the momenta of all the charged tracks),
the invariant mass of the system comprising all the detected
momentum and energy, $M_{\mathrm{vis}}$, and the hadronic mass,
$M_{\mathrm{had}}$, which is similar to $M_{\mathrm{vis}}$ but
excludes tracks and clusters associated with lepton candidates. 
The common preselection consists of the following requirements: 
  \begin{itemize}
  \item less than 2~GeV may be deposited in any of the
  forward detectors, including the silicon tungsten calorimeters, and
  less than 5~GeV may be deposited in any of the ``gamma catchers'';
  \item events are vetoed if they contain activity in the endcap and
    forward detectors consistent with forward-going
    muons, 
    or if the energy deposits in the forward scintillating tile
    counters are large enough to be consistent with a photon,
    or if the fraction of visible energy in the forward region  
    $|\cos{\theta}|>0.9$ exceeds 20\% of $\sqrt{s}$,  
    or if they exhibit evidence of instrumental noise in the jet  
    chamber; 
  \item there must be enough tracks, excluding any lepton
    candidates, to group the tracks and clusters in the event into two jets
    using the Durham algorithm~\cite{durham} and still have at least
    one charged track in each jet; 
  \item at least 20\% of the charged tracks must pass the
  ``good track'' criteria~\cite{trackquality}; 
  \item in events with fewer than 5 ``good'' charged tracks,
  the sum of the charges of good tracks in each jet must be between
  $-1$ and $+1$ 
  and the sum of charges of all good tracks in the event must be zero;
  \item probable cosmic ray interactions are vetoed if any of the
    following conditions are satisfied: 
    \begin{itemize}
      \item time-of-flight information is recorded,
        and differs by more than 10~ns from that expected for
        collision events;
      \item time-of-flight is not recorded, there is at least one hit in the
        muon chambers, only two
        good tracks are present, and their point of nearest approach to the
        interaction point is not within 10~cm; this rejects cosmic rays going
        through the tracking chambers;
      \item there is at least one hit in the muon chambers, there are two or
        fewer good tracks, and ten or more clusters in the electromagnetic
        calorimeters, and clusters in the hadron calorimeters; this rejects
        cosmic rays going through the calorimeters;
     \item there is no hit in the time-of-flight system,
       and both jets
       (when the event is forced into two jets) have polar angles with
       respect to the beam axis of $|\cos\theta_{\mathrm{jet}}|<0.8$; this 
       rejects out-of-time events where a particle goes through the barrel
       of the detector;
%
    \end{itemize}
  \item unmodelled low-energy two-photon backgrounds and some poorly
  modelled forward Bhabha events are removed by requiring
  $p_{\mathrm{T}}^{\mathrm{miss}}$ in excess of 4~GeV, 
  $M_{\mathrm{vis}}$ in excess of 5~GeV  
  and acoplanarity (defined for an event forced
  into two jets as $180^\circ$
  minus the angle between the two jet thrust vectors projected into
  the $x-y$ plane) 
  greater than $5^\circ$;
  \item there must be significant missing energy: the magnitude of the
    vector sum of charged track momenta and the scalar sum of the
    energy of all electromagnetic calorimeter clusters must each be
    less than 65\% of $\sqrt{s}$. 
  \end{itemize}
After this preselection, the accepted cross-sections are reduced to
the levels shown in the final column of Table~\ref{tab:mc}.
Signal efficiencies after the preselection vary from about 8\% for
$\Delta M=3$~GeV to about 60-70\% or more for $\Delta M>10$~GeV.
\subsection{Chargino Pair Production Hadronic Selection \label{sec:hadrsel}}
The event topology for the hadronic channel of the
$\tilde\chi^+_1\tilde\chi^-_1$ search consists of jets
and missing energy, the amount depending on $\Delta M$.
There are four primary quarks in this final state, but the jets
may be highly boosted and difficult to separate, or additional jets
may be present from gluon radiation.
The principal backgrounds are
therefore events with hadronic activity and missing energy:
two-photon ${\e}^+{\e}^- {\q}{\bar{\q}}$, 
${\q}{\bar{\q}}(\gamma)$ in which one or more ISR photons are lost
down the beampipe or the jets are mis-measured,
${\W}^+{\W}^-\to \q {\bar{\q}}{ \q }{\bar{\q}}$ where the jets are
mismeasured, ${\W}^+{\W}^-\to \q {\bar{\q}}\ell\nu$ where the lepton
is not reconstructed and ${\W}{\e}\nu\to \q {\bar{\q}}{\e}\nu$ where the
electron goes down the beampipe.

To avoid overlap with the semileptonic channel, events are vetoed if
they contain a lepton candidate which would be accepted in that
channel.
The overlap between the hadronic channel
and the fully leptonic channel was checked and is negligible.

To reduce some of the main backgrounds, the following additional
preselection cuts are made:
\begin{itemize}
  \item events are required to have at least five charged tracks;
  \item events are removed if the missing momentum vector points to
    the forward region, $|\cos\theta_{\mathrm{miss}}|>0.9$;  
    they are probably due to ISR or two-photon
    interactions or may be missing part of a jet down the beampipe;
  \item the missing transverse momentum,
    $p_{\mathrm{T}}^{\mathrm{miss}}$, is required to be in
    excess of 6~GeV, whether or not information from the hadronic
    calorimeter is used in the calculation;  
  \item the Durham jet finder must find between three and five jets
    with a jet resolution parameter of $y_{\mathrm{cut}}=0.005$; 
  \item when the event is forced into two jets by adjusting the
  resolution parameter $y_{\mathrm{cut}}$:
    \begin{itemize}
      \item each must satisfy the requirement
        $|\cos\theta_{\mathrm{jet}}|<0.9$ on 
        its polar angle with respect to the beam axis; 
      \item to reduce ${\q}{\bar{\q}}$ events, the acoplanarity of the
        jets, $\phi_{\mathrm{acop}}$, is required to exceed $10^\circ$;
      \item the energy of the more energetic jet, $E_{\mathrm{jet}
        1}$, must be less than 100~GeV; 
    \end{itemize}
  \item the invariant visible mass, $M_{\mathrm{vis}}$, must be less
    than 180~GeV to reduce the four-fermion background; 
  \item if there is a ``loose'' lepton candidate, an isolated track
    which does not satisfy the requirements of the semileptonic
    channel (see Section~\ref{sec:semilsel}), its energy,
    $E_{\mathrm{loose}-\ell}$, must be less than 50~GeV.
\end{itemize}
As described in Section~\ref{sec:analysis}, cuts are made on the 
likelihood variables (see below) at each $(M,\Delta M)$ in order to
eliminate events obviously incompatible with signal.
After these mass-dependent cuts, efficiencies for hadronic signal
events vary 
from about 1\% for $\Delta M=3$~GeV to 40-60\% for $\Delta M>5$~GeV.
Two-photon backgrounds are eliminated for $\Delta M$
greater than about 10~GeV and reduced to less than 0.01~pb
for smaller $\Delta M$. 
Four-fermion backgrounds vary  
from a negligible value for $\Delta M<10$~GeV to about 0.7~pb in the
region dominated by ${\W}^+{\W}^-$ background with $M\approx 80$~GeV
and $\Delta M\approx M$, while $\q{\bar{\q}}(\gamma)$ backgrounds vary
from a negligible value for $\Delta M<10$~GeV to about 0.005-0.05~pb for
larger $\Delta M$.  Other backgrounds are negligible.

\begin{figure}[htbp]
  \centering{\Huge OPAL}\\
\resizebox{0.45\textwidth}{!}{\includegraphics{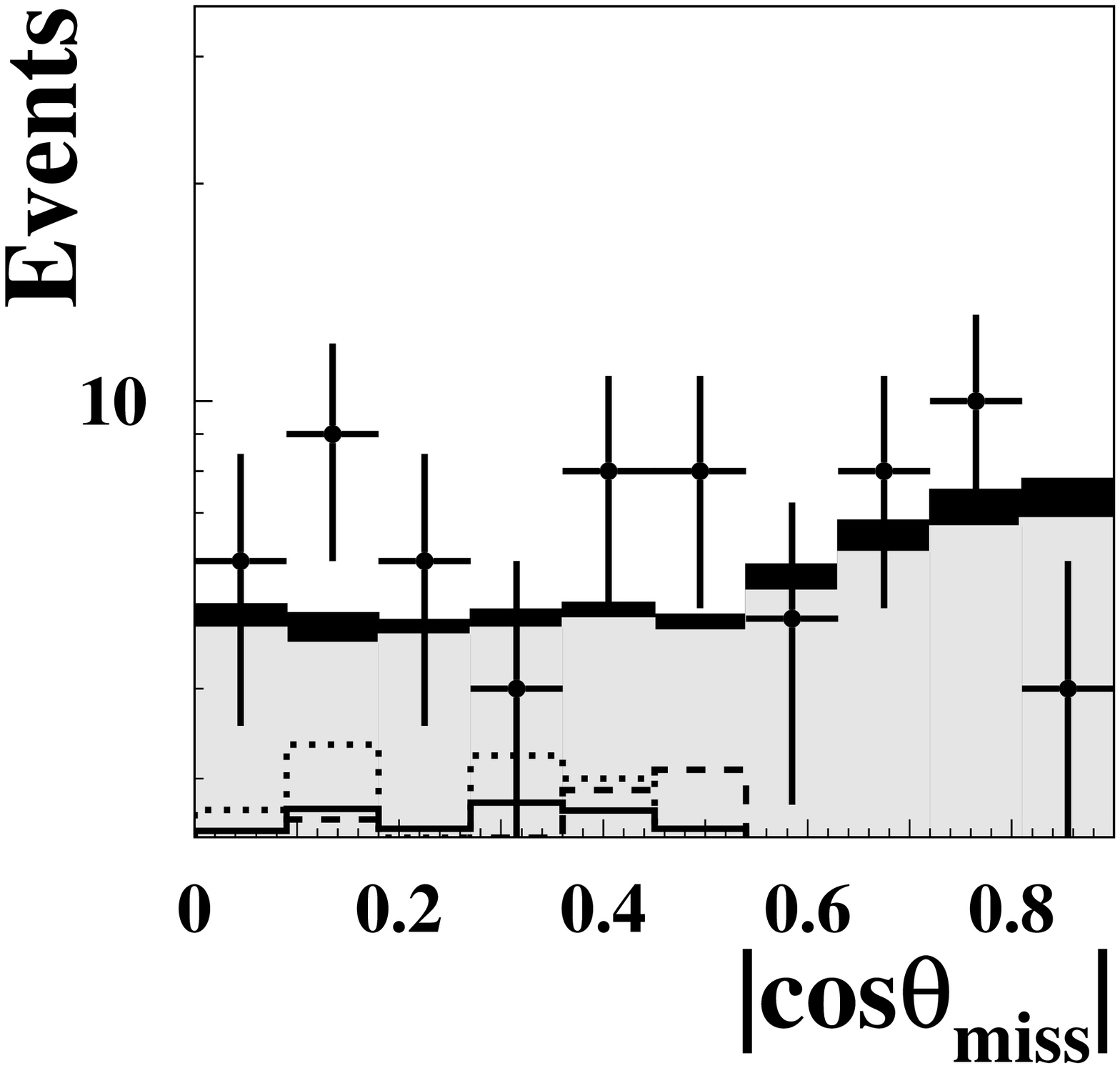}}
\resizebox{0.45\textwidth}{!}{\includegraphics{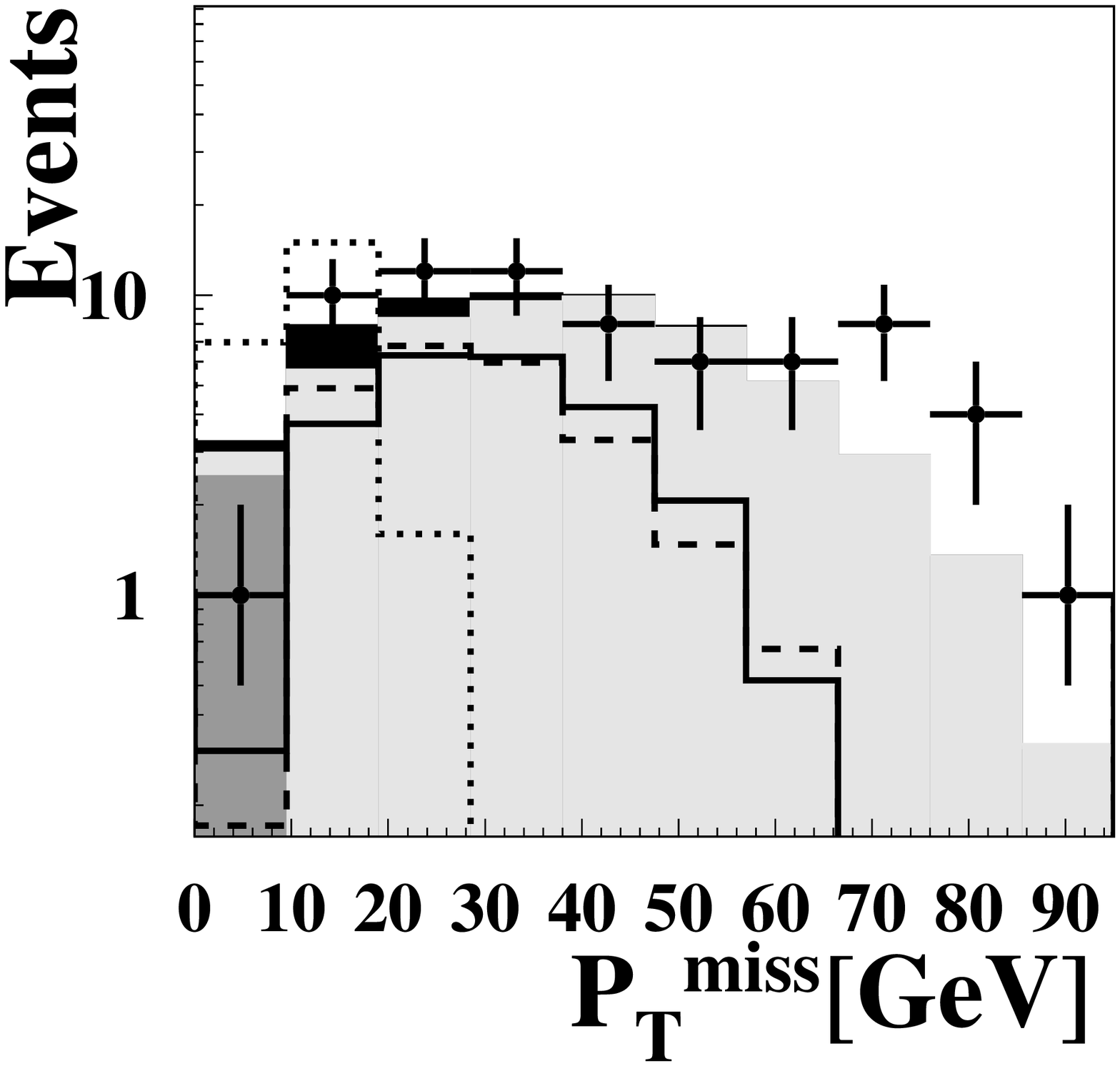}}
\resizebox{0.45\textwidth}{!}{\includegraphics{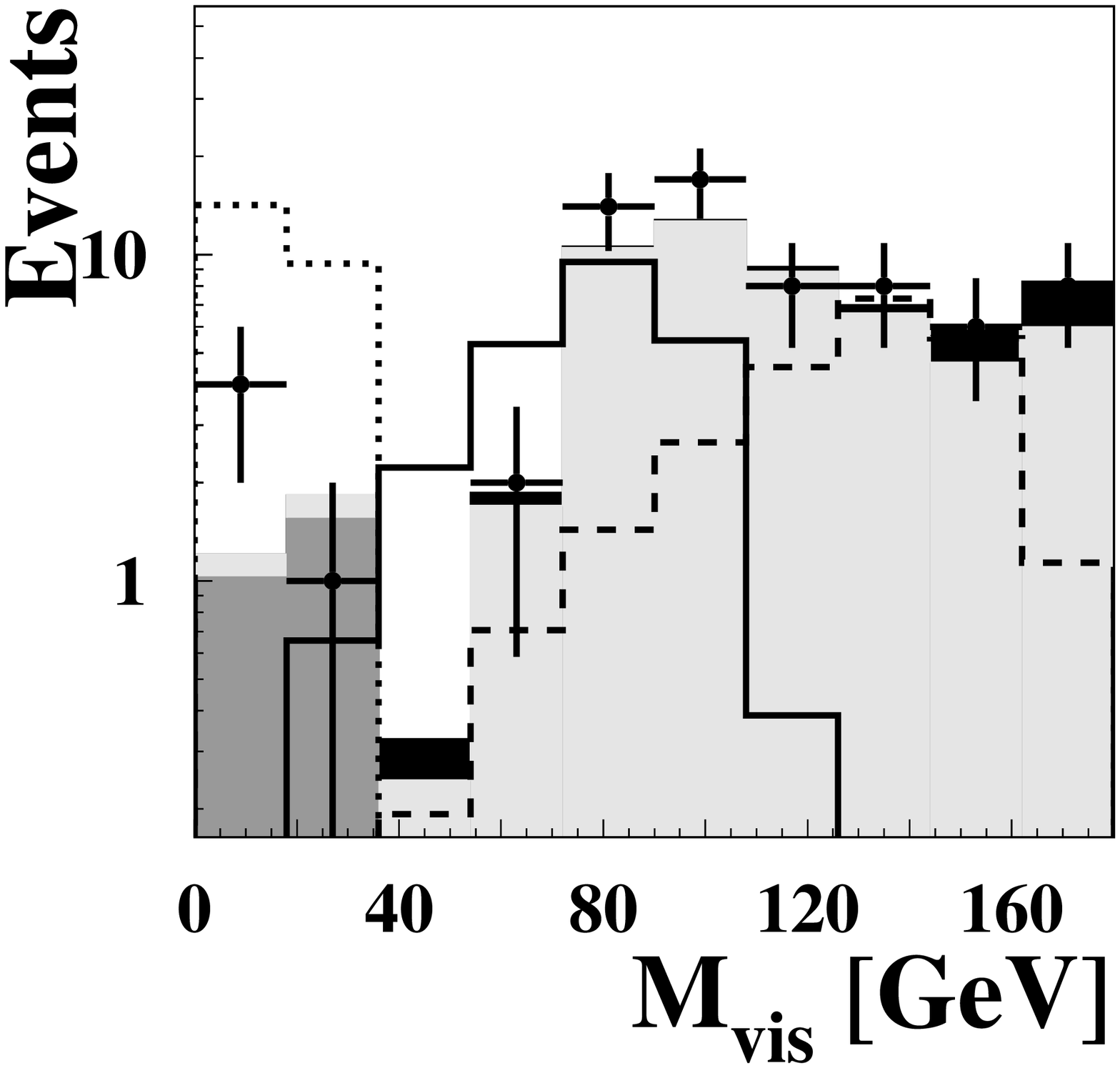}}
\resizebox{0.45\textwidth}{!}{\includegraphics{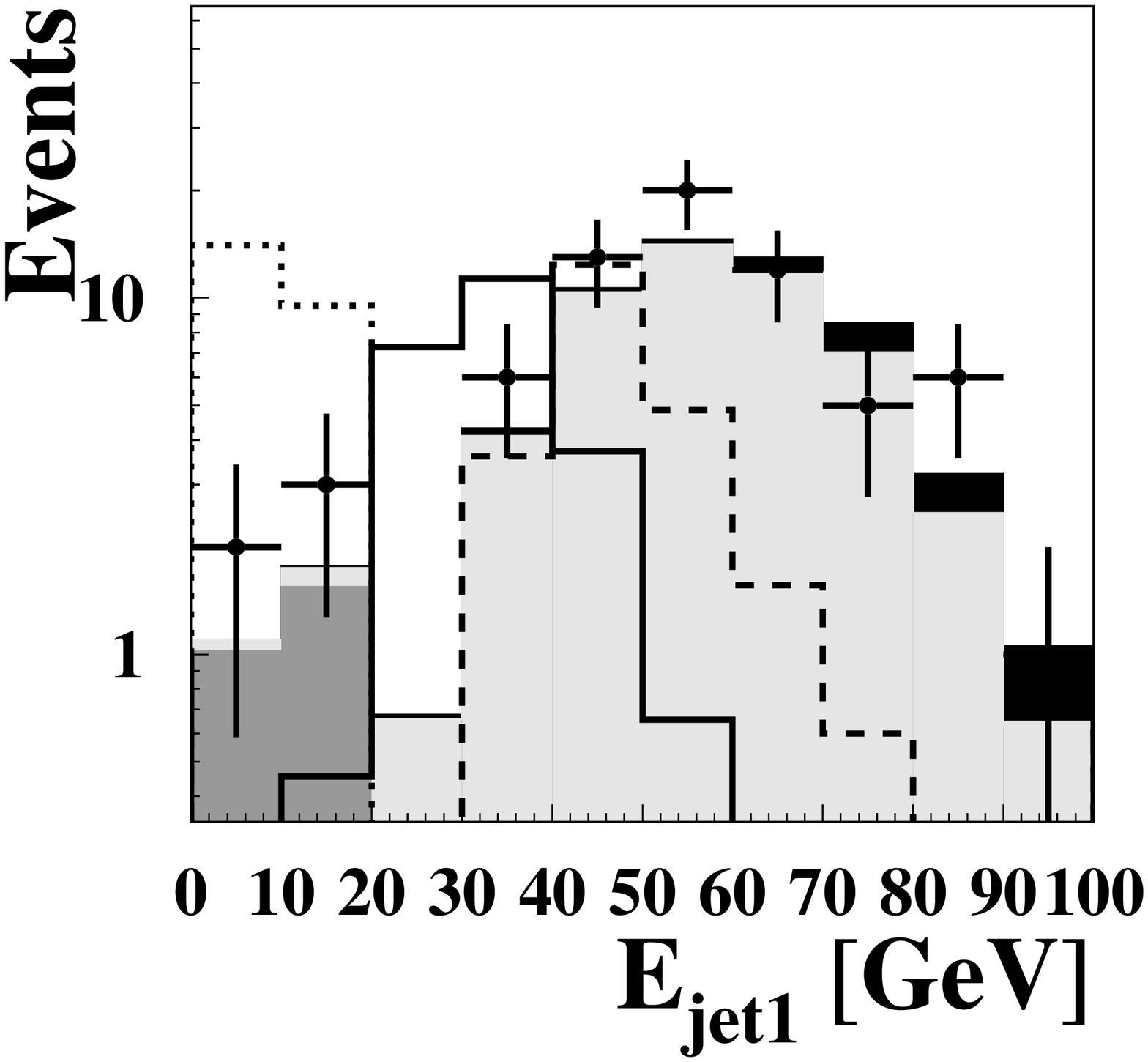}}
\resizebox{0.45\textwidth}{!}{\includegraphics{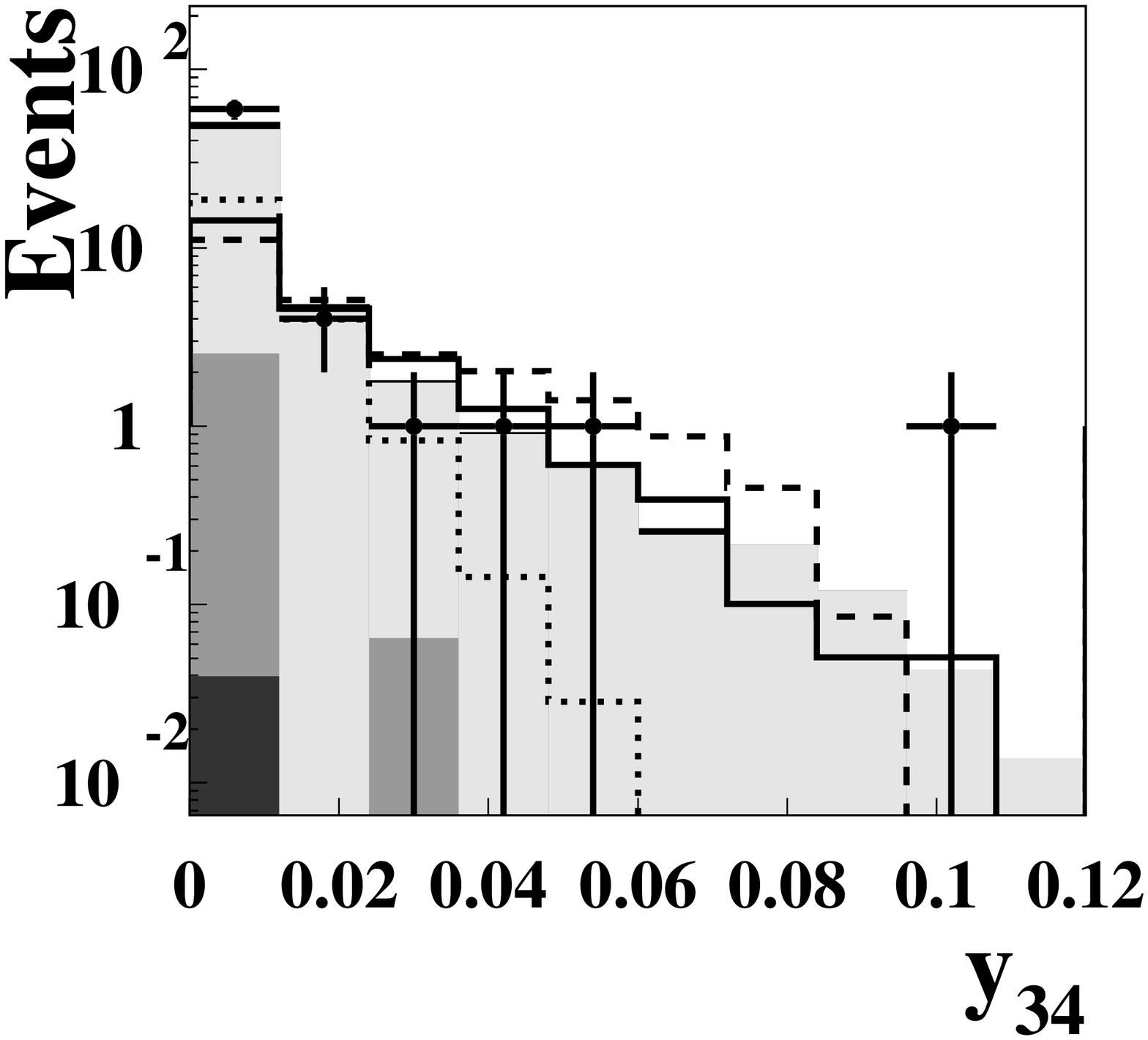}}
\resizebox{0.45\textwidth}{!}{\includegraphics{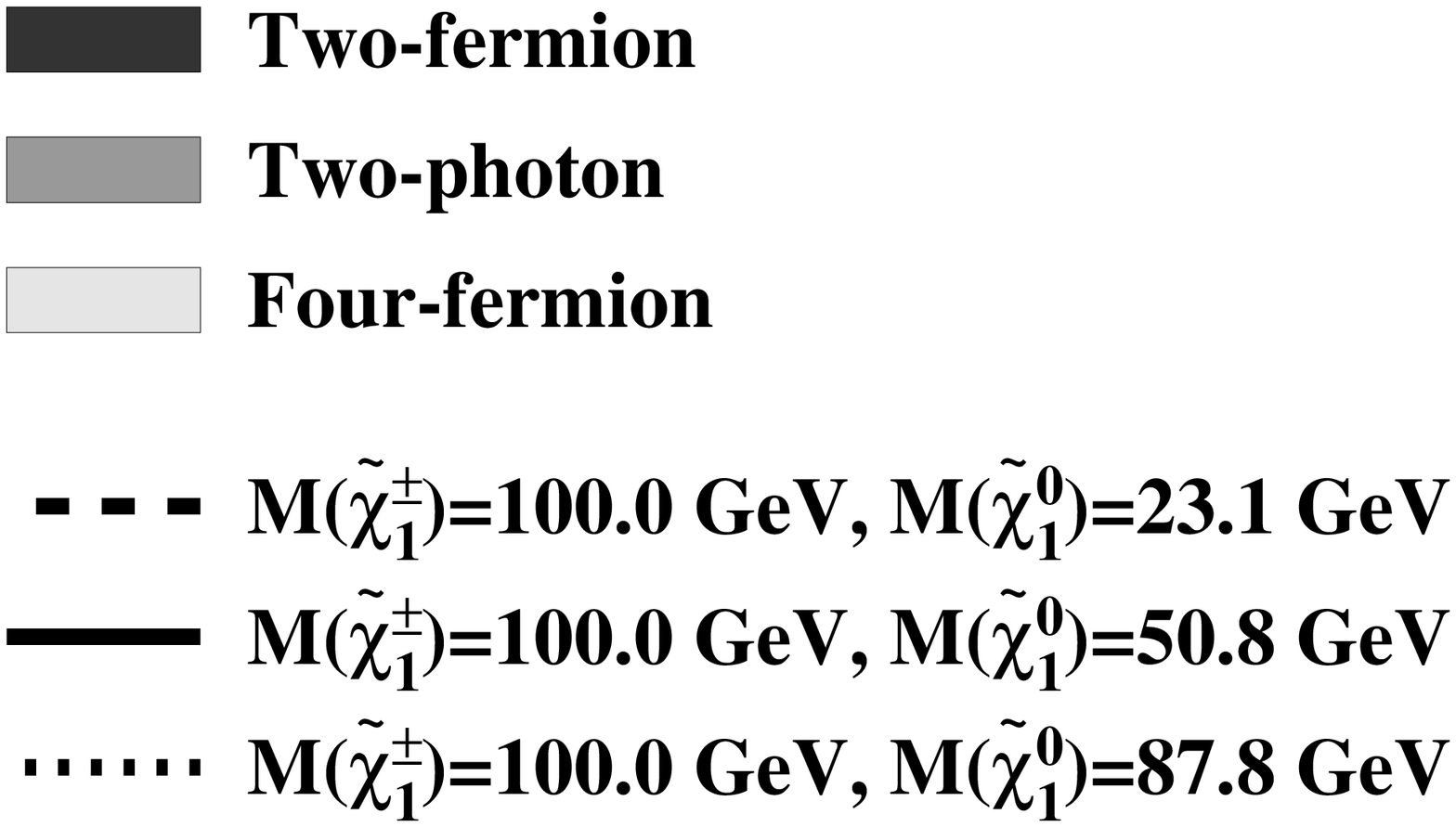}}
\caption{Comparison of data and Monte Carlo distributions of selected
  likelihood variables for the hadronic chargino analysis, shown
  at preselection level for $\sqrt{s}=206$~GeV.  The shaded histograms show
  the various expected backgrounds (the darkest shades are two-fermion
  processes, predominantly $\q {\bar{\q}}$, the intermediate grey
  represents the two-photon contribution and the light grey indicates
  the dominant four-fermion background) and the points with error bars are 
  the OPAL data.  The dashed empty histograms show a sample signal
  with $(M,\Delta M)=$(100~GeV, 76.9~GeV), the empty histograms with
  the solid line show a sample signal with $(M,\Delta M)=$(100~GeV,
  49.2~GeV),
  and the dotted empty histograms show a sample signal with $(M,\Delta
  M)=$(100~GeV, 12.2~GeV),
  all with arbitrary normalization. \label{fig:hadvar}}
\end{figure}

\begin{figure}[htbp]
\resizebox{0.49\textwidth}{!}{\includegraphics{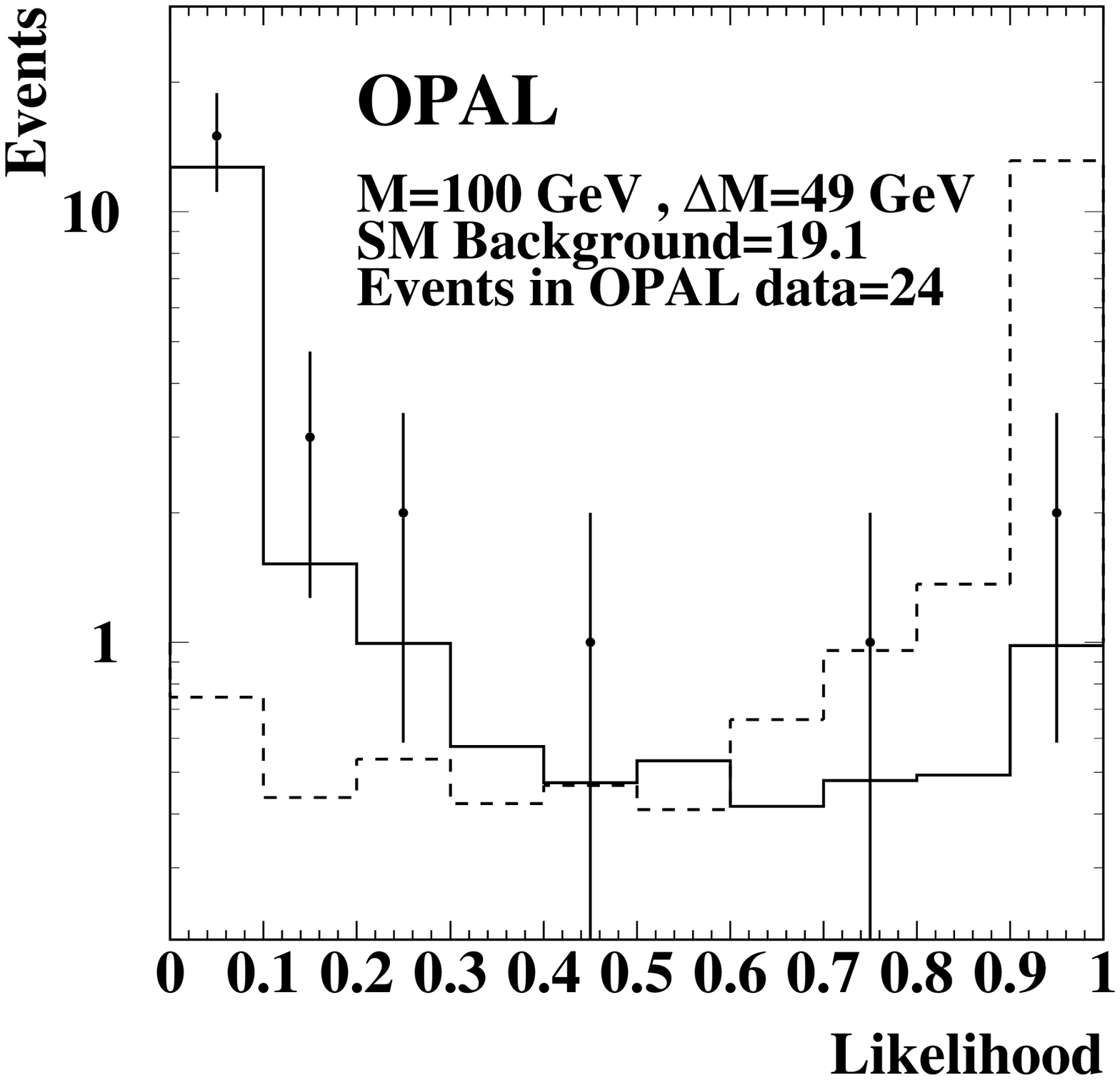}}
\resizebox{0.49\textwidth}{!}{\includegraphics{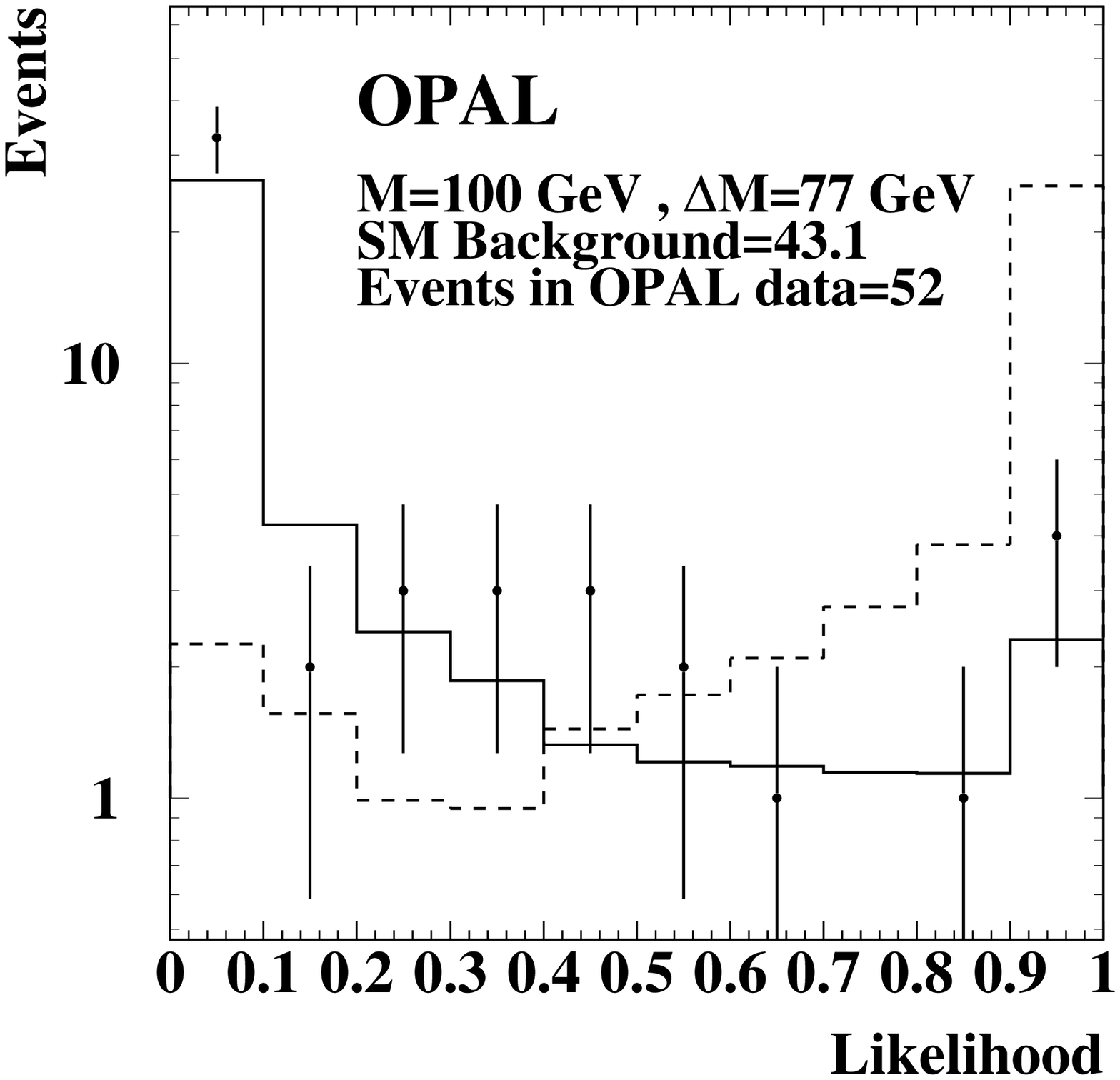}}
  \caption{Sample distributions of $L_R$ for Standard Model background
    in the hadronic channel
    (solid histogram), chargino signal with arbitrary normalization
    (dashed histogram) and OPAL 
    data at $\sqrt{s}=207$~GeV (points with error bars) at $(M,\Delta
    M)=$ (100~GeV, 49~GeV) and (100~GeV,
    77~GeV).
    \label{fig:lhhad}}
  
\end{figure}

The likelihood discriminant is then constructed using the following variables:
\begin{itemize}
  \item  $|\cos\theta_{\mathrm{miss}}|$, the polar angle of the
    missing momentum vector, discriminates against two-photon events
    where one or both initial state electrons are lost down the
    beampipe and events with ISR; 
  \item  $p_{\mathrm{T}}^{\mathrm{miss}}$ also rejects two-photon and
  ISR events; 
  \item  $M_{\mathrm{vis}}$ reduces four-fermion backgrounds; 
  \item $y_{45}$, $y_{34}, y_{23}$, the values of the Durham jet
  resolution parameter for which the event passes from five jets to 
  four, four to three, and three to two, discriminate
  against two jet events from ${\W}^+{\W}^-\to {\q}{\bar{\q}}\ell\nu$ with a
  poorly reconstructed lepton and ${\q}{\bar{\q}}\gamma$; 
  \item $E_{\ell'}$, the energy of the lowest multiplicity jet (where
    the number of jets is determined by the default Durham resolution
    parameter), has some
    discrimination against backgrounds containing $\tau$ leptons; 
  \item  when the event is forced into two jets,
      $\phi_{\mathrm{acop}}$, the acoplanarity angle 
      between the jets, discriminates against Standard Model
      two-fermion events and photon conversions; 
  \item  $E_{\mathrm{jet 1, 2}}$, the energies of the two jets, reject
    four-fermion backgrounds such as ${\W}^+{\W}^-$ and ${\W}{\e}\nu$;
  \item $\cos\theta_{\mathrm{jet}\,1,2}$, the cosines of the polar
    angles of the two jets with respect to the beam axis, reject
    two-photon and ISR events; 
  \item $N_{\mathrm{loose}-\ell}$, the number of lepton candidates
    passing loose requirements, less stringent than those
    of the semileptonic channel, helps to reject ${\W}^+{\W}^-\to 
    {\q}{\bar{\q}}\ell\nu$ events; 
  \item $E_{\mathrm{loose}-\ell}$, the energy of the most energetic
    such loose lepton candidate,  also helps to reject ${\W}^+{\W}^-\to
    {\q}{\bar{\q}}\ell\nu$ events.
\end{itemize}
Several of these variables are substantially correlated, but the effect
of these correlations is minimized by the projection method
used to construct the likelihood, as stated in Section~\ref{subsec:lh}.
Distributions of these variables are shown in Figure~\ref{fig:hadvar}
after the preselection cuts have been applied.
Likelihood distributions are shown for the signal, expected background
and observed data for two pairings of $(M,
\Delta M)$ in Figure~\ref{fig:lhhad}.
\subsection{Chargino Pair Production Semileptonic Selection
  \label{sec:semilsel}}
The event topology for the semileptonic channel consists of two jets,
a lepton and an amount of missing energy which is dependent on
$\Delta M$.  The principal backgrounds are ${\W}^+{\W}^-\to
{\q}{\bar{\q}}\ell\nu$, two-photon processes leading to ${\e}^+{\e}^-
{\q}{\bar{\q}}$ and ${\e}^+{\e}^-\tau^+\tau^-$ final states,
$\tau^+\tau^-$ with at least one of the  
taus decaying hadronically, and ${\q} {\bar{\q}}(\gamma)$ events with
a lepton, usually fake, in one of the jets. 

Overlap with the hadronic channel is avoided by requiring
the presence in the event of an isolated lepton
candidate with the same isolation criteria as in searches at
lower energies~\cite{172opal}. 
Overlap with the leptonic channel is almost eliminated, as for the
hadronic channel, by vetoing events which pass the selection for
${\W}^+{\W}^-\to\ell^+\nu\ell^-{\bar \nu}$ events of the ${\W}^+{\W}^-$
analysis.

After applying the general preselection and the veto cuts just
described, the following additional cuts are made.
To remove poorly reconstructed events, it is required that the event
visible mass, $M_{\mathrm{vis}}$, and hadronic mass,
$M_{\mathrm{had}}$, as well as the invariant mass of the
system consisting of the lepton candidate and the missing momentum
vector, $M_{\ell\mathrm{miss}}$, all be less than or equal to the
centre-of-mass energy.
Cuts are made at each $(M,\Delta M)$, as described in
Section~\ref{sec:analysis}, to limit the accepted variable ranges to
roughly the range of the signal distribution.
These cuts are applied to all of the variables used for likelihood
distributions, which helps to ensure that the signal reference
histograms used in the likelihood have data in all bins and a smooth
distribution.  The same type of cut is applied to some additional
variables which are not used in the likelihood because they are less
powerful discriminants or because they are too strongly correlated
with better variables that are already in the likelihood.

\begin{figure}[htbp]
  \centering{\Huge OPAL}\\
  \vspace{-2ex}
\resizebox{0.45\textwidth}{!}{\includegraphics{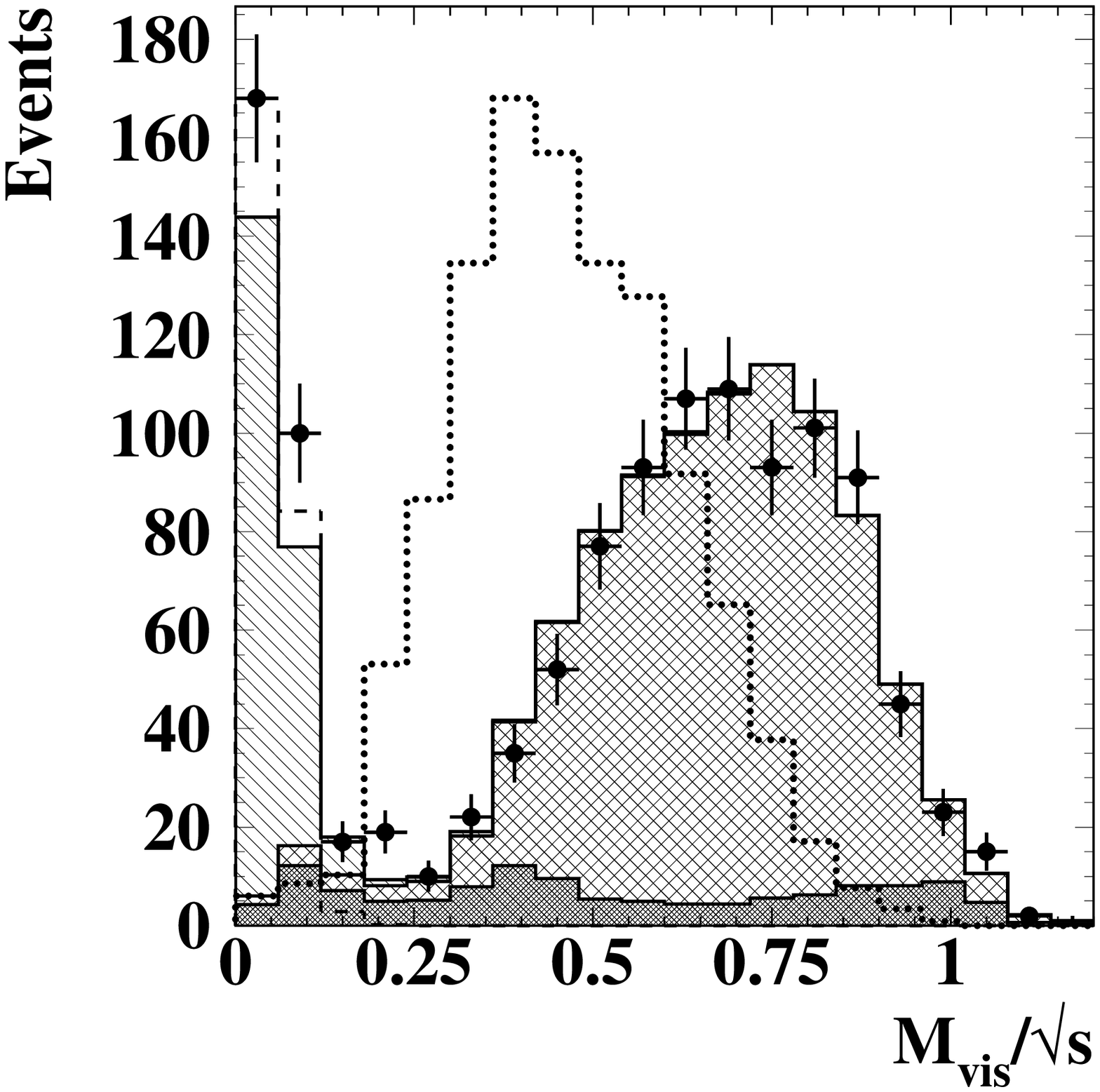}}
\resizebox{0.45\textwidth}{!}{\includegraphics{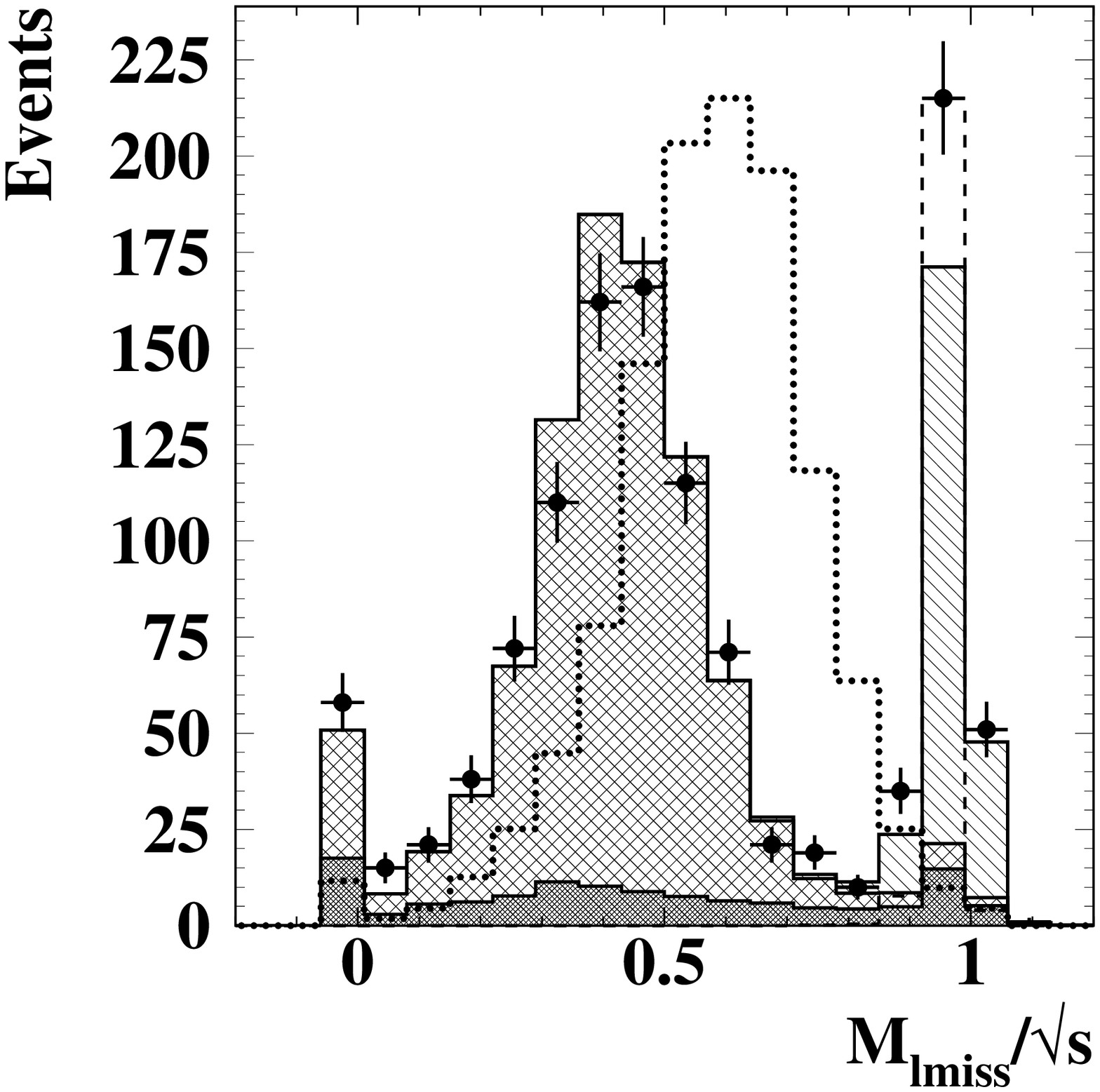}}\\
\vspace{-2ex}
\resizebox{0.45\textwidth}{!}{\includegraphics{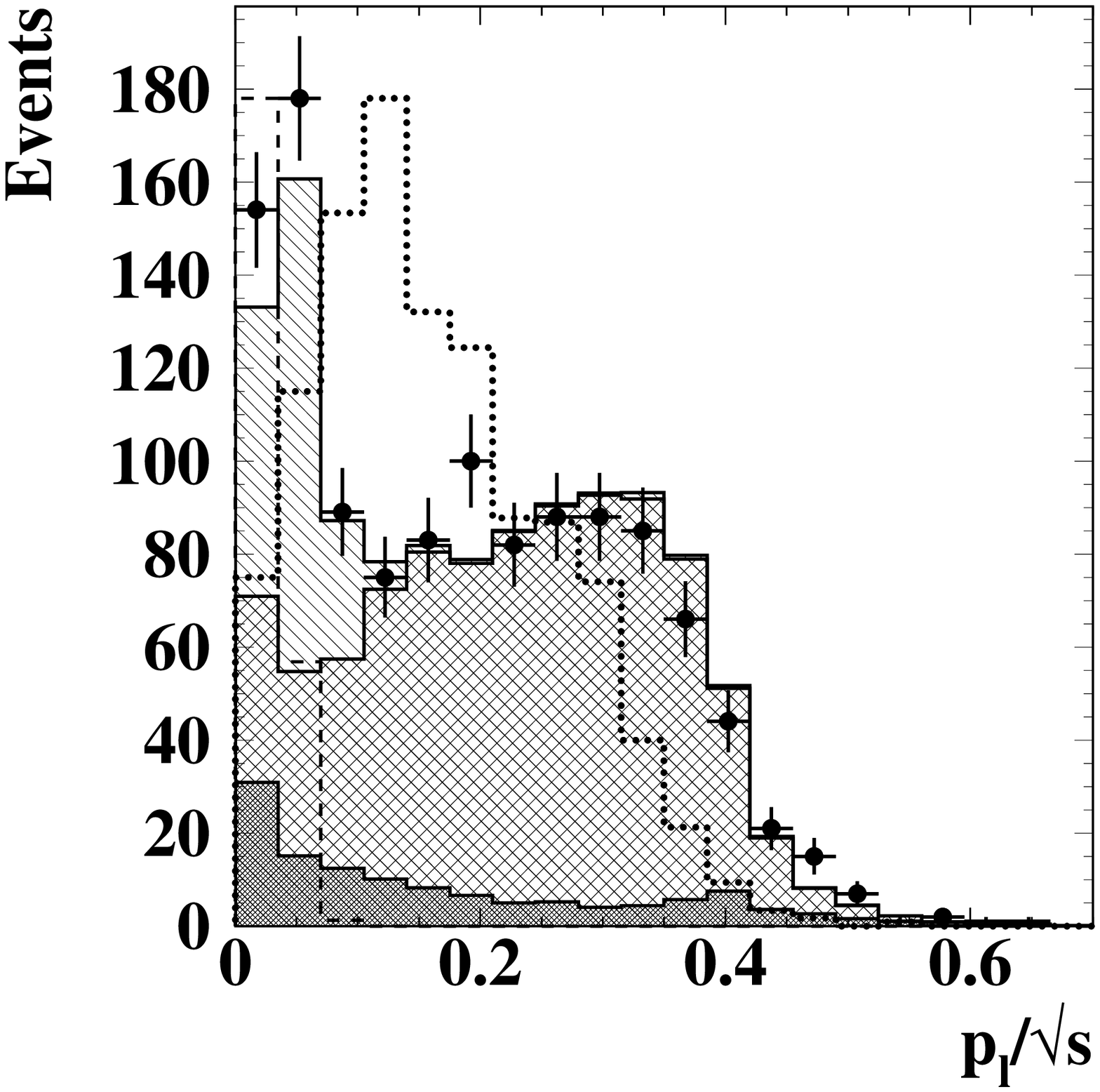}}
\resizebox{0.45\textwidth}{!}{\includegraphics{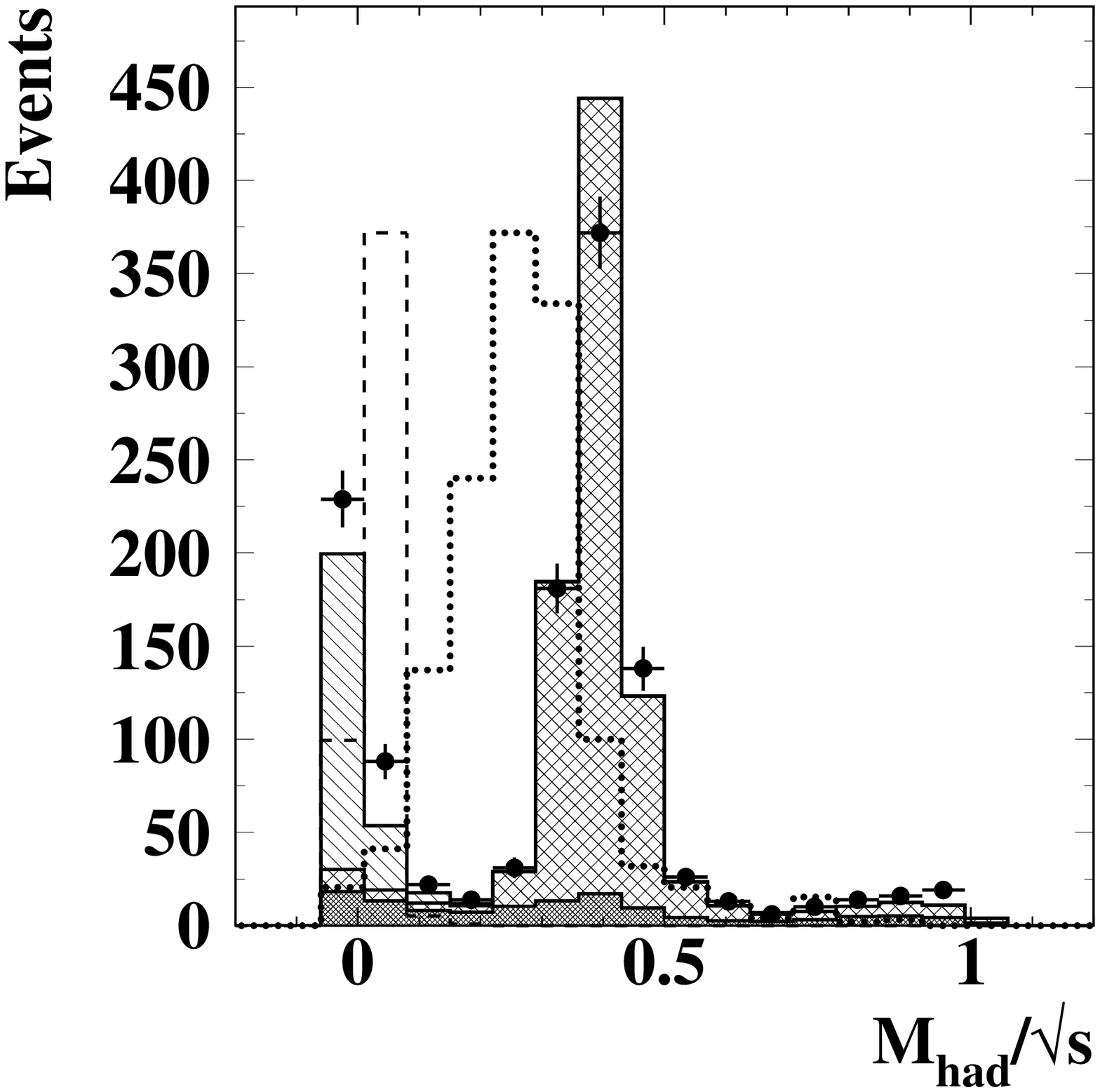}}\\
\vspace{-2ex}
\resizebox{0.45\textwidth}{!}{\includegraphics{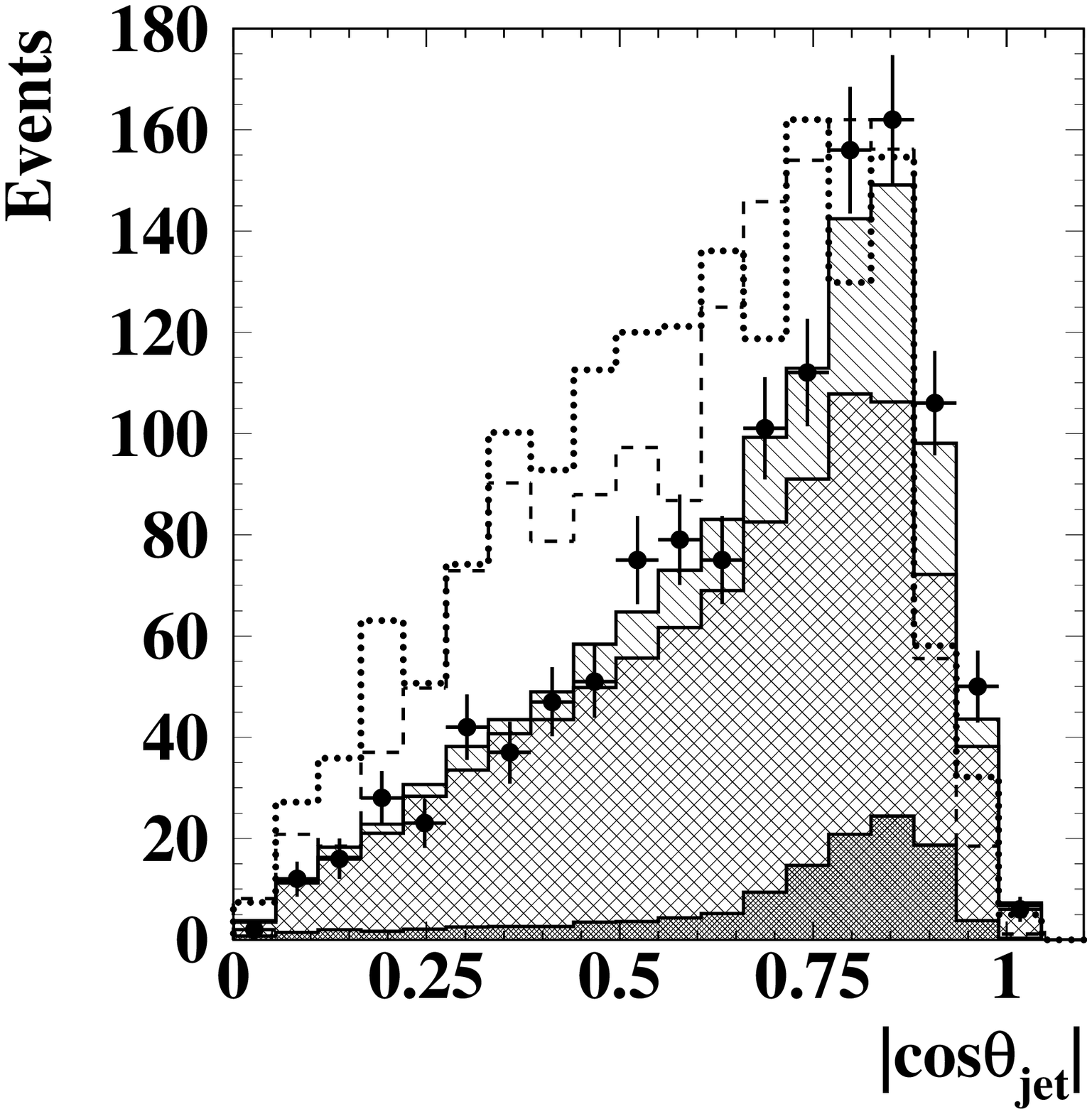}}
\resizebox{0.45\textwidth}{!}{\includegraphics{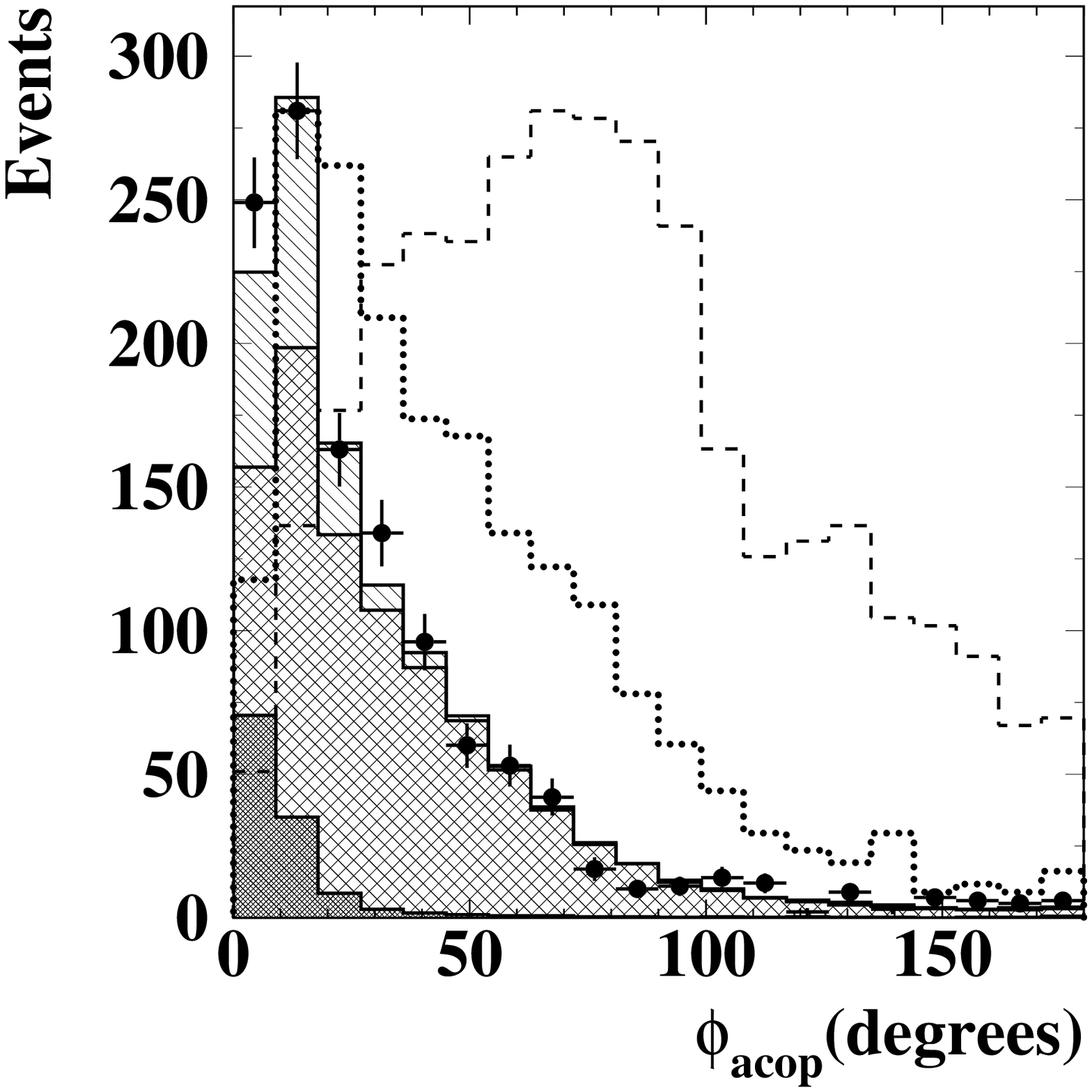}}
\caption{Comparison of data and Monte Carlo distributions of selected
  likelihood variables for the semileptonic chargino analysis, shown
  at preselection level for $\sqrt{s}=207$~GeV.  The solid histograms
  represent the expected background, with dark grey for
  two-fermion events, cross-hatching for four-fermion events and
  single-hatching for two-photon events.  The points with error bars are
  the OPAL data taken in the year 2000 at an average centre-of-mass
  energy of 206.1~GeV.  Dotted histograms show a ``$\W^+\W^-$-like''
  signal with $(M,\Delta 
  M)=$(80~GeV, 76.3~GeV) and dashed histograms show a
  ``two-photon--like'' signal
  with $(M,\Delta M)=$(85~GeV, 6.9~GeV), both with arbitrary
  normalization. \label{fig:semivar}}
\end{figure}

\begin{figure}[htbp]
\resizebox{0.49\textwidth}{!}{\includegraphics{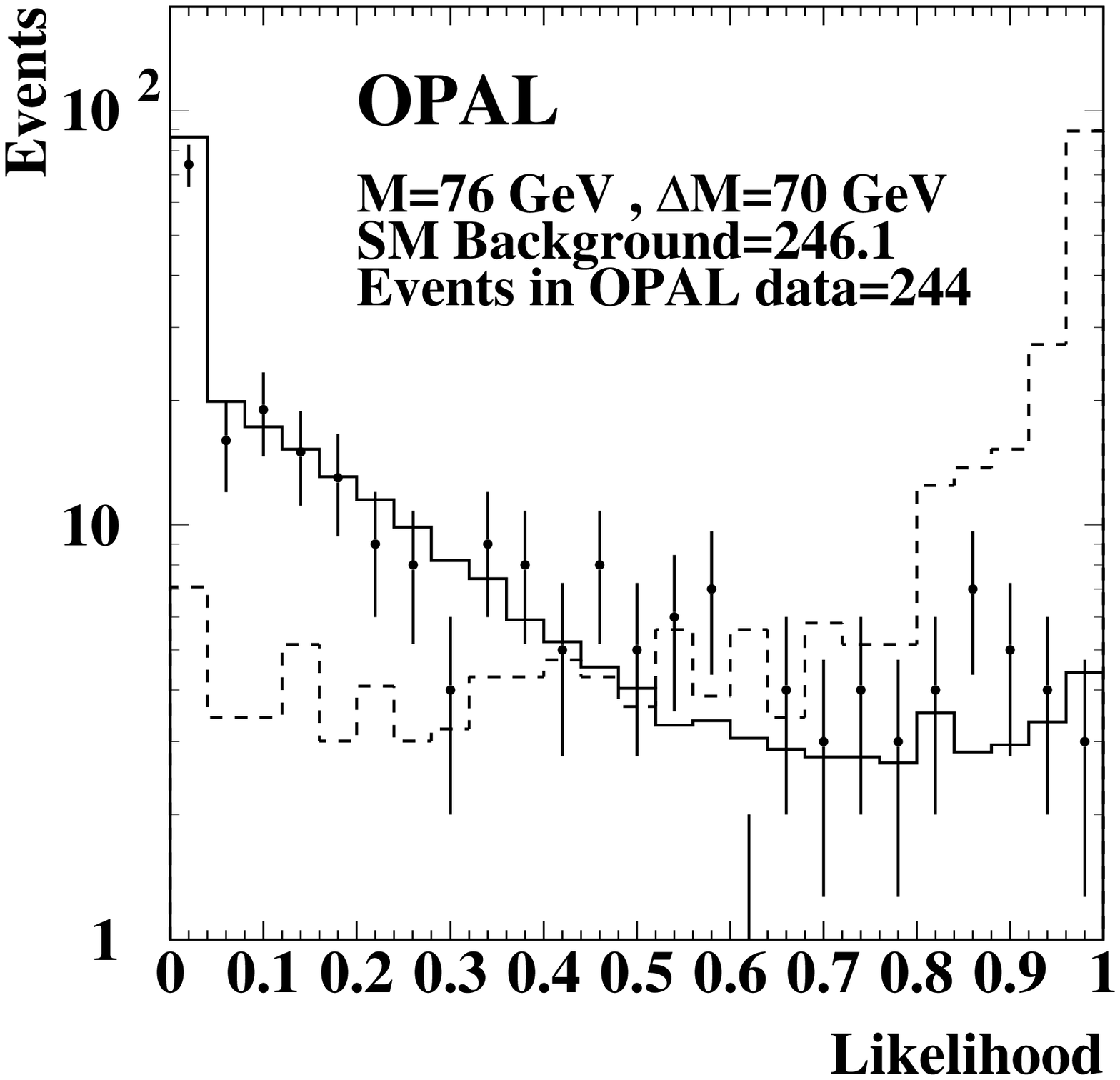}}
\resizebox{0.49\textwidth}{!}{\includegraphics{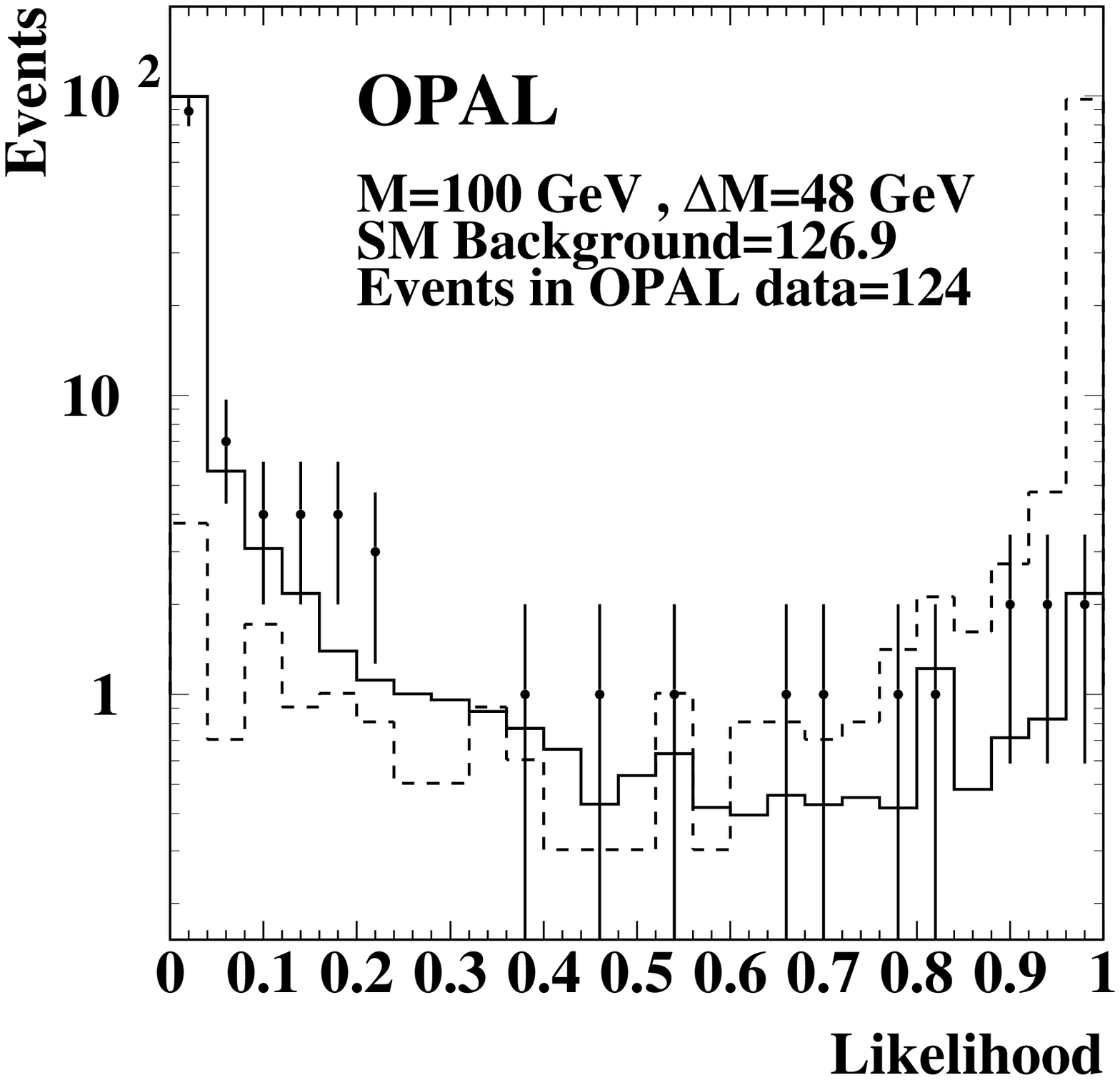}}
  \caption{Sample distributions of $L_R$ for Standard Model background
    in the semileptonic channel
    (solid histogram), chargino signal with arbitrary normalization
    (dashed histogram) and OPAL 
    data at $\sqrt{s}=207$~GeV (points with error bars) at $(M,\Delta
    M)=$ (76~GeV, 70~GeV) and (100~GeV,
    48~GeV).\label{fig:lhsemi}} 
\end{figure}

The following variables are used in the likelihood.
Distributions of these variables are shown at preselection level in
Figure~\ref{fig:semivar}.
\begin{itemize}
  \item $M_{\mathrm{vis}}/\sqrt{s}$ discriminates against
    four-fermion background, especially ${\W}{\e}\nu\to
    {\q}{\bar{\q}}{\e}\nu$ with 
    a fake lepton in one of the two jets; 
  \item $M_{\ell\mathrm{miss}}/\sqrt{s}$ helps to reject
${\W}^+{\W}^-\to {\q}{\bar{\q}}\ell\nu$; 
  \item the scaled magnitude of the momentum of the lepton candidate, 
$p_\ell/\sqrt{s}$, also helps to reject ${\W}^+{\W}^-\to
{\q}{\bar{\q}}\ell\nu$ ; 
  \item $M_{\mathrm{had}}/\sqrt{s}$ also helps to reject ${\W}^+{\W}^-$
    events; 
  \item $a_T$, the magnitude of the component of the missing momentum
perpendicular to the event thrust axis, scaled to the beam energy,
discriminates against two-photon and $\tau^+\tau^-$ events; 
  \item when the event, excluding the lepton candidate, is forced into
    two jets, $\phi_{\mathrm{acop}}$, the acoplanarity angle of the two
    jets, rejects two-photon backgrounds and events with ISR;
  \item $|\cos\theta_{\mathrm{jet}}|$, where $\theta_{\mathrm{jet}}$
    is the polar angle of the jet closer to the beam axis,
    rejects two-photon and two-fermion backgrounds;
  \item the scaled energy of the less energetic jet,  
    $E_{\mathrm{j}}/\sqrt{s}$, 
    helps to remove some two-fermion backgrounds and events with
    poorly reconstructed jets.
\end{itemize}
The following variables are used only for additional
mass-dependent cuts.
\begin{itemize}
  \item the energy of the more energetic jet, scaled to the beam
energy, discriminates against four-fermion backgrounds; 
  \item $p_{\mathrm{T}}^{\mathrm{miss}}/\sqrt{s}$, discriminates
                                against two-photon backgrounds; 
  \item $|\cos\theta_a|$, where $\theta_a$ is the polar angle of the
    $a_T$ vector, rejects two-photon and $\tau^+\tau^-$ backgrounds;
  \item $|\cos{\theta_{\mathrm{miss}}}|$ rejects two-photon and ISR events.
\end{itemize}
Signal efficiencies for semileptonic events after these additional cuts,
including the cuts on the likelihood variables,
vary from about 0.08\% at $\Delta M=3$~GeV to about 30\% for
$5\mbox{ GeV}<\Delta M<10\mbox{ GeV}$ to about 60\% for $\Delta M>10$~GeV.
The main background surviving after the cuts in the region
where $\Delta M>10$~GeV comes from four-fermion processes with a
cross-section of about
1~--~2.5~pb.  About 0.5~--~0.7~pb of ${\e}^+{\e}^-\tau^+\tau^-$ two-photon
events survive for all values of $\Delta M$ and
this is the dominant background for $\Delta M<10$~GeV.  There are also
about 0.05~--~0.1~pb of ${\e}^+{\e}^- {\q}{\bar{\q}}$ two-photon events for
all $\Delta M$ and about 0.1~pb each of tau pairs and
${\q}{\bar{\q}}(\gamma)$ events for $\Delta M>10$~GeV. 
Some examples of likelihood distributions are shown in
Figure~\ref{fig:lhsemi} for signal, data and expected Standard Model
background. 
\subsection{Neutralino Associated Production Hadronic Selection
  \label{sec:nnsel}}
The final state topology for $\tilde\chi^0_2\tilde\chi^0_1$ production
with a hadronic decay of the virtual ${\Z}^0$ contains two hadronic
jets and missing energy
from the two stable lightest neutralinos in the final state.
The analysis is very similar to the hadronic decay channel analysis 
for chargino pair production.
Some changes are made to reflect the difference in event topology.
The following variables are used:
\begin{itemize}
  \item Cut Variables --- as for the chargino analysis except that
    no cuts are made on the number of jets in the event, on
    $E_{\ell'}$ nor on $E_{\mathrm{jet} 1}$;
  \item Likelihood Variables:
\begin{itemize}
  \item $|\cos\theta_{\mathrm{miss}}|$, $p_{\mathrm{T}}^{\mathrm{miss}}$,
    $\phi_{\mathrm{acop}}$ and  $M_{\mathrm{vis}}$ are used, as for the
    hadronic channel of the chargino analysis;
  \item $R_{\mathrm{fwd}}^{\mathrm{vis}}$, the ratio of energy
    deposited in the forward region, 
      $|\cos\theta |>0.9$, to the total visible energy, helps to remove
      ISR and two-photon events; 
  \item $M_{\mathrm{vis}}/E_{\mathrm{vis}}$, the ratio of the visible
    invariant mass to the total visible 
    energy gives a measure of the ``softness'' and helps to remove
    ${\q}{\bar{\q}}\nu{\bar \nu}$ events;
  \item $d_{23} \equiv \sqrt{y_{23}E_{\mathrm{vis}}^2}$,  helps to remove
    backgrounds containing more than two jets.
\end{itemize}
\end{itemize}
\section{Extended Likelihood Method, Combination of Channels and
Energies}
  Once the discriminant value $L_R$ has been calculated for all
  events, as described in Section~\ref{sec:analysis}, it can be used
  either in an optimized cut analysis or in an ``extended'' likelihood
  analysis, as described in~\cite{acoplanarold}.
  The optimized cut analysis has the disadvantage of being difficult
  to use effectively in a search such as this one, where the signal
  cross-section and hence the signal purity are unknown, and the
  number of selected events is sensitive to the position of the cut so
  that small modifications to the analysis can occasionally have a
  very large effect on the limits obtained.

  In the case of an unknown signal cross-section, an
  ``extended'' likelihood technique is more appropriate, as it is
  optimal for all luminosities and signal cross-sections.  In this
  case, the distributions of the discriminant $L_R$ are evaluated for
  signal and background samples and the value of $L_R$ is evaluated
  for each data event.  A single likelihood $L$ can then
  be constructed which describes the probability, given
  the signal and background $L_R$ distributions $S$ and $B$, of
  observing $N$ events passing all cuts with likelihoods $L_{R_i}$:
  \begin{equation}
    L=\frac{e^{-\nu}\nu^{N}}{N!}\prod^N_{i=1}P(L_{R_i}|B,S)  \label{eqlh}
  \end{equation}
  where $\nu\equiv\mu_B+\epsilon{\cal L}\omega\sigma_S$ is the expected
  number of background events passing the cuts, $\mu_B$, plus the
  expected number of preselected signal events for a signal
  cross-section $\sigma_S$, and integrated luminosity ${\cal L}$,
  assuming signal efficiency $\epsilon$.
  In the extended likelihood method, efficiencies are evaluated
  at the end of the cut-based selection.  All candidates passing
  the preselection and additional cuts contribute to the final limit
  calculation, with weights given by their likelihood of being signal.
  There is no cut on the likelihood value, and therefore no
  sensitivity to cut position, making the limit very robust.

  A weighting factor,
  $\omega=\frac{\sigma(\sqrt{s_i})}{\sigma(\sqrt{s_{\mathrm
  {limit}}})}$, is included to allow the cross-section limit to be set
  at a particular centre-of-mass energy using data taken
  at many different energies.  Since charginos and neutralinos are
  fermions, $\sigma$ is assumed to scale as $\beta/s$.
  With some algebraic manipulation~\cite{acoplanarold}
   Equation~\ref{eqlh} can be rewritten in the form: 
  \begin{equation}
    \ln{L(\sigma_S)}=
     -(\mu_B+\epsilon{\cal L}\omega\sigma_S)+\sum^N_{i=1}
     \ln[\mu_B B(L_{R_i})+\epsilon{\cal L}\omega\sigma_S S(L_{R_i})]
  \end{equation}
  This can be calculated separately for each channel
  with data taken at several centre-of-mass energies.
  The 95\% confidence level limit on the signal cross-section
  $\sigma_S(\sqrt{s_{\mathrm{limit}}})$ at
  $\sqrt{s_{\mathrm{limit}}}=208$~GeV is 
  computed using Bayes' theorem with a uniform prior in $\sigma_S$ by
  solving: 
  \begin{equation}
    0.95=\frac{{\displaystyle\int_0^{\sigma_{95}^{\mathrm{limit}}}} \prod_{i=1}^{N_{{\mathrm{ channels}},
  \sqrt{s}}} L_i(\sigma_S^{\mathrm{limit}}) \mathrm{d}\sigma_S^{\mathrm{limit}}
              }{{\displaystyle\int_0^{\infty}} \prod_{i=1}^{N_{{\mathrm{channels}},
  \sqrt{s}}} L_i(\sigma_S^{\mathrm{limit}}) \mathrm{d}\sigma_S^{\mathrm{limit}}
              }\label{eq:limit}
  \end{equation}

Limits on the signal cross-sections are calculated according to
Equation~\ref{eq:limit} separately  
for each channel in the chargino analysis and for all channels
combined assuming 100\% ${\W}^{\ast}$ branching ratios.
In all cases, all the energy bins from Table~\ref{tab:lumi} are
combined.
Data from a particular centre-of-mass energy contribute to the
combination only up to the kinematic cut-off for that energy, about
$(\sqrt{s}-1\mbox{ GeV})/2$.
These cut-offs are then rounded down to the nearest 0.5~GeV.
Thus, all data with energies above 192~GeV contribute up to a
chargino mass of 95.5~GeV, only those with energies above 196~GeV
contribute up to 97.5~GeV and so on until the limit for chargino
masses between 103 and 103.5~GeV comes only from the data taken at an
average centre-of-mass energy of 208~GeV.
The limit is calculated on the
cross-section at $\sqrt{s_{\mathrm{limit}}}=208$~GeV.

For points where the background likelihood histograms contain empty
bins, the extended likelihood method cannot be used and a simple
Poisson probability is evaluated, based on the observed number of
events passing the preselection and additional cuts and the number
expected from background Monte Carlo.  This happens for most of the
points with $\Delta M< 15$~GeV and for some points with $\Delta
M$ between 15 and 30~GeV. 

\section{Systematic Errors}
There are numerous possible sources of systematic errors for these
analyses.  In general, however, it is found that even large
errors on the efficiencies and backgrounds have rather a small effect
on the limits, so only the larger effects are considered.
Some sources of potentially large systematic uncertainties are:
\begin{itemize}
  \item limited Monte Carlo statistics for signal efficiency evaluation;
  \item limited Monte Carlo statistics for background evaluation;
  \item mismodelling of likelihood variables from incomplete or
    incorrect Monte Carlo simulation and detector calibration effects; 
    this includes, among other things, the effect of ignoring
    interference between multiperipheral and other 
    four-fermion processes containing an electron and a positron in the
    final state, for which a specific check was also done, and the
    effect was found to be negligible for the background, given the
    statistical errors on the multiperipheral background samples;
    a relative error of 5~\% was assigned to the signal efficiencies;
  \item a further 1~\% relative systematic error was assigned to the
    signal efficiencies to account for the uncertainty in the
    smoothing procedure applied to the reference histograms;
  \item interpolation of signal histograms to obtain a finer grid in
    $\Delta M$ than was generated; a test was done using Monte Carlo
    samples generated at intermediate values of $\Delta M$ and a
    relative error of 1~\% was assigned to the efficiencies;
  \item effects of having signal generated only at three $\sqrt{s}$ to
    evaluate the efficiency at all energies and using rather crude
    $\beta/s$ models to evaluate the cross-sections of these
    fermionic signals at other $\sqrt{s}$ were evaluated by repeating
    the whole analysis using reference histograms created from signals
    and backgrounds generated at 196~GeV instead of 207~GeV (in the
    kinematic range where this was possible) and
    comparing the results with the standard analysis; a conservative
    relative error of 5~\% was assigned to the efficiencies;
  \item uncertainties in the estimation of the integrated luminosity
    of about 0.2--0.3\%.
\end{itemize}
The evaluation of the uncertainties due to the first three items in
the list is described in detail in
subsections~\ref{subsec:sigeff}-\ref{subsec:mismod}.  Detector
efficiency losses due to occupancy or due to the cuts on energy in the
forward detectors, which reduce the effective integrated luminosity by
about 3\% were included simply by reducing all integrated luminosities by
3\%.  The following potential sources of systematic uncertainty were
also considered and found to be negligible compared with those
listed above:
\begin{itemize}
  \item effects of having two-photon background samples generated only
    at some of the $\sqrt{s}$ used and simply assuming that the
    cross-sections of these processes are proportional to
    $\log{s}$ to evaluate them at other energies.
  \item effects of using reference histograms generated at a single
    $\sqrt{s}$ for evaluating the likelihood at all energies; a
    cross-check was done using reference histograms generated at
    $\sqrt{s}=196$~GeV;
\end{itemize}

All errors are evaluated by estimating the
uncertainties on the rates of events passing a cut, since evaluating 
changes in the shapes of the $L_R$ distributions at every point on the 
$(M,\Delta M)$ grid for every possible systematic 
is not a practical approach.

\subsection{Statistical uncertainties  \label{subsec:sigeff}}
The statistical uncertainties on the signal efficiency and on the
background estimates due to the finite number of Monte Carlo events
generated are evaluated after the preselection and additional cuts for each
mass grid point.  

The most important sources of background are four-fermion final
states, ${\q}{\bar{\q}}(\gamma)$ and $\tau^+\tau^-$ two-fermion final
states, and two-photon processes, principally  ${\e}^+{\e}^-\tau^+\tau^-$
and untagged ${\e}^+{\e}^-{\q}{\bar{\q}}$.  The statistical errors 
on the two- and four-fermion final states are very small, since the
integrated luminosity of the generated Monte Carlo is about 50 to 100
times the integrated data luminosity at most centre-of-mass energies;
however, the two-photon processes have very large cross-sections and
it was only possible to generate about 8 to 15 times the integrated
data luminosity.  
The statistical errors on each background are then added in
quadrature to obtain the statistical error on the total background at
each mass grid point for each centre-of-mass energy and in each
analysis channel.
The statistical errors on the background are
therefore relatively high in the low $\Delta M$ region, which is
dominated by two-photon processes, and almost negligible for $\Delta M
> 10$~GeV.

\subsection{Uncertainties due to variable mismodelling\label{subsec:mismod}}
Systematic uncertainties due to differences between the
data from the OPAL detector and the Monte Carlo models are
estimated after an optimized cut on $L_R$.
The $L_R$ cut is optimized by selecting the bin boundary on
the $L_R$ distributions for signal and background which gives the
smallest expected value of $N_{95}$ (the most signal events there
could be, at a confidence level of 95\%, given the expected signal and
background distributions and assuming a number of observed
events equal to the expected number of background events).  
The cut
is optimized for the default $L_R$ distributions, and the number of
events passing it is evaluated with modified $L_R$ distributions made
from reference histograms where variables have been shifted or
smeared, to determine the systematic effects on the expected background
levels if the background Monte Carlo differed from the data.  

For the evaluation of systematic effects due to variable mismodelling,
reference histograms of the likelihood variables are constructed
after the general preselection, with none of the additional
mass-dependent cuts.  Histograms for the sum of all the expected
backgrounds are compared with histograms of the data.  The peaks in
the histograms are fitted with Gaussians for both the data and the Monte
Carlo.  The differences between the peaks are compared with the
statistical uncertainties on each background.
If a distribution has two peaks and one is due mainly to
two-photon events and the other mainly to four-fermion events,
which can be determined by looking at the Monte Carlo data for each
background individually, the comparison for each background is made
only for the relevant peak.
No attempt was made to apply this procedure to the variables with
obviously non-Gaussian distributions, such as cosine distributions;
their shapes generally show good agreement between data and Monte Carlo.
If the difference in peak position is larger than the
statistical uncertainty for a particular background, then the
difference minus the statistical uncertainty is taken to be the
systematic uncertainty, and reference histograms are constructed for
each background with the variable shifted by the amount given by this
prescription.  A similar procedure is applied to the comparison of
the widths of the peaks in data and Monte Carlo and used to determine
an amount by which to smear the variables for the reference
histograms.
When these are compared with quantities obtained by comparing OPAL
${\Z}^0$ data with Monte Carlo, which allows a comparison which is not
statistically limited, agreement is reasonably good.  

\subsection{Method for extrapolating systematic uncertainty estimates
  to the full mass grid}
Variations due to systematic uncertainties are evaluated at a
centre-of-mass energy of 207~GeV, which allows mass grid points near
the kinematic limit to be considered, and is an energy at which both
signal and background Monte Carlo samples are available.  It is
assumed that the uncertainties at 207~GeV are typical and can be used at
all centre-of-mass energies.
The errors are evaluated at three mass grid points:
$(M_{\chi^\pm}=80.0\, {\mathrm{GeV}}, \Delta M= 76.3\, {\mathrm{GeV}})$,
where signal events give a final state topology almost identical to
that of on-shell ${\W}$-pair production, $(M_{\chi^\pm}=85.0\, {\mathrm
{GeV}}, \Delta M= 6.9\, {\mathrm{GeV}})$, where the detected particles
from signal events are very soft and the main background is from
two-photon events, and $(M_{\chi^\pm}=100.0\, {\mathrm
{GeV}}, \Delta M=49.2\,  {\mathrm{GeV}})$, which is near the kinematic
limit and has background coming from both four-fermion and
two-fermion production.
Uncertainties are estimated for
each of the main background sources at each of these points.  The
difference between the result with the smear or shift and the original
result is found and the statistical uncertainty for that background
subtracted off.  These differences are then
summed in quadrature for the shifts and smears of all the variables.
The errors at the three points are then compared and, in general, the
error from the most relevant of the three for the particular
background is chosen to be the error; in other words, the errors on the
four-fermion background are generally taken from the point
$(M_{\chi^\pm}=80.0\, {\mathrm{GeV}}, \Delta M= 76.3\,
{\mathrm{GeV}})$, while the errors on two-photon backgrounds come from
the point $(M_{\chi^\pm}=85.0\, {\mathrm {GeV}}, \Delta M= 6.9\,
{\mathrm{GeV}})$. 
\begin{table}[tbp]
  \begin{center}
\caption{Relative systematic errors on major backgrounds in the
  semileptonic channel for the chargino analysis.\label{tab:relsyst}}
\begin{tabular}{|l|r|}
 \hline
 Background & Relative  \\
 Source & Systematic Error \\
 \hline
 ${\q}{\bar{\q}}$ & 0.48 \\
 $\tau^+\tau^-$ & 0.19 \\
 4-fermion & 0.25 \\
 ${\e}^+{\e}^-\mu^+\mu^-$ & 1.0 \\
 ${\e}^+{\e}^-\tau^+\tau^-$ & 0.95 \\
 ${\e}^+{\e}^-{\q}{\bar{\q}}$ & 1.0 \\
 \hline
\end{tabular}
\end{center}
\end{table}
The results of this study are used to give the relative systematic
errors shown in Tables~\ref{tab:relsyst} and~\ref{tab:systhad}.
\begin{table}[tbp]
  \begin{center}
\caption{Relative systematic errors on major backgrounds in the
 hadronic channel for the chargino and neutralino
 analyses.\label{tab:systhad}} 
\begin{tabular}{|l|r|}
 \hline
 Background & Relative  \\
 Source & Systematic Error \\
 \hline
 ${\q}{\bar{\q}}$ & 0.3 \\
 4-fermion & 0.8 \\
 ${\e}^+{\e}^-{\q}{\bar{\q}}$ & 0.5 \\
 \hline
\end{tabular}
\end{center}
\end{table}
The fractional contribution of each background was then calculated for
the rest of the mass grid and the relative errors on each background
scaled according to the contribution of that background at each point.
The systematic errors for each background at each grid point are then
added in quadrature and eventually summed in quadrature with the
statistical error on the background.
The efficiency errors are summed in quadrature and a total relative
systematic error of 7\% is applied to the signal efficiencies and
summed in quadrature with the statistical uncertainties.

\subsection{Incorporating uncertainties into cross-section limits}
Once the systematic errors have been evaluated for the signal
efficiency and background expectation at every point on the mass grid
for all energies, the limit setting code is run twice more over the
data: once with all the efficiencies decreased by their errors and all 
background expectations increased by their errors to give the worst
expected sensitivity, and once with the efficiencies increased and backgrounds
decreased to give the best expected sensitivity.
There are then three limits for each point in the grid:
the standard one, $L95_0$, the one standard deviation ``worst case''
one, $L95_+$, and the one standard deviation ``best case'' one,
$L95_-$.  All are evaluated at a centre-of-mass energy of 208~GeV.
A Gaussian-weighted average of these gives 
the limit convolved with the systematic errors:
\begin{equation}
  L95=\frac{L95_0+\frac{1}{\sqrt{e}}(L95_++L95_-)}{1+\frac{2}{\sqrt{e}}}.
\end{equation}
This is a three-point integral; it is a simple and
robust way to deal with large errors, which tend to make more complex
methods unstable.
Despite the large relative errors on the background expectation, the
limits typically change by only about 0.001--0.02~pb, with an increase
of about 0.1~pb for the hadronic channel in the ``${\W}^+{\W}^-$''
region where $(M,\Delta M)\approx (80\, \mathrm{GeV}, 80\, \mathrm{GeV})$
and an increase of several picobarns along the $\Delta M=3$~GeV
``two-photon'' strip for the semileptonic channel.

\section{Final cross-section limits}
In order to give a visual impression of what the limits might look
like in the absence of a signal, ``expected'' limits are calculated.
For each point on the mass grid, an average is made of
the cross-section limits from a hundred toy Monte Carlo experiments.
In each experiment, a random number of ``events'' is generated using a
Poisson distribution with a mean value of $\mu_B$, the expected number
of background events for the mass grid point.  Each ``event'' is
assigned a random likelihood $L_{R_i}$ from the background likelihood
distribution $B(L_{R_i})$.  The signal cross-section
limit is then calculated just as it is for the data.

\subsection{Chargino Pair Production, Hadronic channel \label{sec:hadr}}
\begin{figure}[htbp]
\resizebox{0.49\textwidth}{!}{\includegraphics{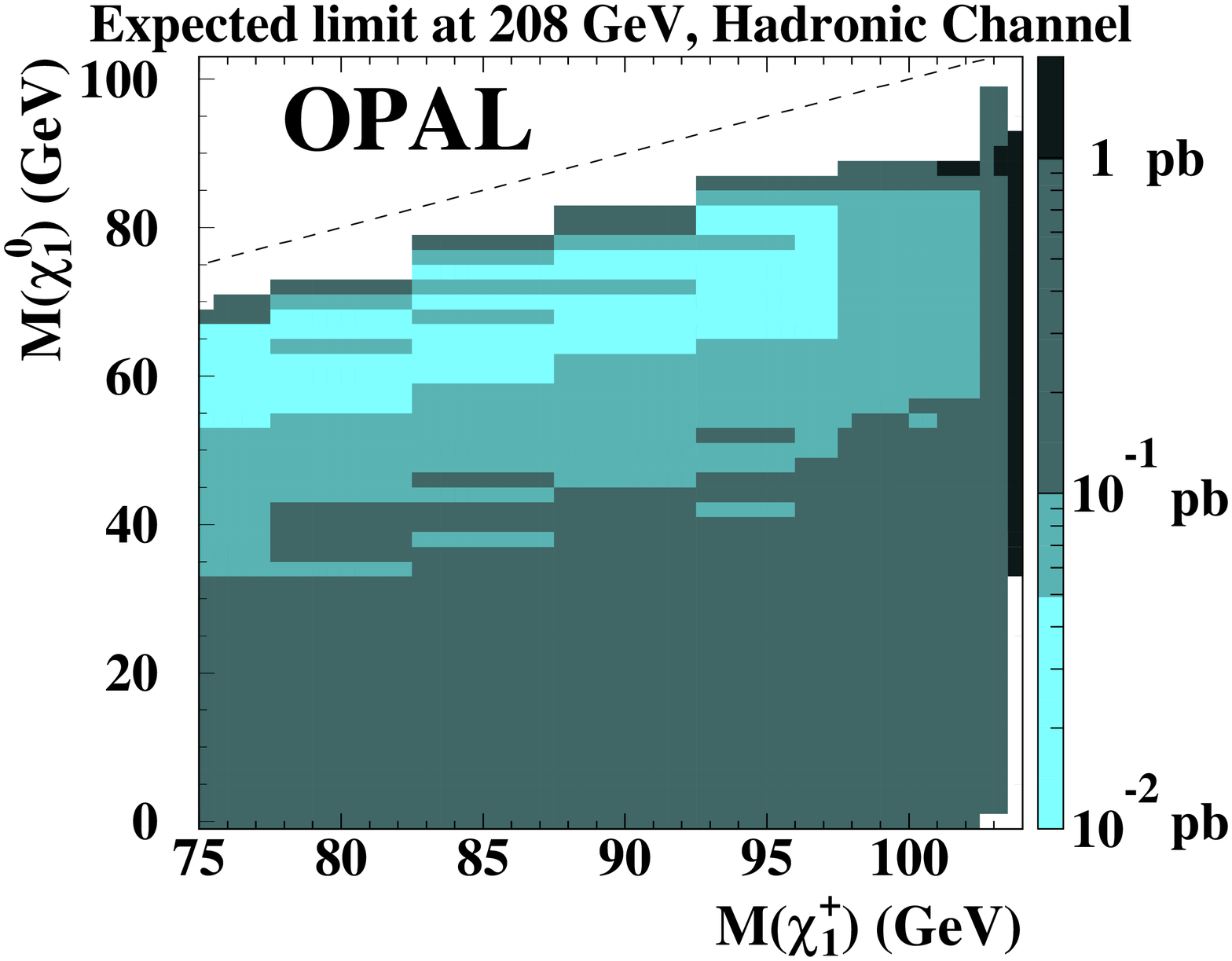}}
\resizebox{0.49\textwidth}{!}{\includegraphics{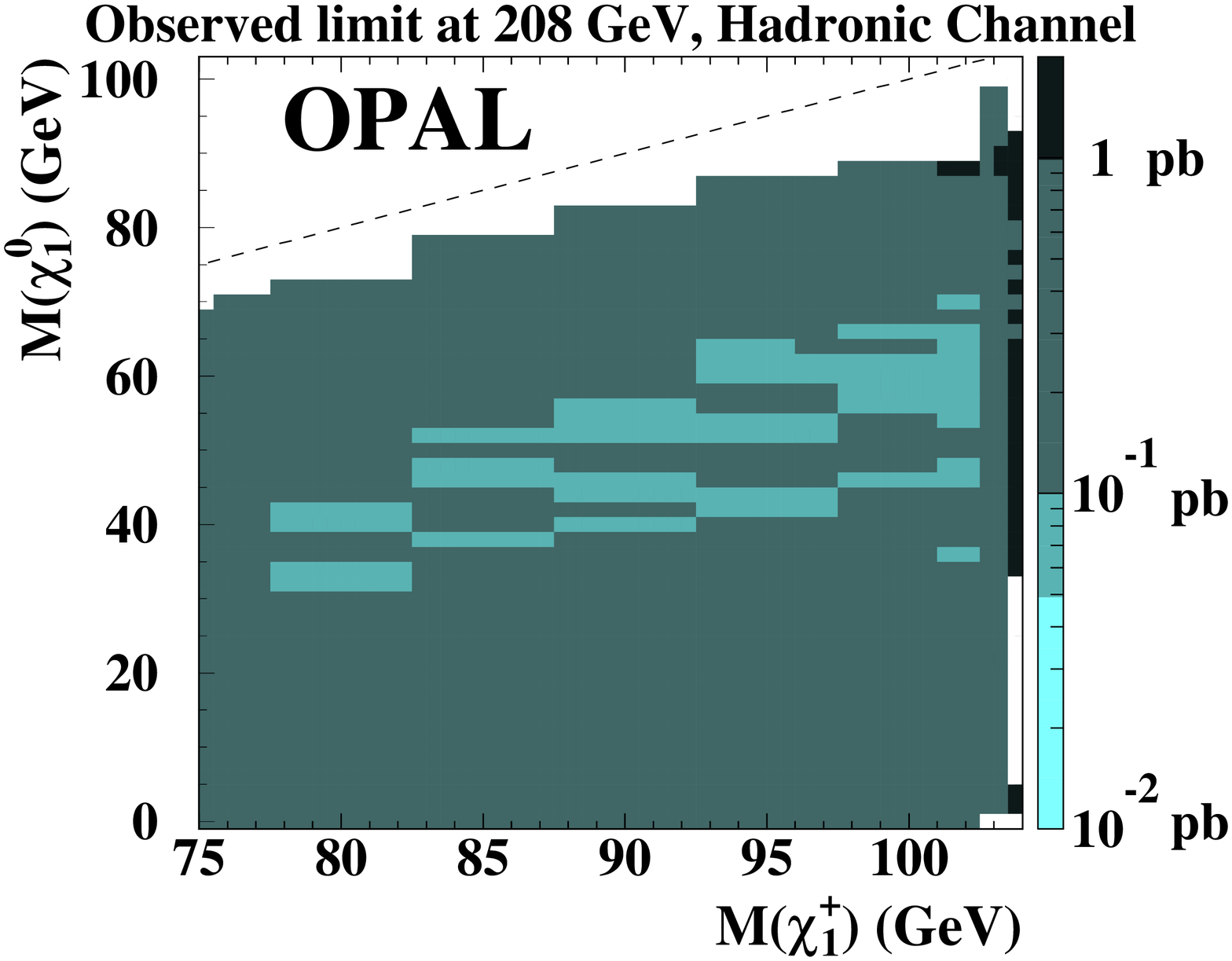}}
\caption{The expected and observed 95\% confidence level limits on
  $\sigma({\e}^+{\e}^-\to\tilde\chi^+_1\tilde\chi^-_1)
  BR^2(\tilde\chi^\pm\to q{\bar  q}\tilde\chi^0_1)$, using
  data taken at 192-209~GeV. White areas indicate that the limit is
  greater than 2~pb or that no limit could be set.\label{fig:hadron}} 
\end{figure}

The limit observed is in general slightly higher than the expected
limit; however, the 95\% confidence level 
limit on $\sigma_{\tilde\chi^+_1\tilde\chi^-_1}
BR^2(\tilde\chi^\pm_1\to {\q}{\bar{\q}}\tilde\chi^0_1)$
(shown in Figure~\ref{fig:hadron}) is below about 0.3~pb almost up to the
kinematic limit for $\Delta M < 3$~GeV and well below 0.1~pb in most of
the parameter space explored, whereas a chargino signal would be
expected to have a cross-section of several picobarns.
\subsection{Chargino Pair Production, Semileptonic channel \label{sec:semil}}
\begin{figure}[htbp]
\resizebox{0.49\textwidth}{!}{\includegraphics{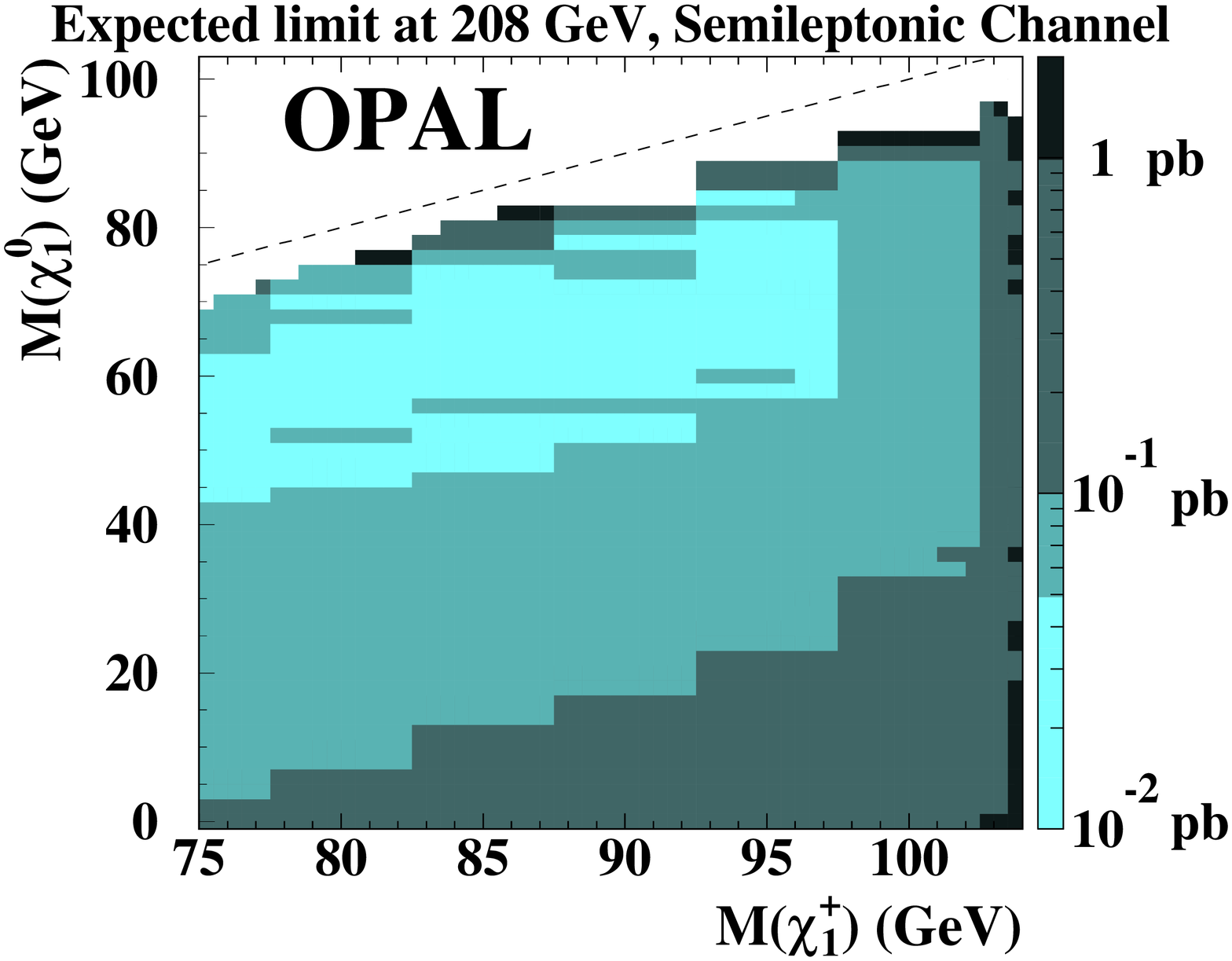}}
\resizebox{0.49\textwidth}{!}{\includegraphics{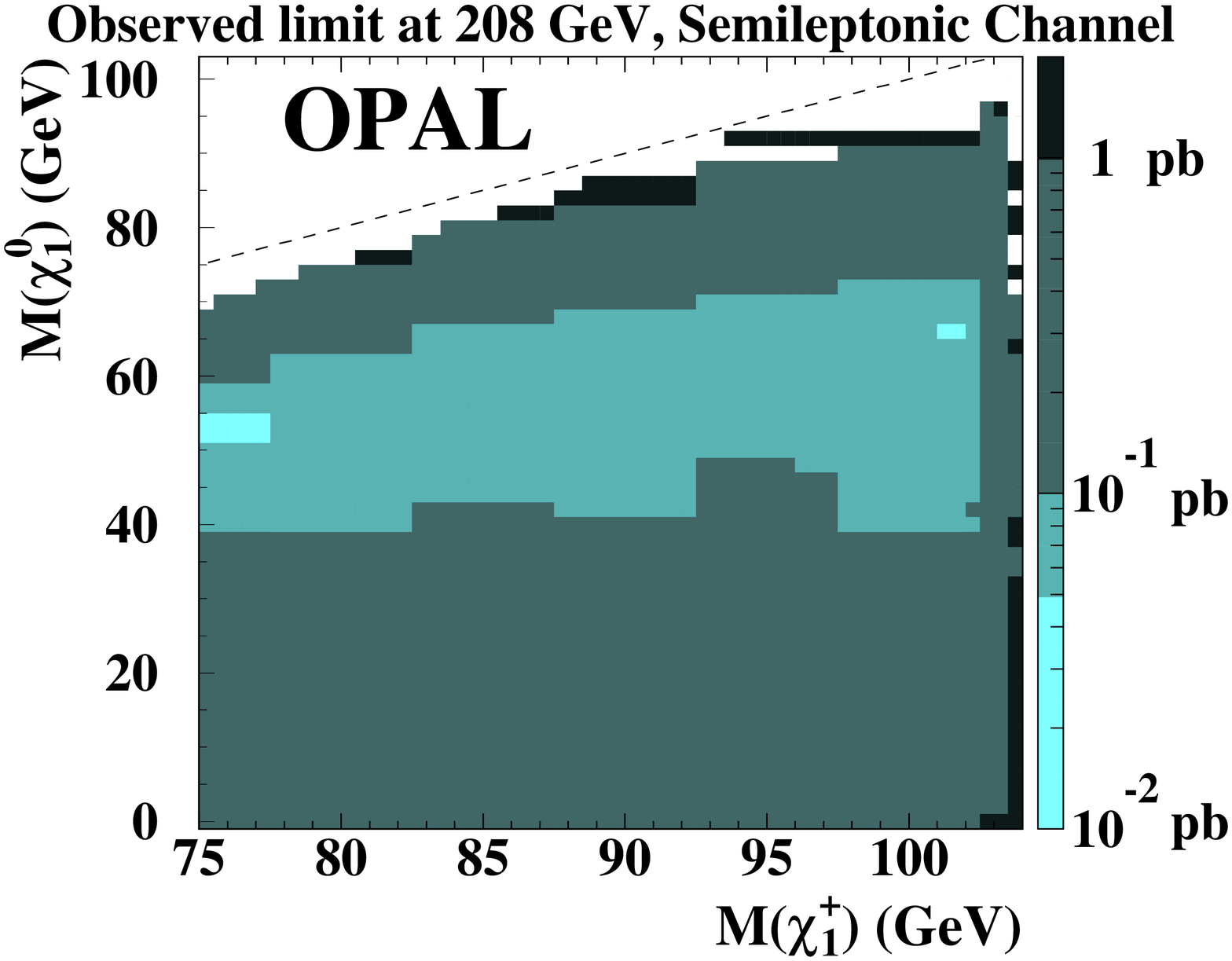}}
\caption{The expected and observed 95\% confidence level limits on the
  cross-section for the final state containing jets and a lepton with
  missing energy, using data taken at 192-209~GeV. White areas
  indicate that the limit is 
  greater than 2~pb or that no limit could be set.\label{fig:semilep}}
\end{figure}

The observed limits are not as stringent as the expected ones;
however, the 95\% confidence 
level limit on $\sigma_{\tilde\chi^\pm_1\tilde\chi^\pm_1}
BR(\tilde\chi^\pm_1\to {\q}{\bar{\q}}\tilde\chi^0_1)BR(\tilde\chi^\pm_1\to
\ell\nu\tilde\chi^0_1)$ (shown in Figure~\ref{fig:semilep}) is below
about 0.3~pb almost up to the 
kinematic limit for $\Delta M < 3$~GeV and well below 0.1~pb in much of
the parameter space explored.
\subsection{Chargino Pair Production, Leptonic channel}
\begin{figure}[htbp]
\resizebox{0.49\textwidth}{!}{\includegraphics{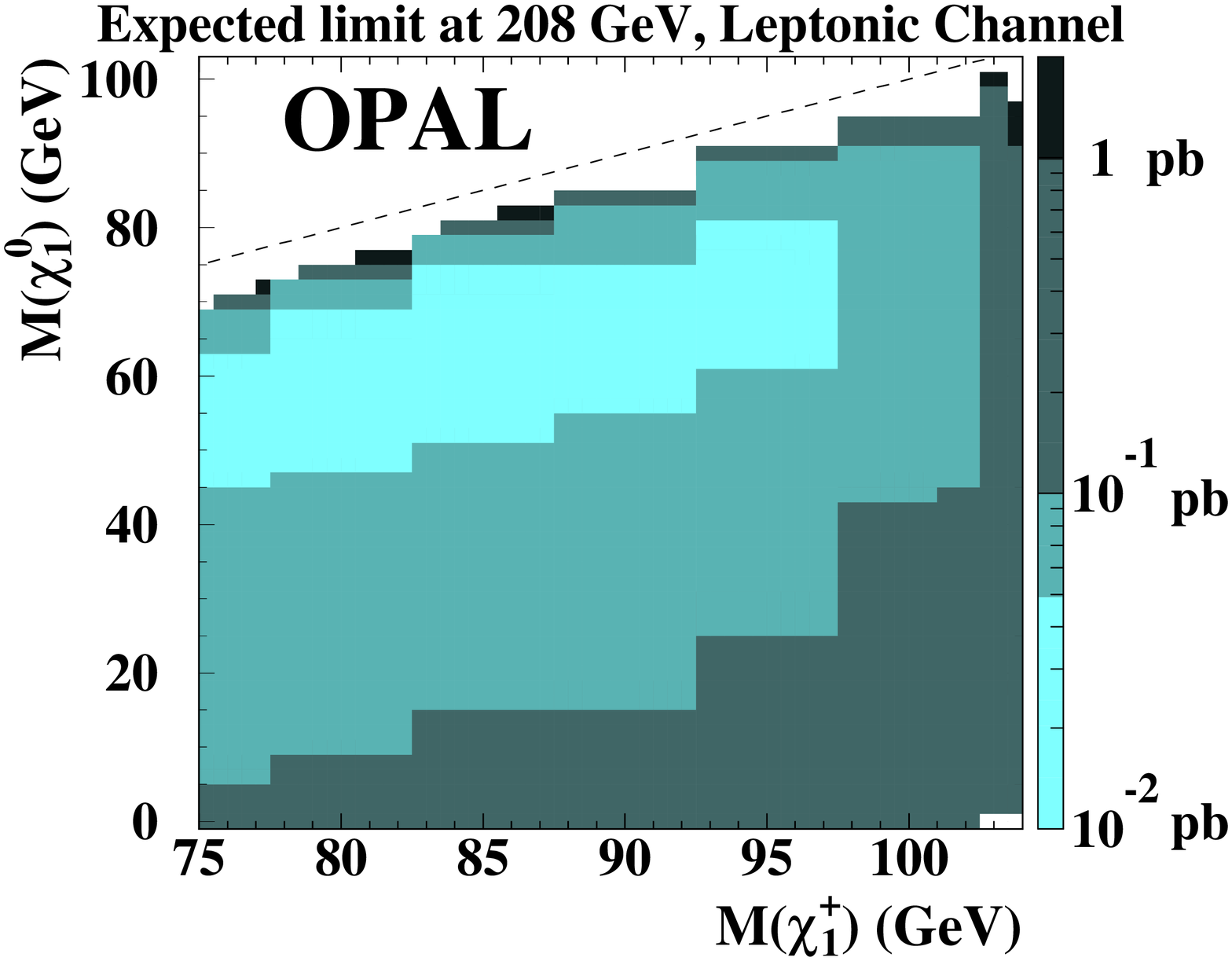}}
\resizebox{0.49\textwidth}{!}{\includegraphics{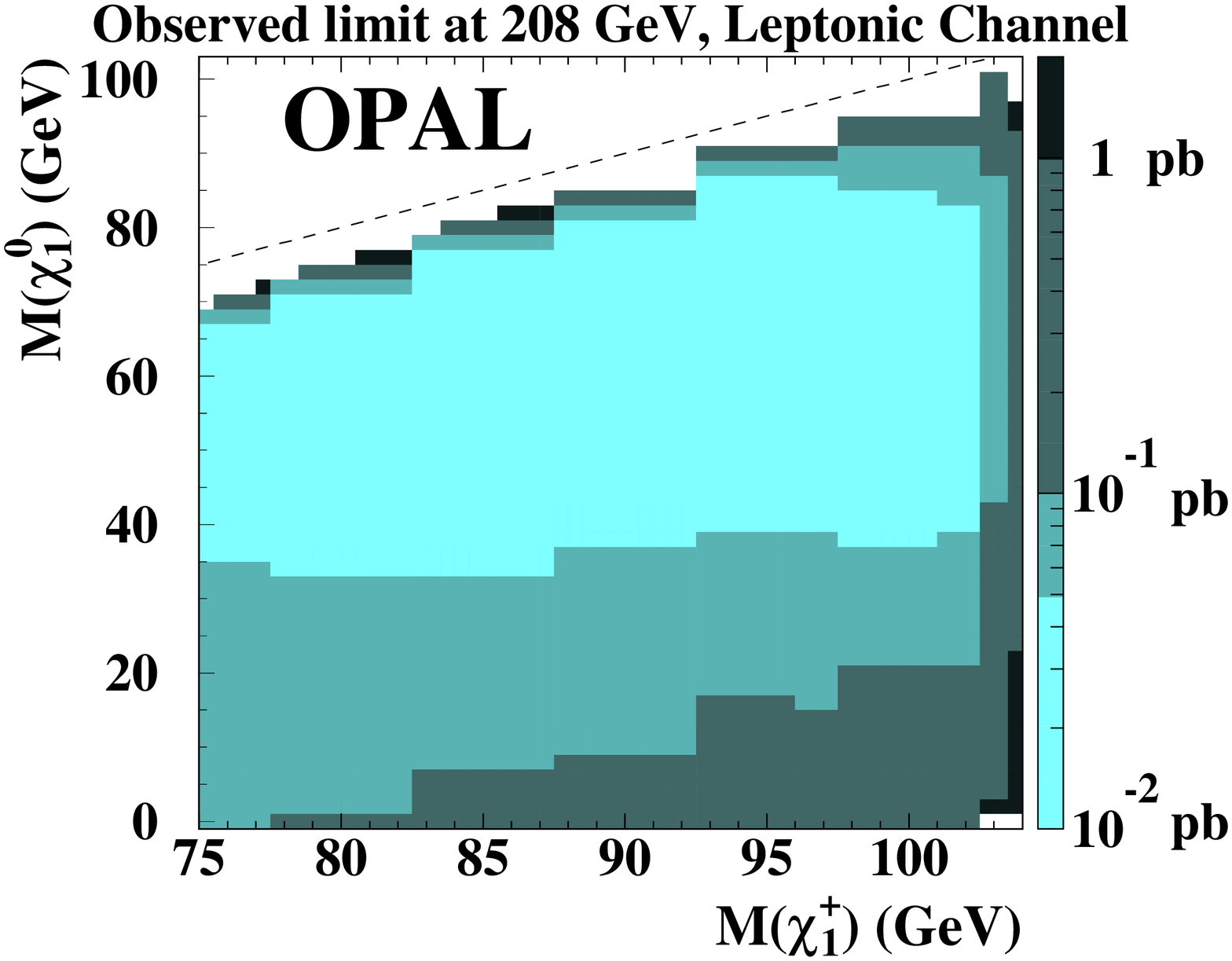}}
\caption{The expected and observed 95\% confidence level limits on the
  product of the cross-section for chargino pair production and the
  square of the branching ratio for $\tilde\chi^\pm$ to leptons and
  missing energy, using data taken at 192-209~GeV. White areas
  indicate that the limit is 
  greater than 2~pb or that no limit could be set. \label{fig:lepton}}
\end{figure}

The observed 95\% confidence level
limits on $\sigma_{\tilde\chi^\pm_1\tilde\chi^\pm_1}
BR^2(\tilde\chi^\pm_1\to \ell\nu\tilde\chi^0_1)$ taken
from~\cite{acoplanar} are shown in
Figure~\ref{fig:lepton} and are slightly better than the expected
limits. 
\subsection{\boldmath Chargino Pair Production, Channels Combined
  Assuming 100\% ${\W}^\pm$ Branching Ratios}
\begin{figure}[htbp]
\resizebox{0.49\textwidth}{!}{\includegraphics{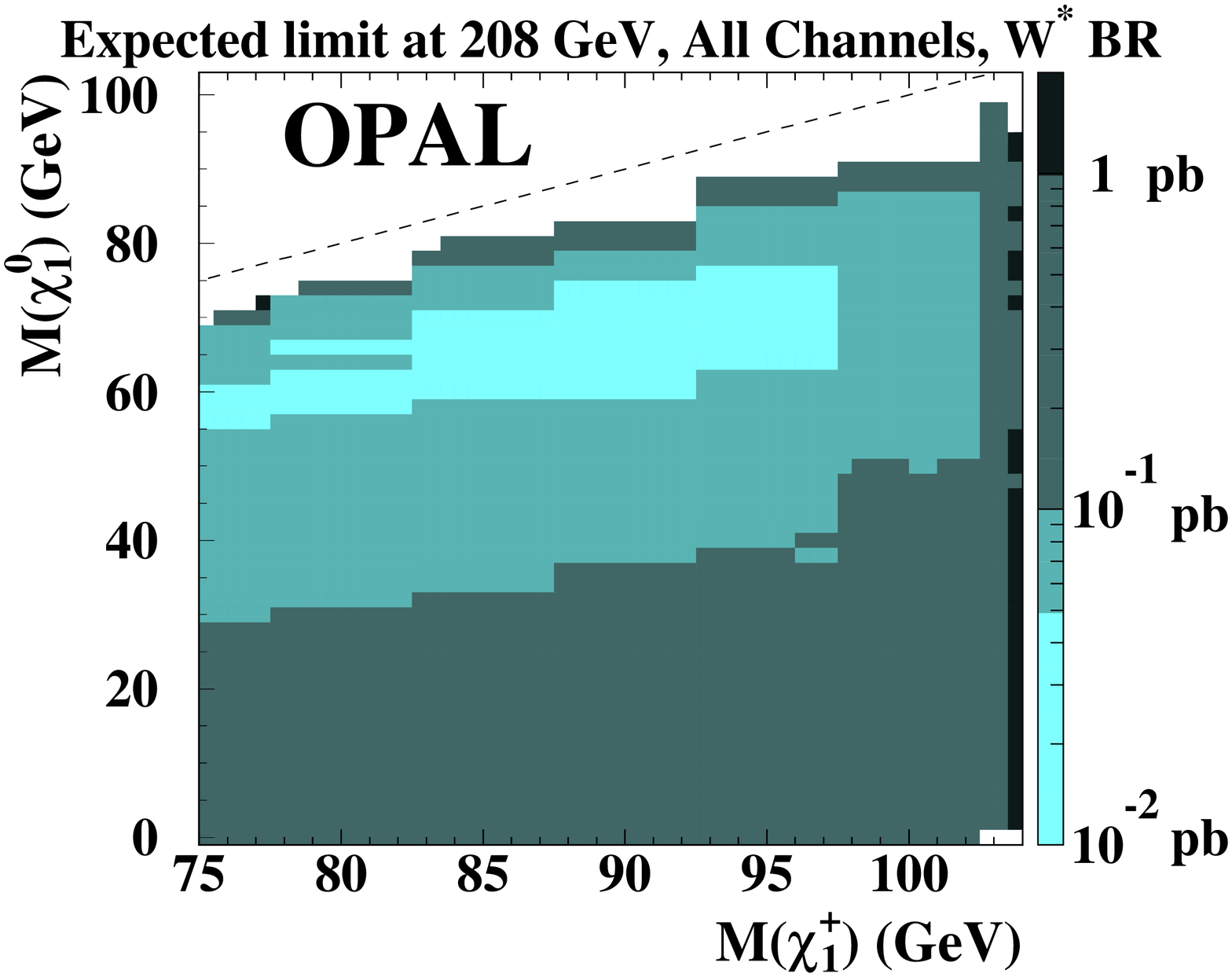}}
\resizebox{0.49\textwidth}{!}{\includegraphics{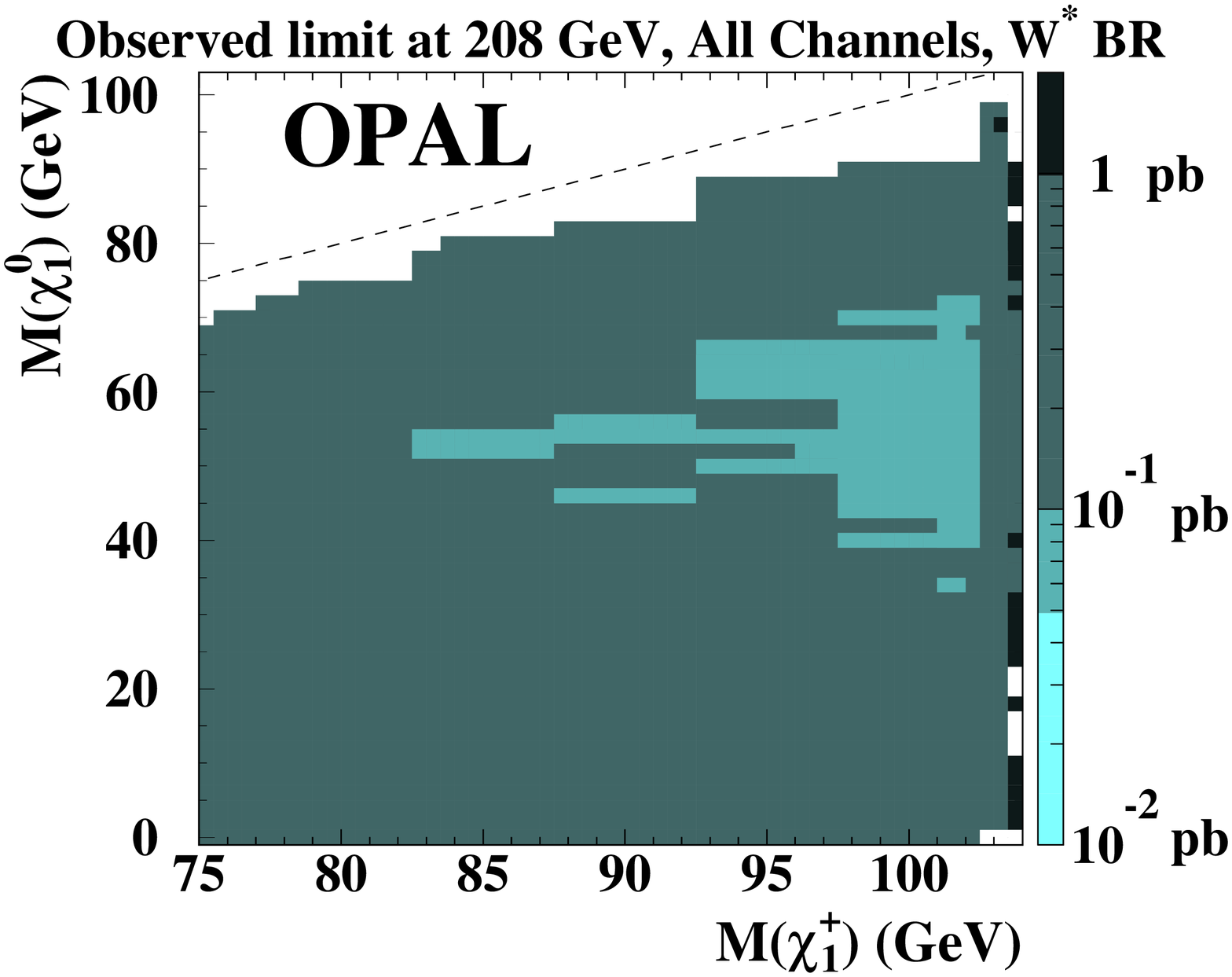}}
\caption{The expected and observed 95\% confidence level limits on the
  cross-section for chargino pair production, assuming 100\% ${\W}^\ast$
  branching ratios for the chargino decays and using all the data sets
  analyzed. White areas indicate that the limit is
  greater than 2~pb or that no limit could be set.\label{fig:wstar}}
\end{figure}

If the lightest neutralino is the LSP, chargino branching ratios to
final states containing ${\q}{\bar{\q}}$ and $\ell\nu$ are the
${\W}^\pm$ decay branching ratios, except for very low values of
$\Delta M$ where the leptonic branching ratio increases rapidly.
The three channels are combined
according to ${\W}^\pm$ branching ratios and these results, which are
generally valid in the absence of light scalar leptons and scalar
neutrinos, are shown in  Figure~\ref{fig:wstar}. 
\subsection{Neutralino Associated Production, Hadronic Channel}
\begin{figure}[htbp]
\resizebox{0.49\textwidth}{!}{\includegraphics{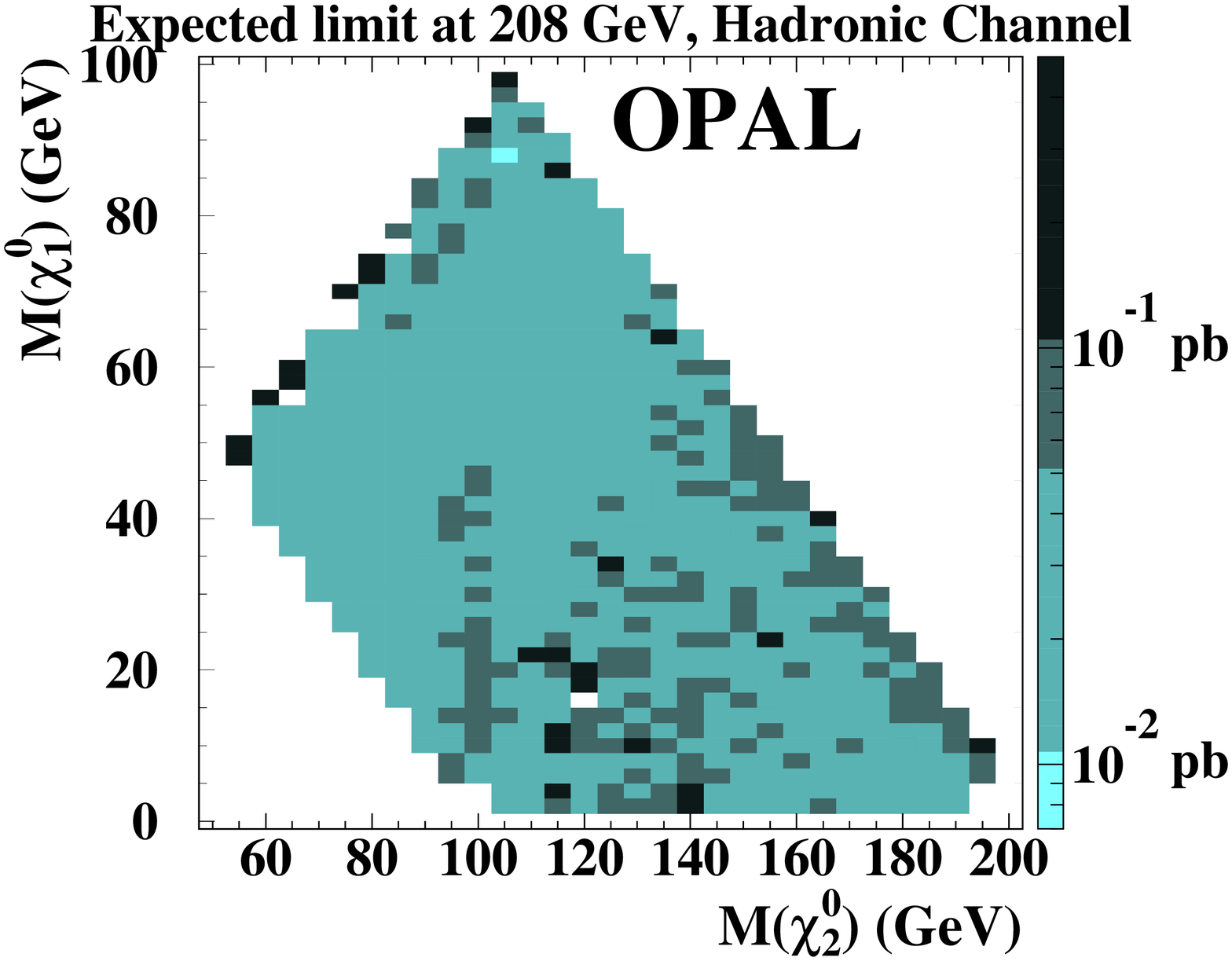}}
\resizebox{0.49\textwidth}{!}{\includegraphics{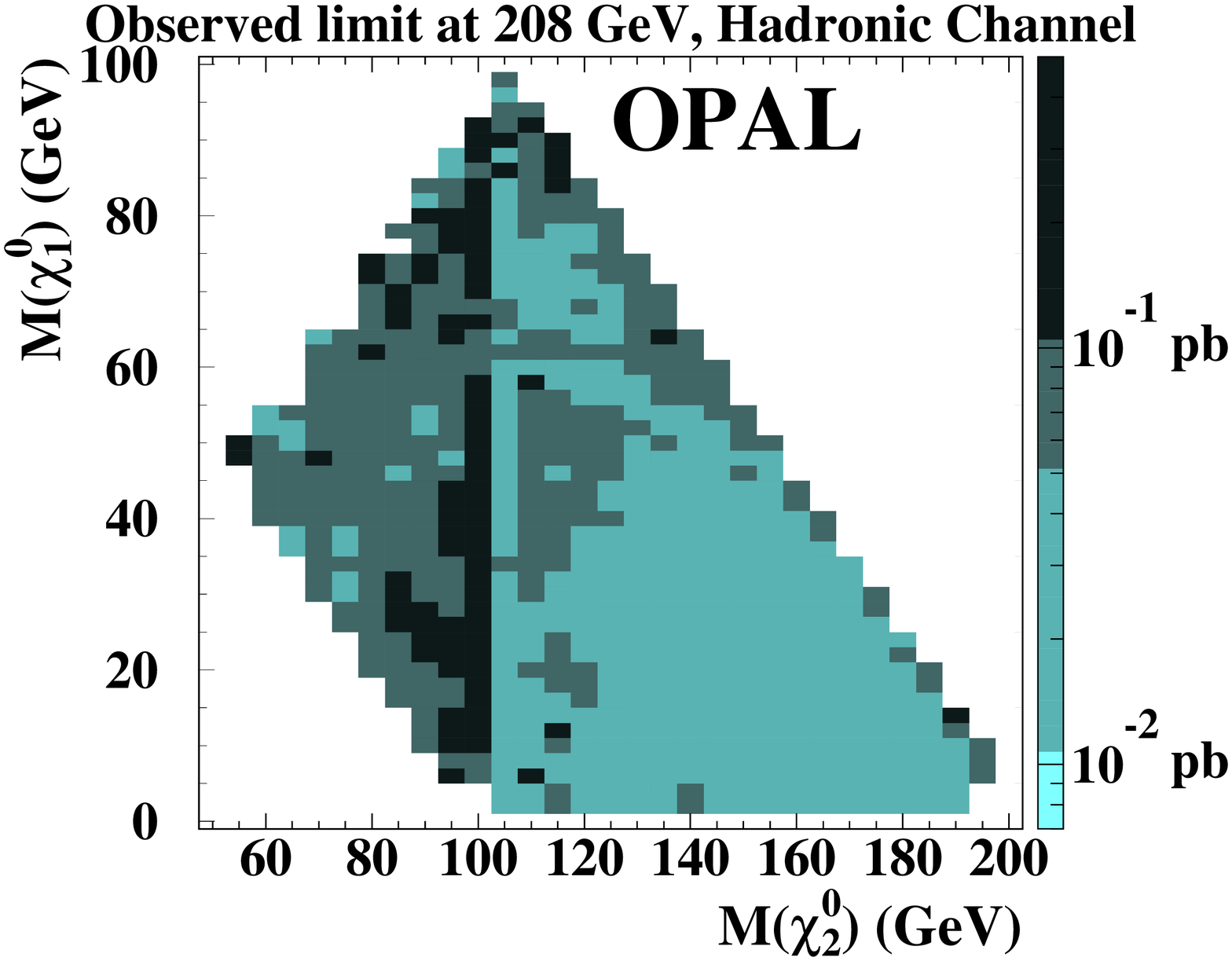}}
\caption{The expected and observed 95\% confidence level limits on the
  product of the cross-section for $\tilde\chi^0_2\tilde\chi^0_1$
  associated production and the branching ratio for $\tilde\chi^0_2\to
  {\q}{\bar{\q}}\tilde\chi^0_1$ decays. No signal Monte Carlo events
  were generated with $M_{\tilde\chi^0_2}+M_{\tilde\chi^0_1}<100$~GeV,
  so no limit is evaluated in this region. White areas indicate that
  the limit is 
  greater than 0.5~pb or that no limit could be set.\label{fig:neuthad}
  }
\end{figure}

\begin{figure}[htbp]
\resizebox{0.49\textwidth}{!}{\includegraphics{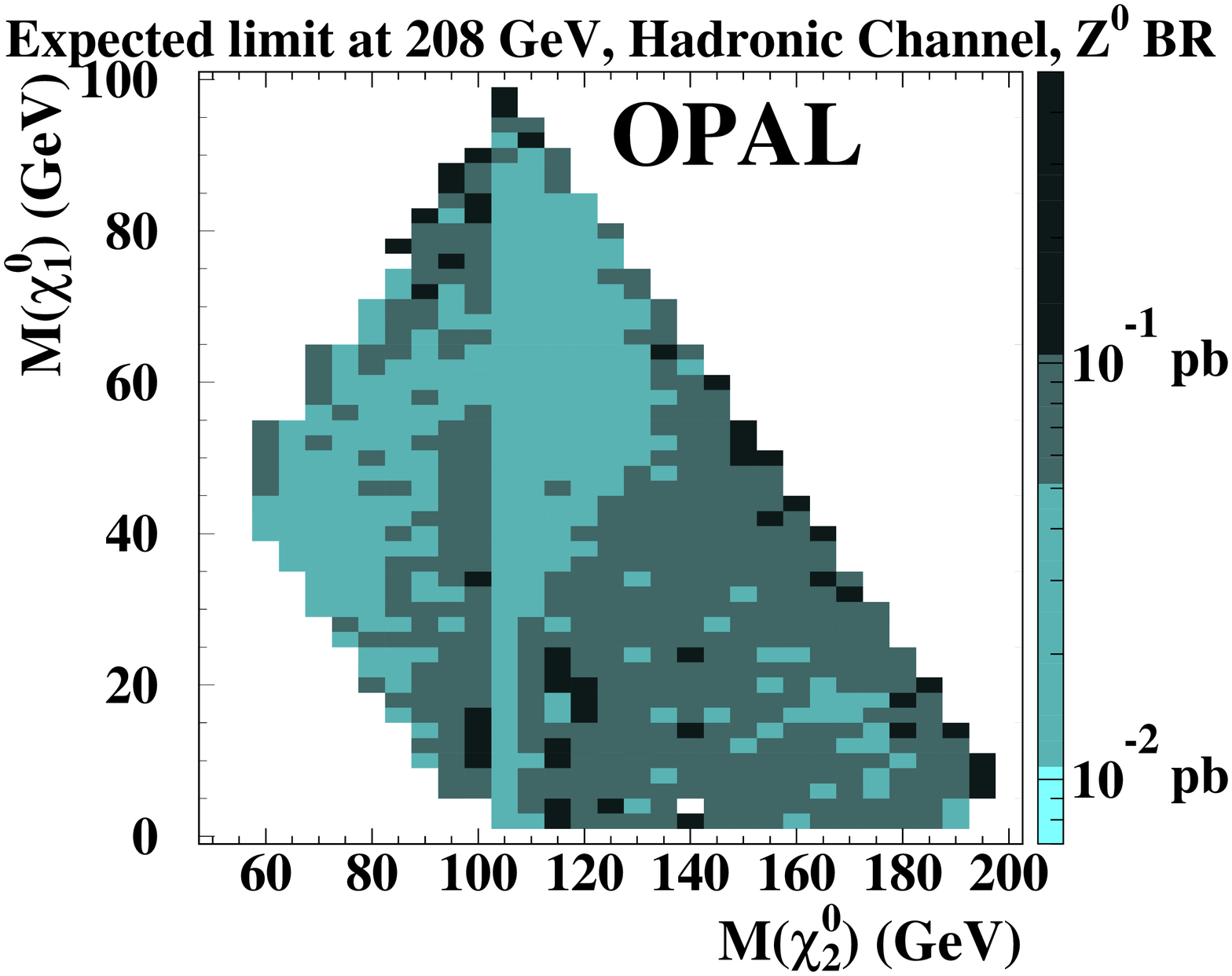}}
\resizebox{0.49\textwidth}{!}{\includegraphics{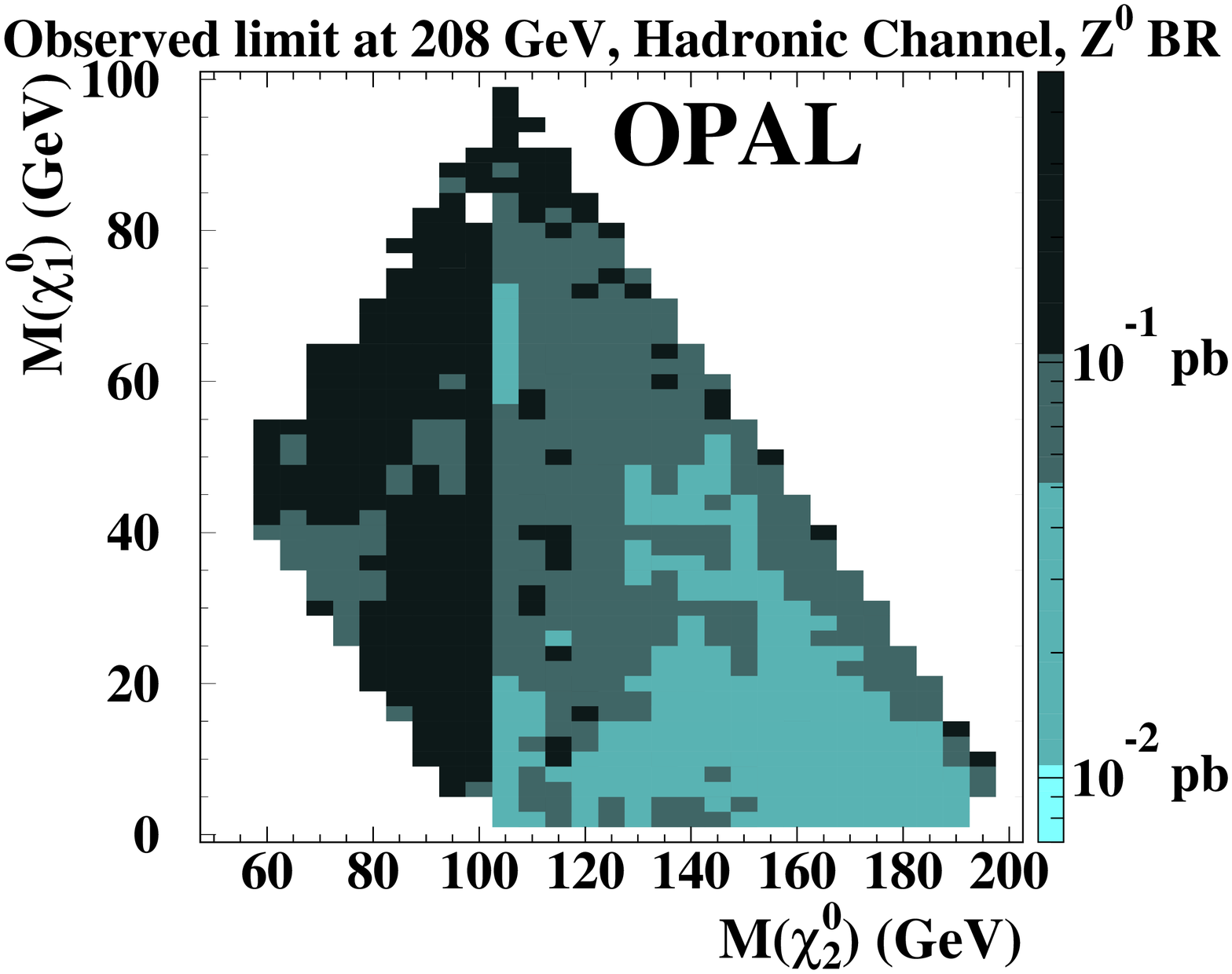}}
\caption{The expected and observed 95\% confidence level limits on the
  cross-section for $\tilde\chi^0_2\tilde\chi^0_1$ associated
  production, based on decays to hadronic final states and assuming
  100\% ${\Z}^{0\ast}$ branching ratios for the $\tilde\chi^0_2$
  decays.  No signal Monte Carlo events
  were generated with $M_{\tilde\chi^0_2}+M_{\tilde\chi^0_1}<100$~GeV,
  so no limit is evaluated in this region. White areas indicate that
  the limit is 
  greater than 0.5~pb or that no limit could be set.\label{fig:neutral}}
\end{figure}

Since only the hadronic channel was analyzed for neutralino associated
production, the results are presented both as a limit on
$\sigma_{\tilde\chi^0_2\tilde\chi^0_1} 
BR(\tilde\chi^0_2\to {\q}{\bar{\q}}\tilde\chi^0_1)$
(Figure~\ref{fig:neuthad}) and as a limit on 
$\sigma_{\tilde\chi^0_2\tilde\chi^0_1}$ assuming 100\% ${\Z}^{0\ast}$
branching ratios for $\tilde\chi^0_2$ decay
(Figure~\ref{fig:neutral}).
The 95\% confidence level limit on the cross-section is less than
0.1~pb for $\Delta M>5$~GeV to within about a GeV of the kinematic
limit.  
There is no value of $(M(\tilde\chi^0_2),M(\tilde\chi^0_1))$ for which
a significant excess is observed.
\clearpage

Despite the small general excess which results in larger than expected
cross-section limits in the hadronic and
semileptonic channels, there is no indication of chargino or
neutralino production in the OPAL data.

\section{Interpretation in the Constrained MSSM}
The experimental
information required to obtain limits in any particular model is
contained in the cross-section limit for chargino
production as a function of the leptonic branching ratio and in the
cross-section limit for neutralino associated production as a function
of the invisible branching ratio.  Comparing the limits with the
cross-section and leptonic branching fraction predicted by any
specific model allows that model to be tested with the OPAL data.  

In this section, the results are interpreted in the context of
the Contrained Minimal Supersymmetric Standard Model (CMSSM).
This is a model inspired by Minimal Supergravity, in which it is
assumed that at some large Grand Unification (GUT) scale all of the
gauginos have a common mass $m_{1/2}$.  This implies that there is a
fixed relation between the three gaugino masses at the electroweak
scale, $M_1$ for $U(1)$ partners, $M_2$ for $SU(2)$ partners, and
$M_3$ for $SU(3)$ partners, so a single free parameter, taken
to be $M_2$, generates all the gaugino masses. It is also assumed that the 
sfermions (but not all scalars as in true Minimal Supergravity) have a
common mass at the SUSY-breaking scale, denoted by $m_0$.  With these
constraints, there are only six free parameters not present
in the SM: $M_2$, $m_0$, the common trilinear sfermion-Higgs coupling
at the Planck scale $A_0$, the ratio of the two vacuum expectation
values for the Higgs doublet fields coupling to down- and up-type
quarks $\tan\beta$, the mass mixing parameter of the Higgs fields
$\mu$, and the mass of the pseudoscalar Higgs boson $m_A$.  With these
six parameters, it is possible to calculate cross-sections for
$\chi^+_1\chi^-_1$ and $\chi^0_2\chi^0_1$ production.

The properties of the charginos and neutralinos depend chiefly on $M_2$,
$\mu$ and $\tan\beta$.  If $M_2\gg|\mu|$, then the lightest neutralino is
higgsino-like and $t$-channel contributions are negligible; the
$\chi^{\pm}_1, \chi^0_1$ and $\chi^0_2$ are all close in mass
and the cross-section for $\chi^0_2\chi^0_1$ production is
similar to that of $\chi^+_1\chi^-_1$ production.  If $M_2\ll|\mu|$ then
the $\chi^0_1$ is gaugino-like; $t$-channel processes are important
and may interfere destructively with the $s$-channel.  Scans
over $M_2$ and $\mu$ are done at several values of $\tan\beta$.
The $m_0$ and $A_0$ parameters determine the
masses of the sfermions.  Results here are shown for $m_0=500\mbox{
  GeV}$, implying that there are no light sleptons or sneutrinos,
and two-body decays of charginos to $\ell^\pm\tilde\nu$
can be ignored, and for $A_0=0$, which generally means that mixing of
left- and right-handed sfermion eigenstates is assumed to be small.
A scan is performed with
SUSYGEN~\cite{susygen} over the region $0 < M_2 < 5000$~GeV,
$-1000\mbox{ GeV}<\mu < 1000\mbox{ GeV}$ at $\tan\beta=1,2,5,10,40$.
The granularity of the scan depends on the region and is generally
between 1 and 5~GeV in $M_2$ and $\mu$.

\begin{figure}[htbp]
\centering{
\resizebox{0.45\textwidth}{!}{\includegraphics{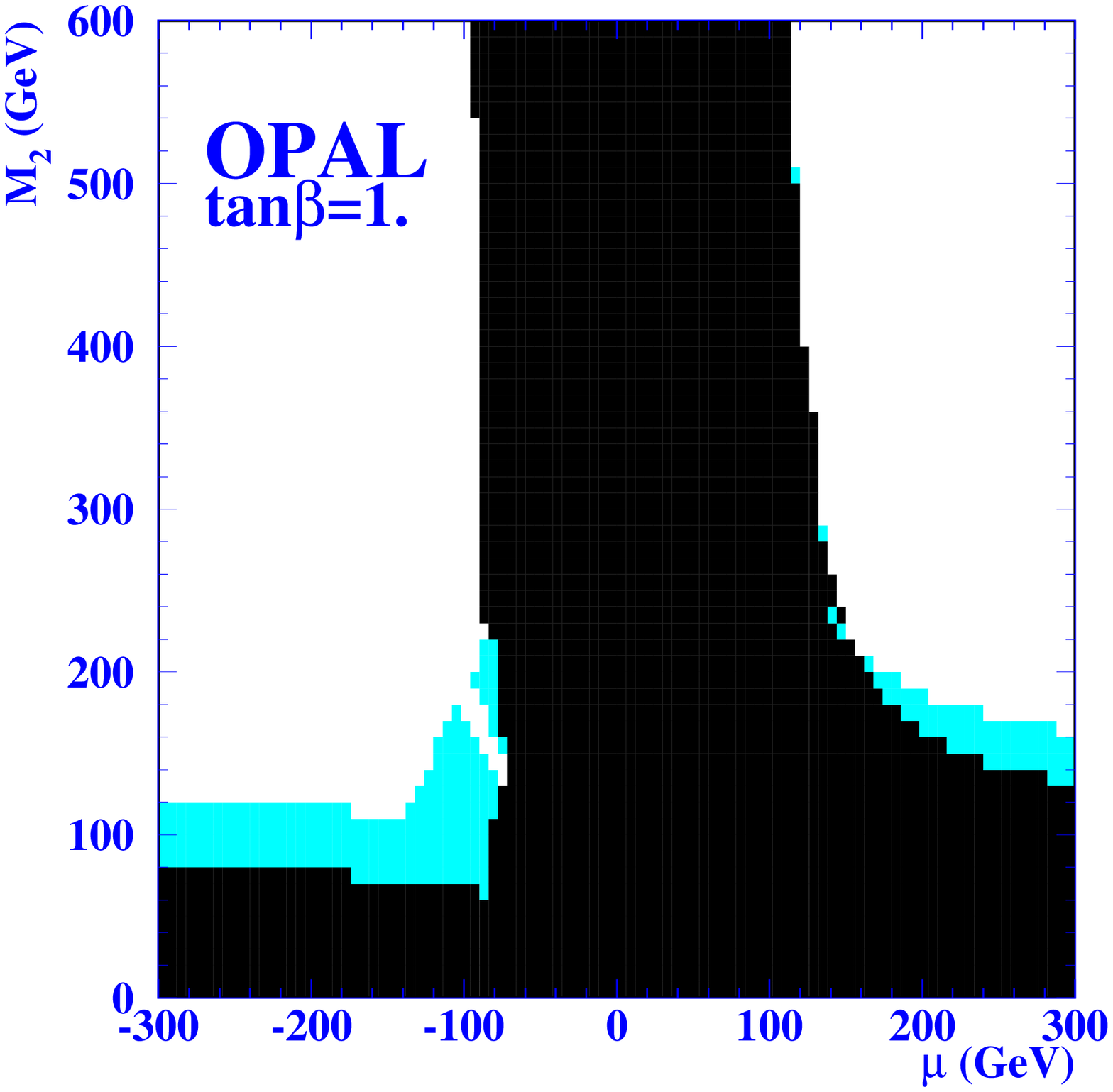}}
\resizebox{0.45\textwidth}{!}{\includegraphics{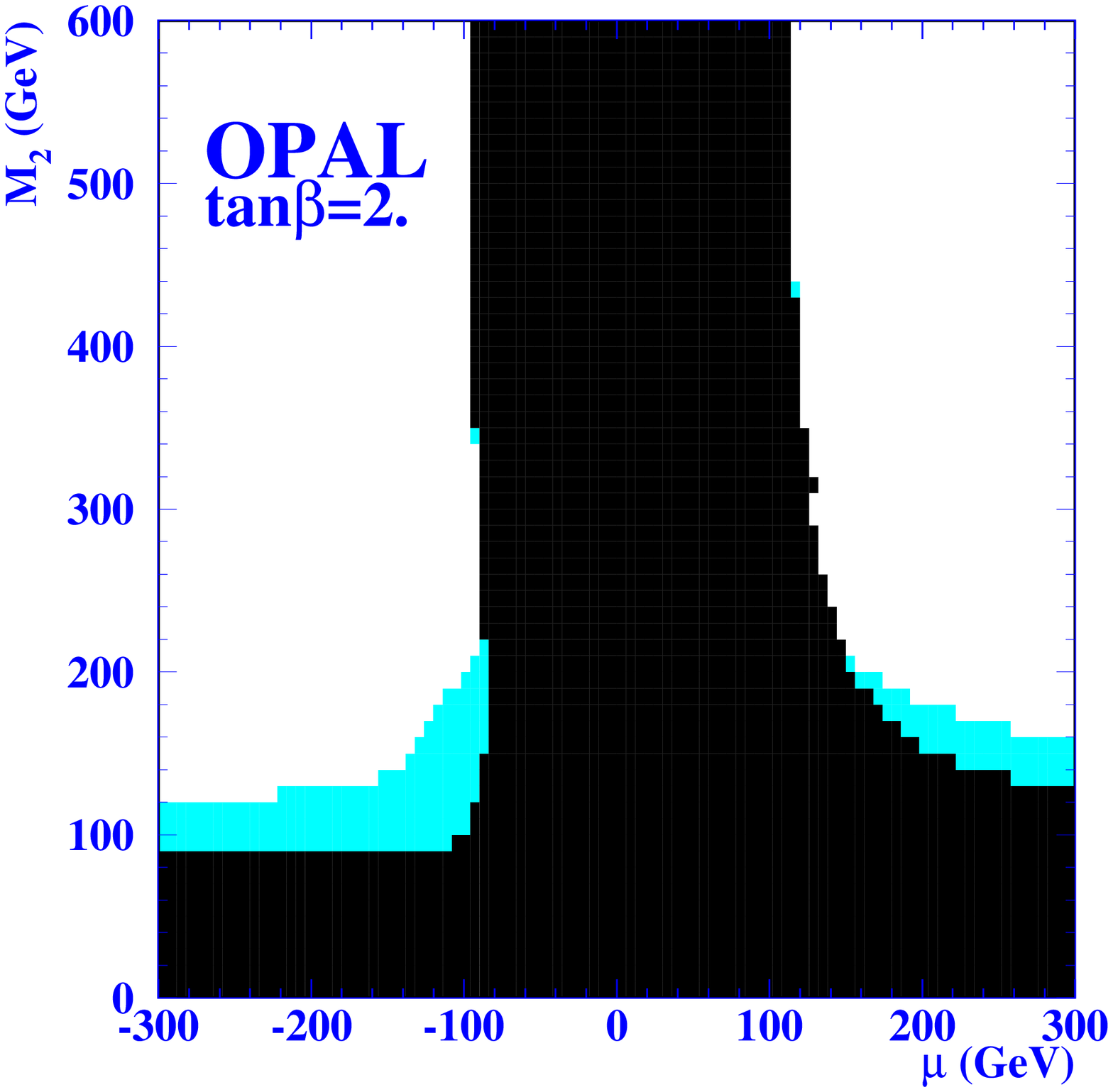}}
\resizebox{0.45\textwidth}{!}{\includegraphics{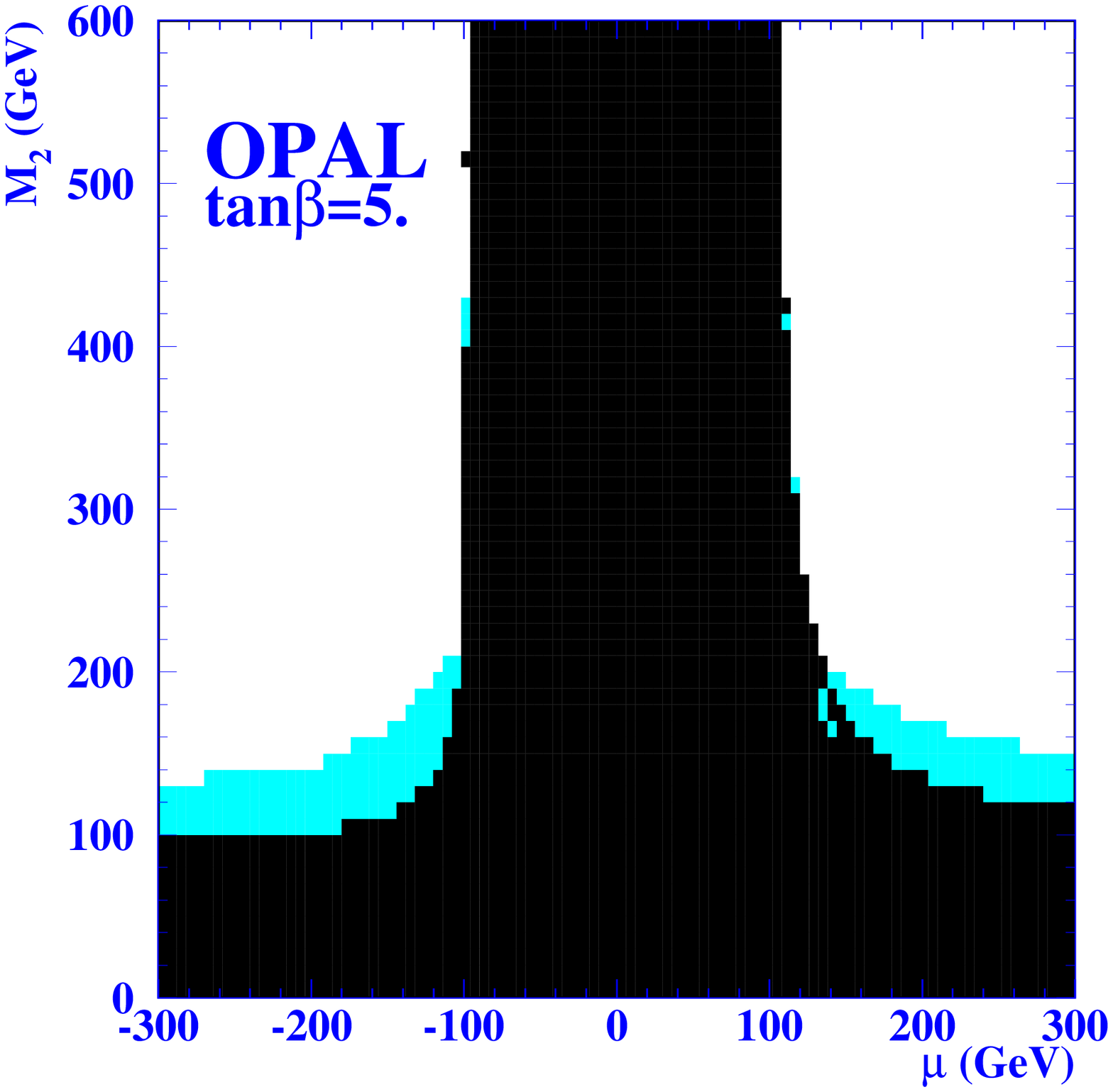}}
\resizebox{0.45\textwidth}{!}{\includegraphics{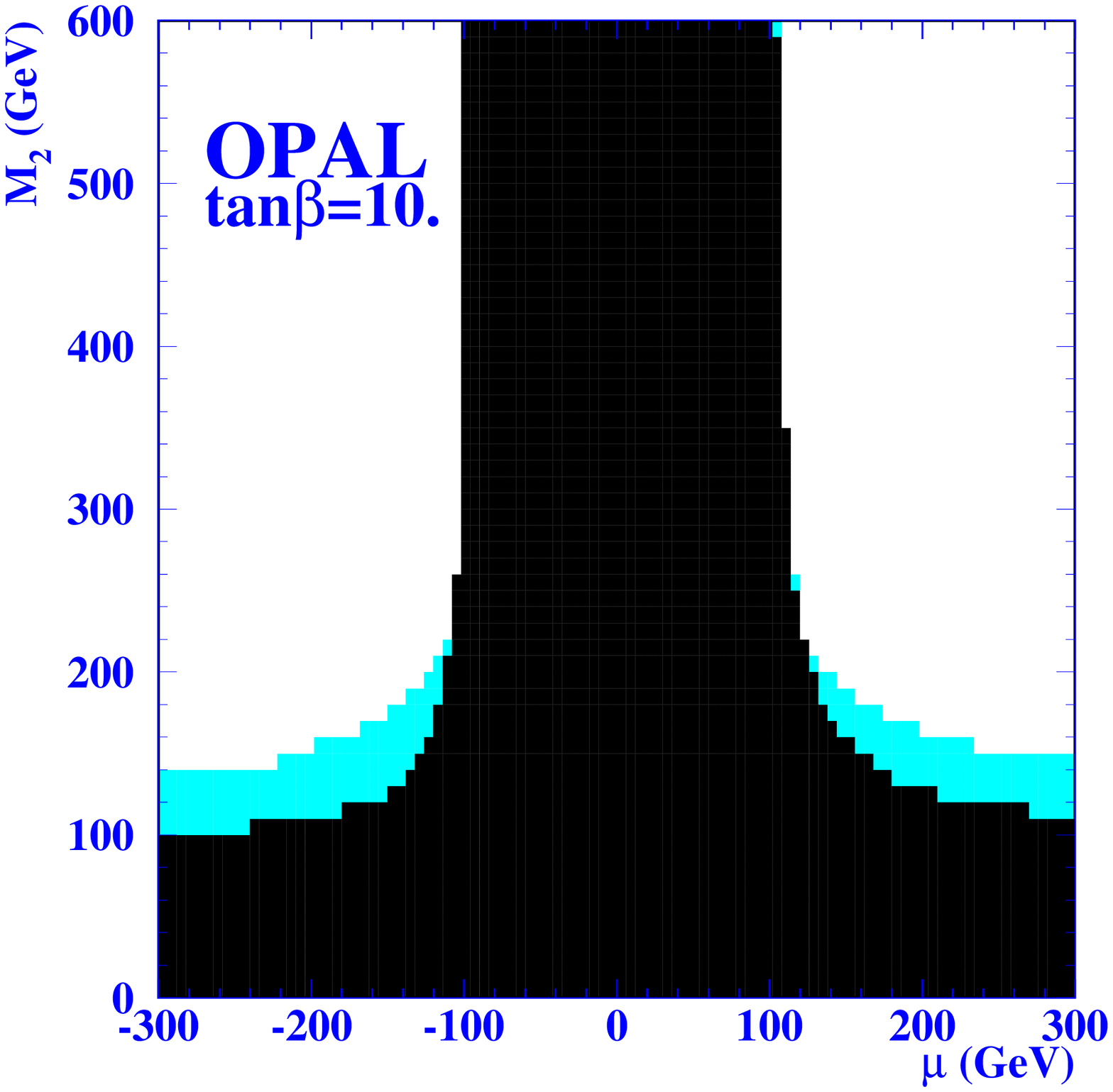}}
\resizebox{0.45\textwidth}{!}{\includegraphics{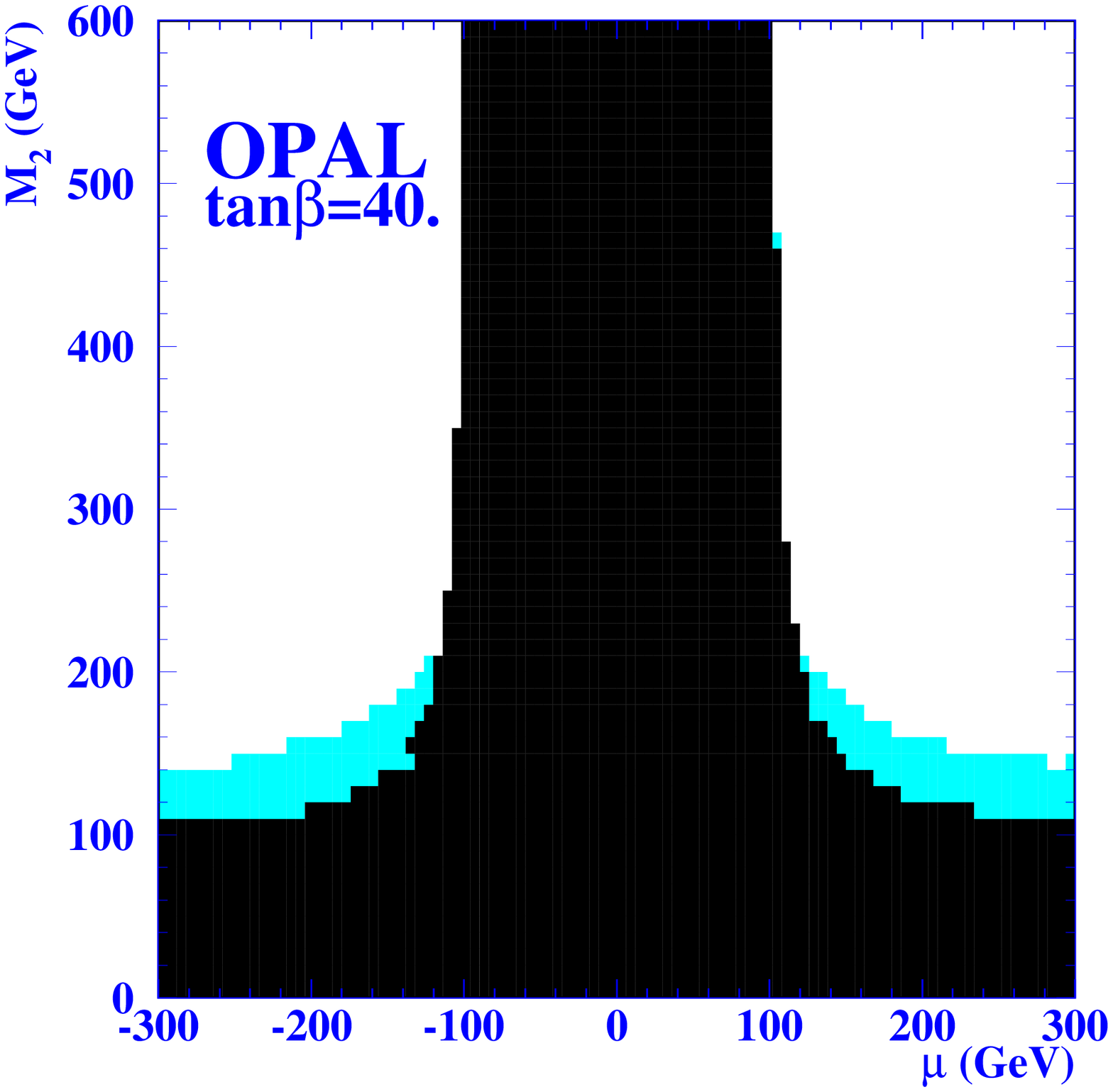}}
\resizebox{0.45\textwidth}{!}{\includegraphics{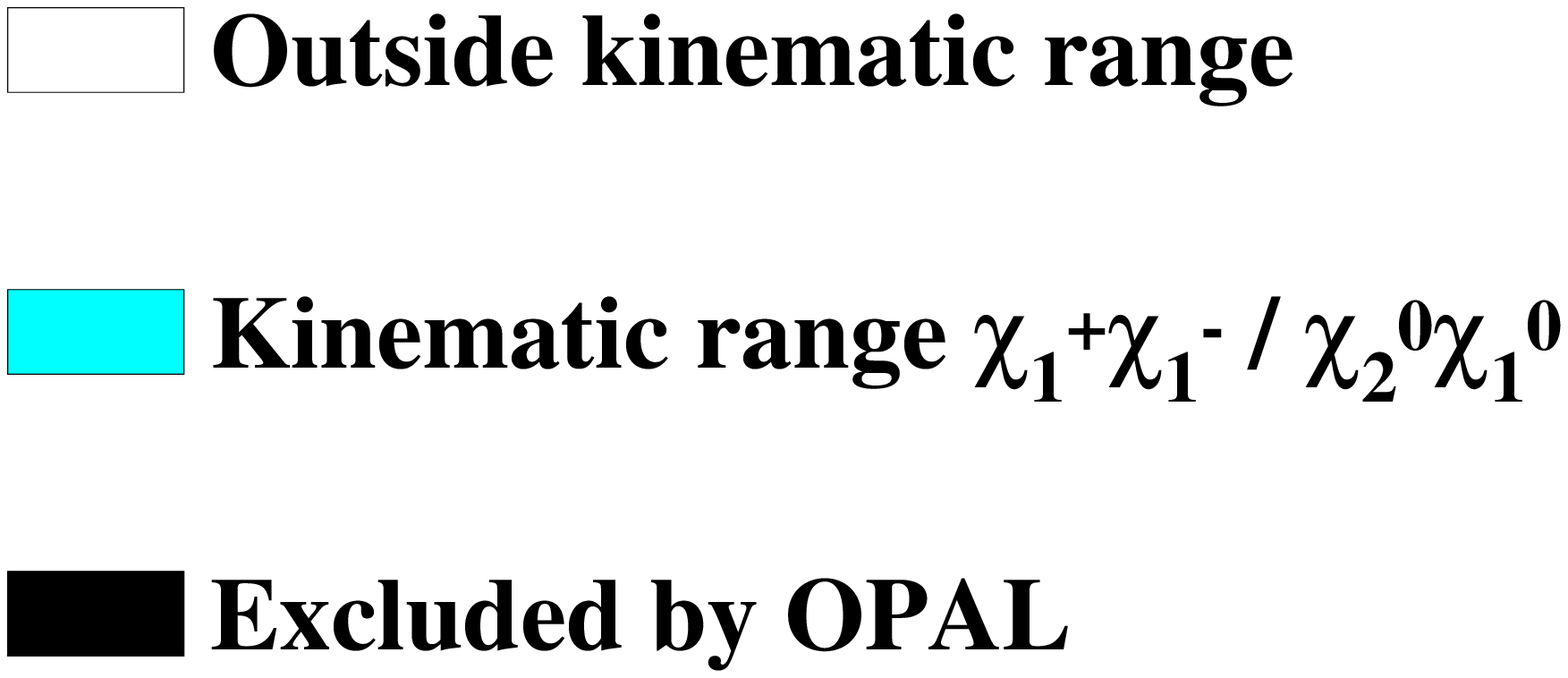}}
}
\caption{Regions of the CMSSM parameter space excluded by the OPAL search
   are shown in black. These include
  regions excluded by the OPAL data taken at energies up to
  189~GeV~\cite{189opal} as well as those excluded with the present dataset
  and analysis. 
  The regions with the lighter shading are kinematically accessible for
  either $\tilde\chi^+_1\tilde\chi^-_1$ or
  $\tilde\chi^0_2\tilde\chi^0_1$ production at 
  $\sqrt{s}=208$~GeV, but are not excluded.
  These results are valid for $m_0\ge 500$~GeV and $A_0=0$ and for
  the values of $\tan\beta$ indicated on the individual plots.
  \label{fig:m2mu}}
\end{figure}

The simplest thing to do with the cross-section limits obtained in the
previous section and the results of the CMSSM scan is to see what
regions of the parameter space are excluded.  By construction, the
three chargino channels do not use the same events, so the best way to
use this information is to sum the limits for the three channels
according to the branching ratios predicted by SUSYGEN for a
particular set of $M_2,\mu,\tan\beta$ values and compare with the
SUSYGEN prediction for the chargino total cross-section.  The events
selected in the neutralino associated production search may overlap with the
events selected in the chargino search, so the optimal way to include
both searches is to calculate whether the expected limit for the
chargino or the neutralino search would give the better exclusion for
a particular $M_2,\mu,\tan\beta$, and use the channel which is
expected to be better.  Exclusions in the $M_2,\mu$ plane are shown
in figure~\ref{fig:m2mu} for the five values of $\tan\beta$ studied.
These exclusions are valid only for $m_0\ge 500$~GeV and for $A_0=0$.
The regions which are shown as kinematically accessible but not
excluded correspond to regions where $\tilde\chi^0_2\tilde\chi^0_1$
production is possible, but with an expected cross-section too small
to be excluded.

\begin{figure}[htbp]
\centering{
\resizebox{0.40\textwidth}{!}{\includegraphics{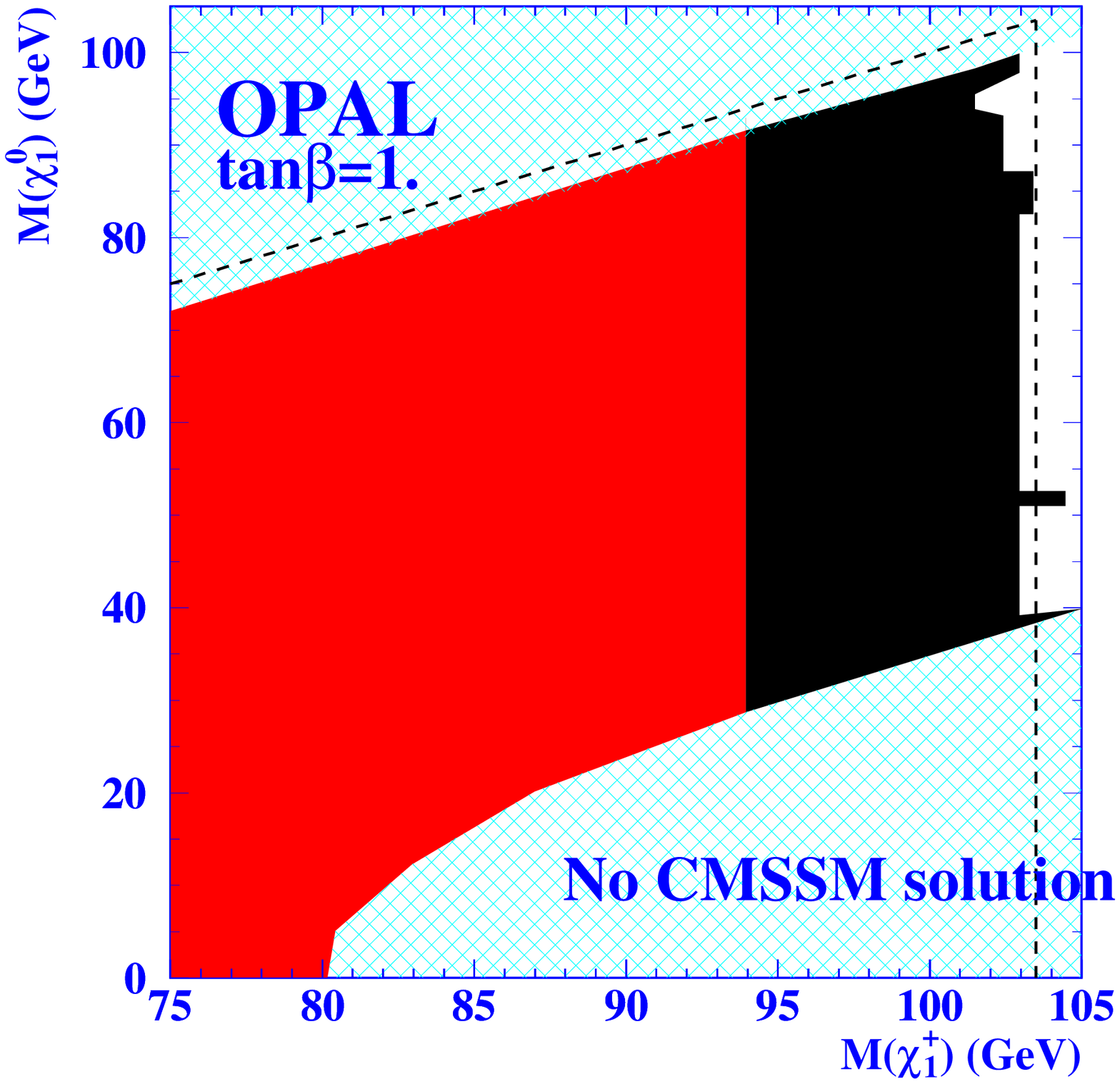}}
\resizebox{0.40\textwidth}{!}{\includegraphics{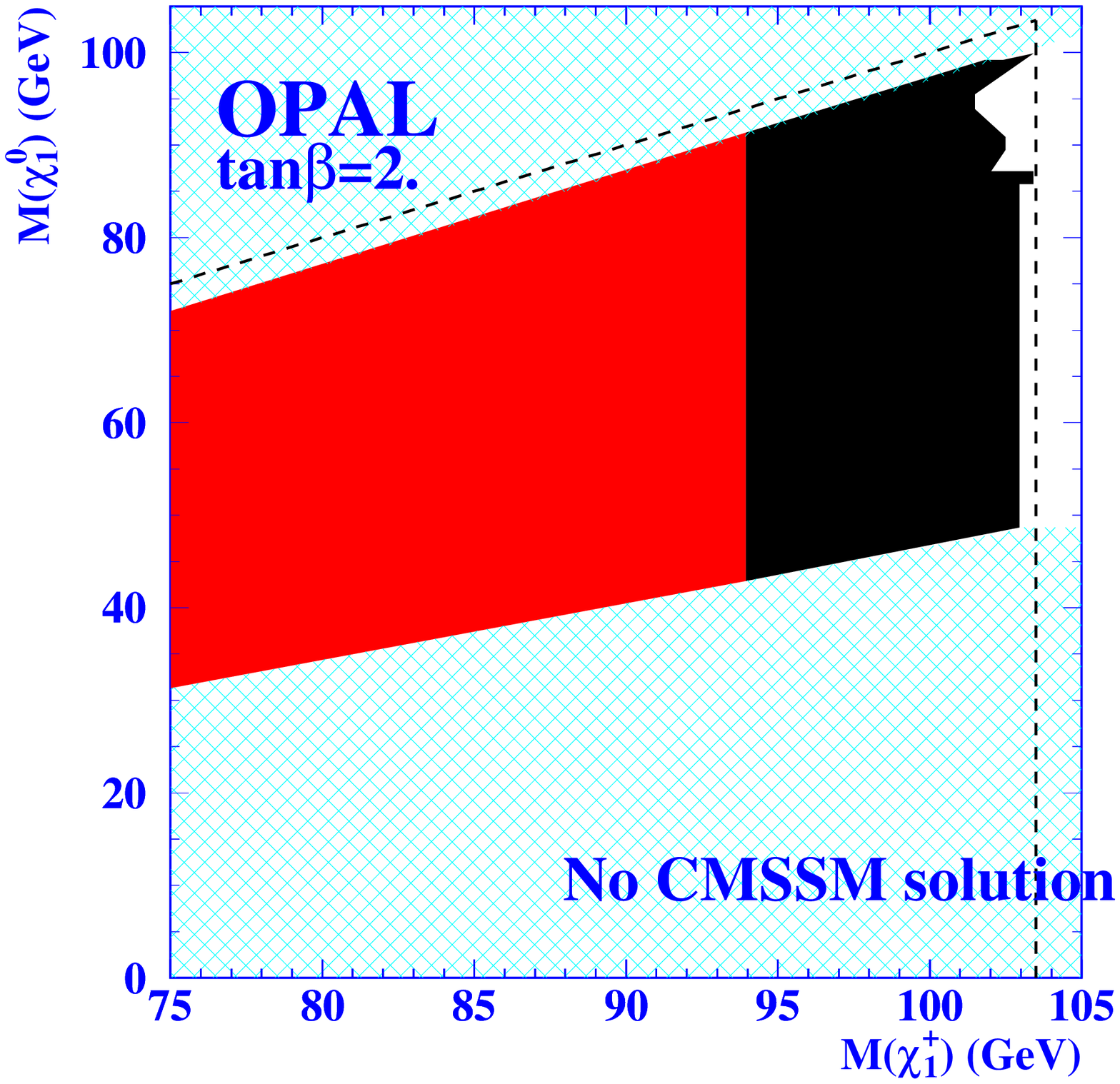}}
\resizebox{0.40\textwidth}{!}{\includegraphics{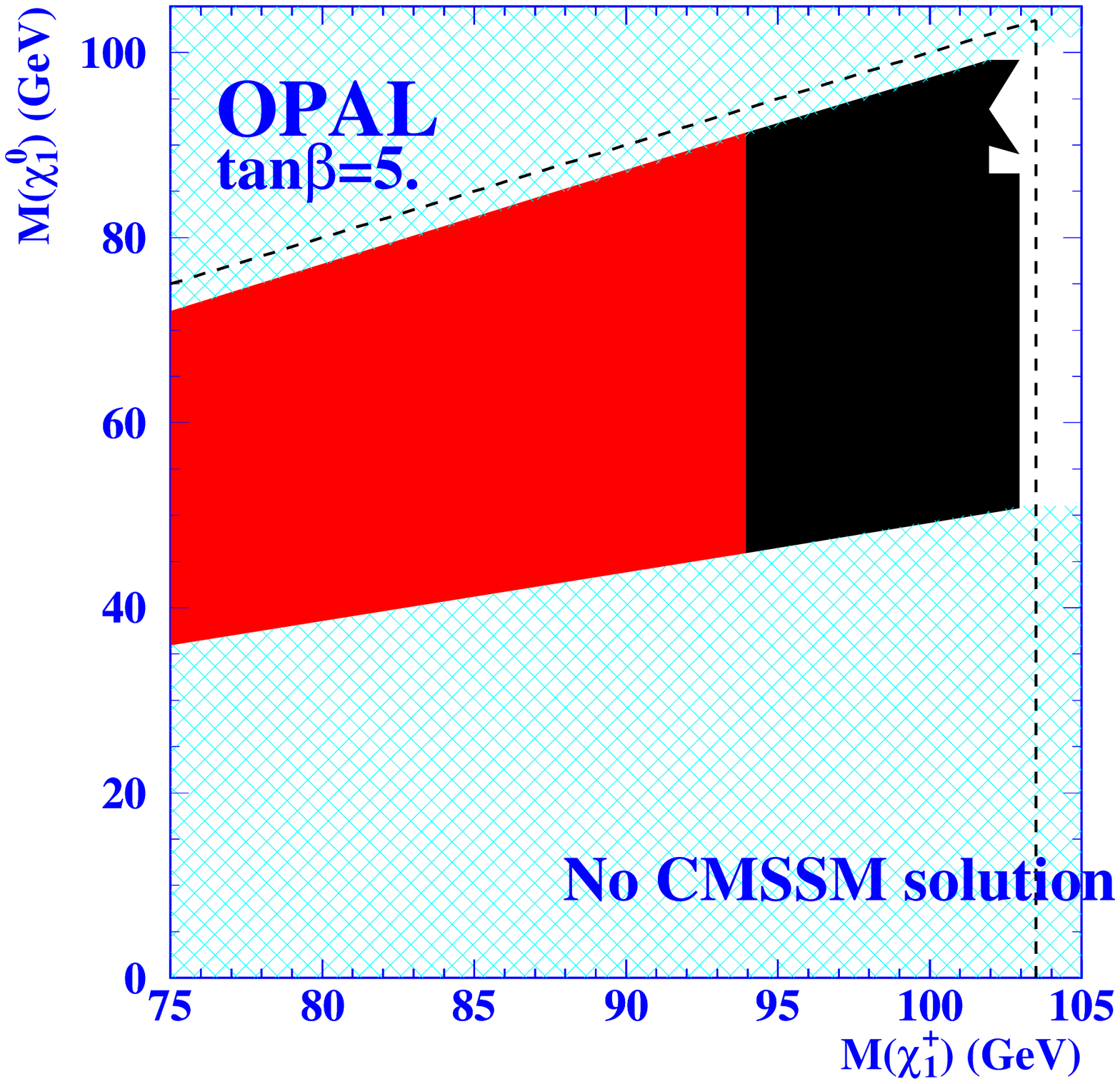}}
\resizebox{0.40\textwidth}{!}{\includegraphics{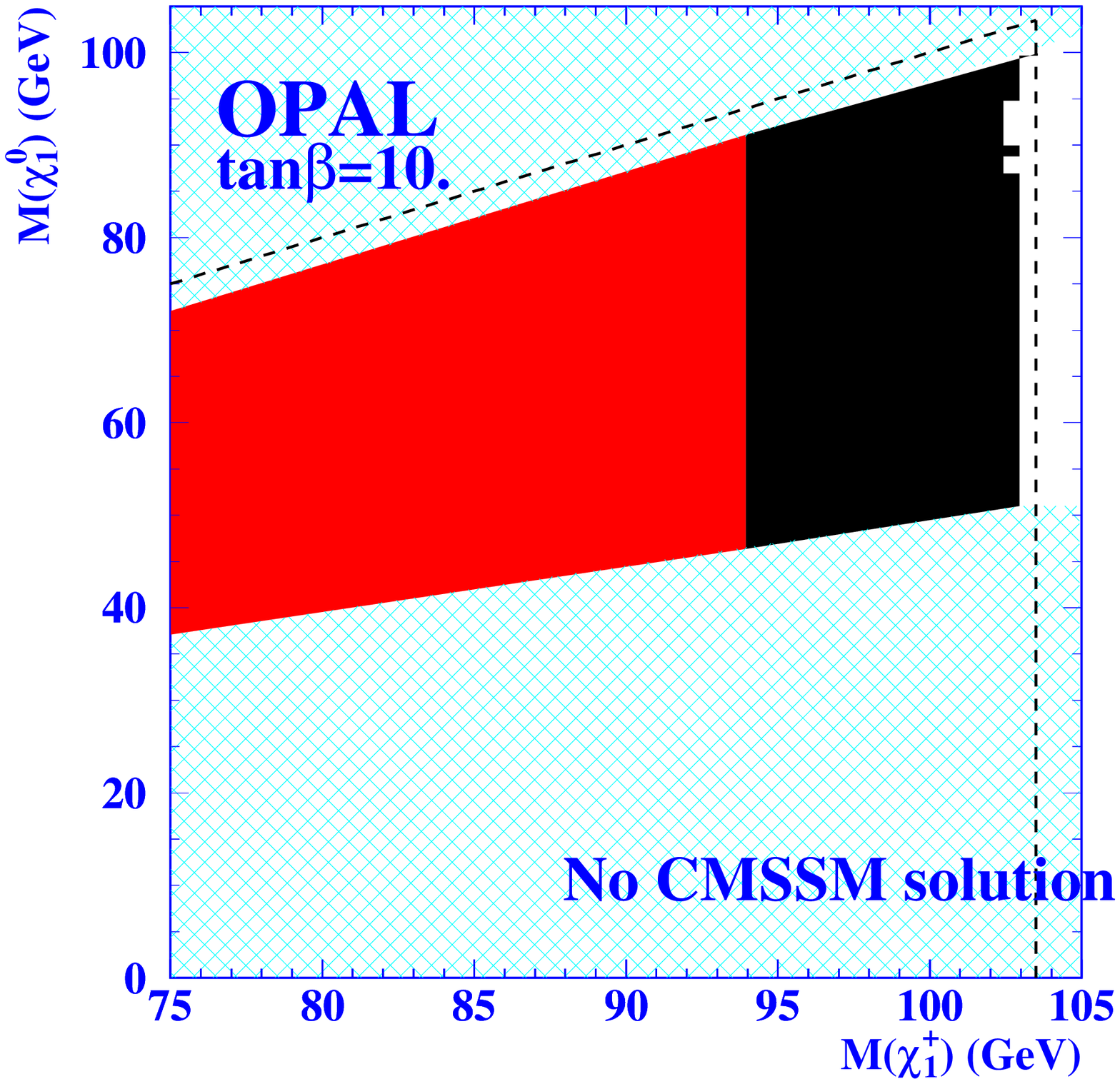}}
\resizebox{0.40\textwidth}{!}{\includegraphics{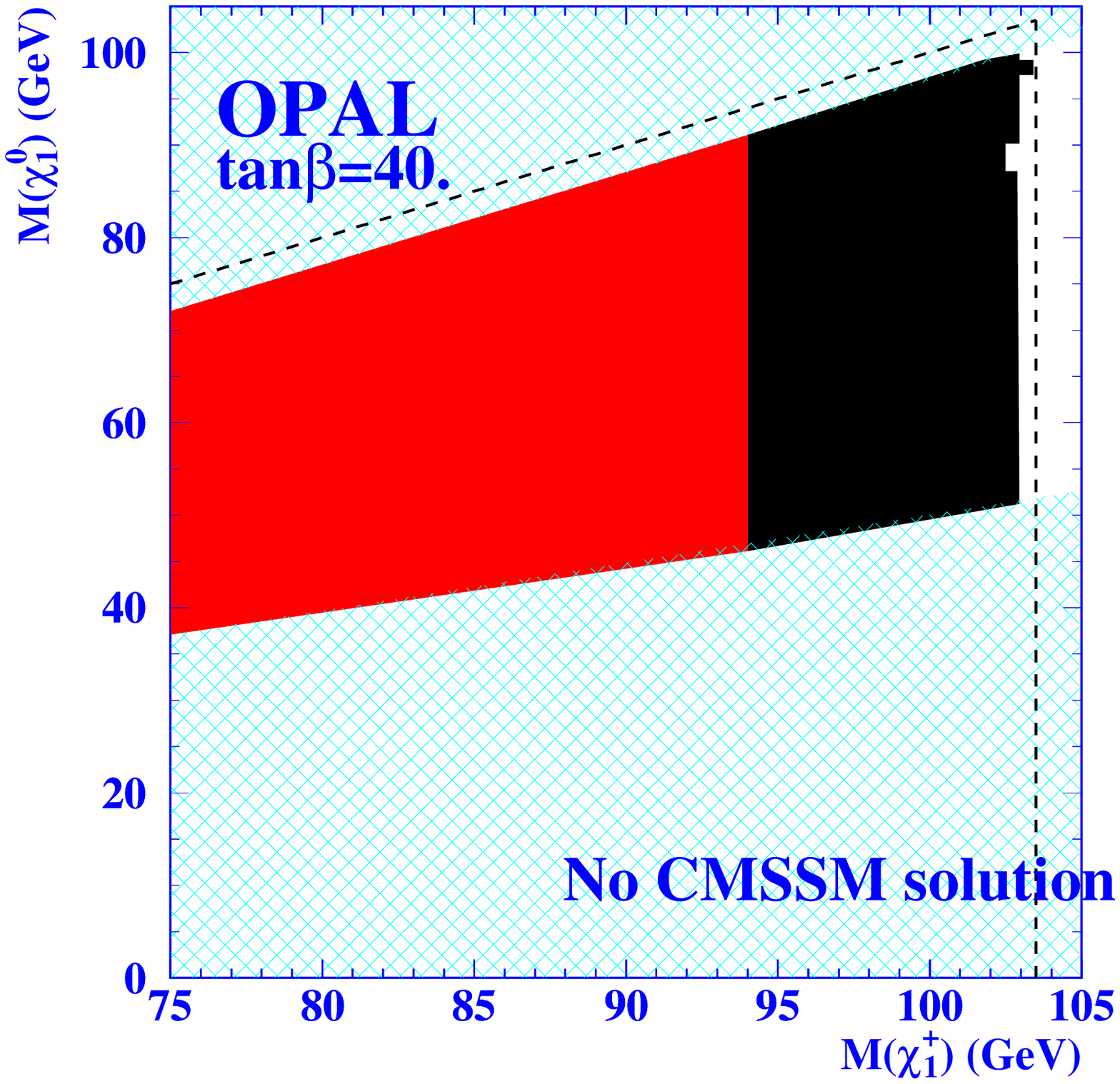}}
\resizebox{0.40\textwidth}{!}{\includegraphics{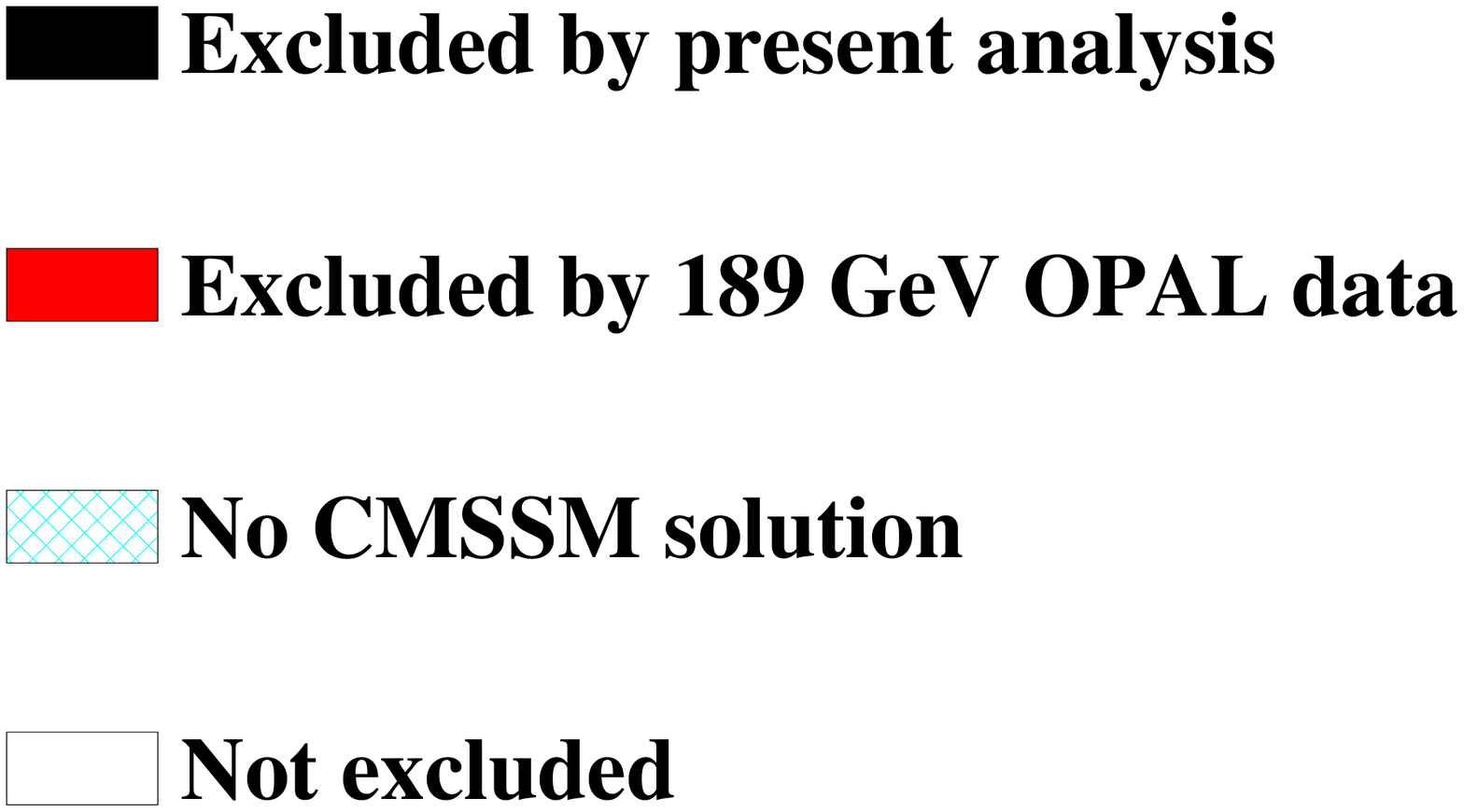}}
}
\caption{The shaded regions of the
  $\tilde\chi^{\pm}_1 - \tilde\chi^0_1$ mass plane are excluded by
  OPAL data, with the lighter colour showing the region excluded by
  previously published searches~\cite{189opal} at centre-of-mass
  energies up to 189~GeV, and the black region being excluded by the
  present search.  The kinematic limits of the chargino search are 
  indicated by dotted lines; the very small excluded regions outside
  these limits are due to the neutralino search.  The cross-hatched
  regions labelled ``No CMSSM solution'' 
  correspond to regions of the parameter space having no solution in
  the CMSSM. The regions left white are not excluded.\label{fig:charmassexcl}}
\end{figure}

\begin{figure}[htbp]
\centering{
\resizebox{0.40\textwidth}{!}{\includegraphics{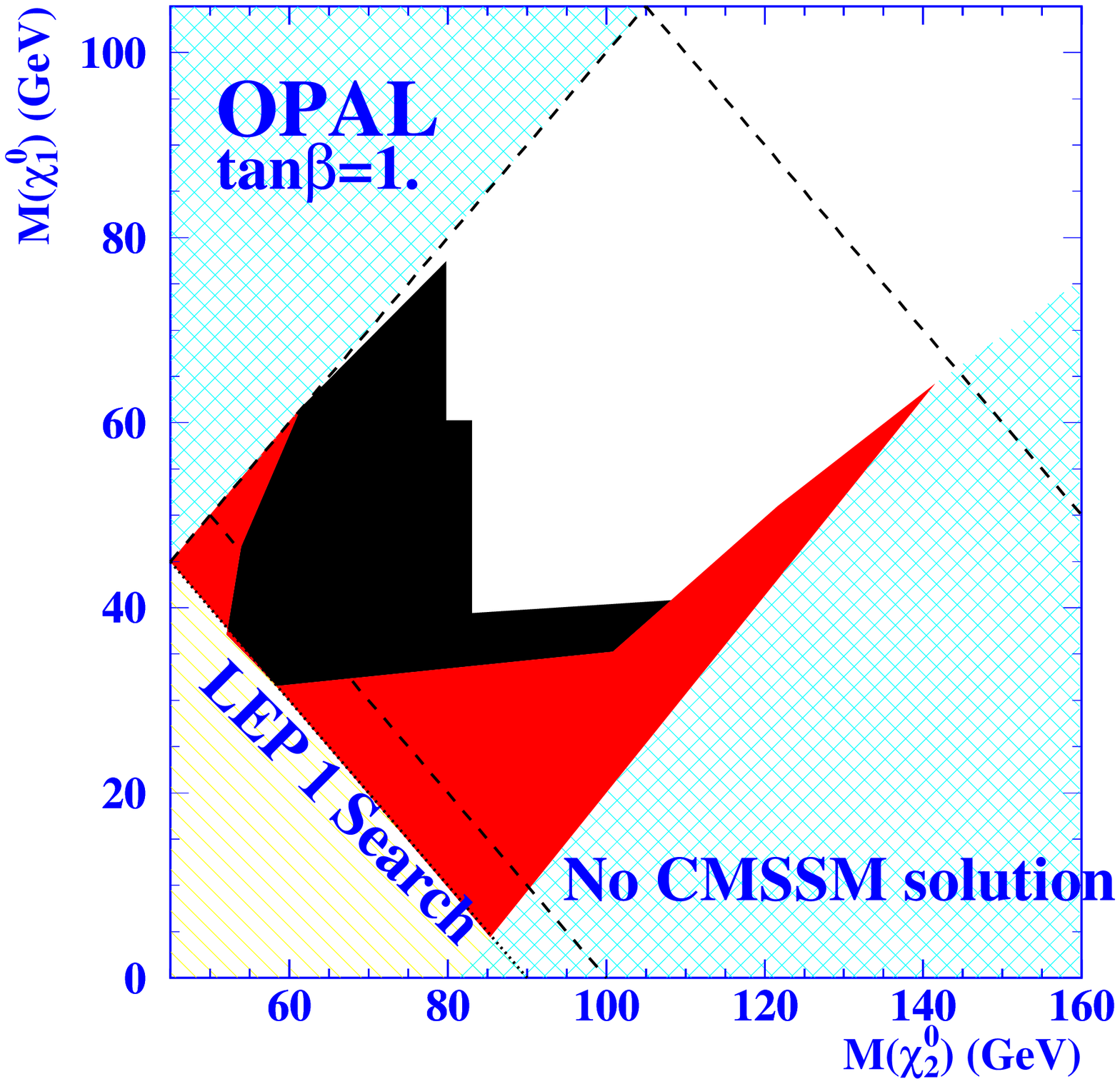}}
\resizebox{0.40\textwidth}{!}{\includegraphics{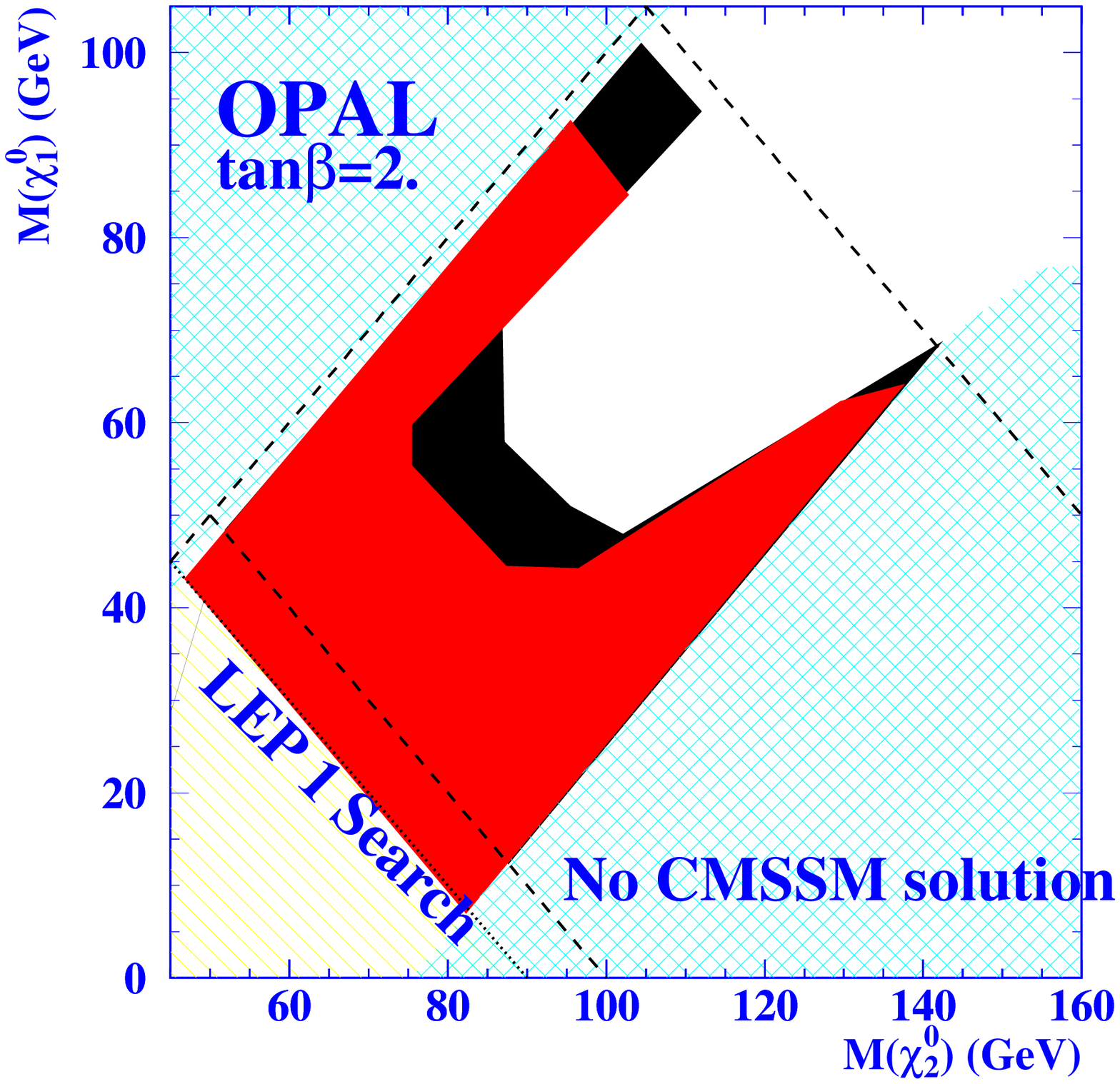}}
\resizebox{0.40\textwidth}{!}{\includegraphics{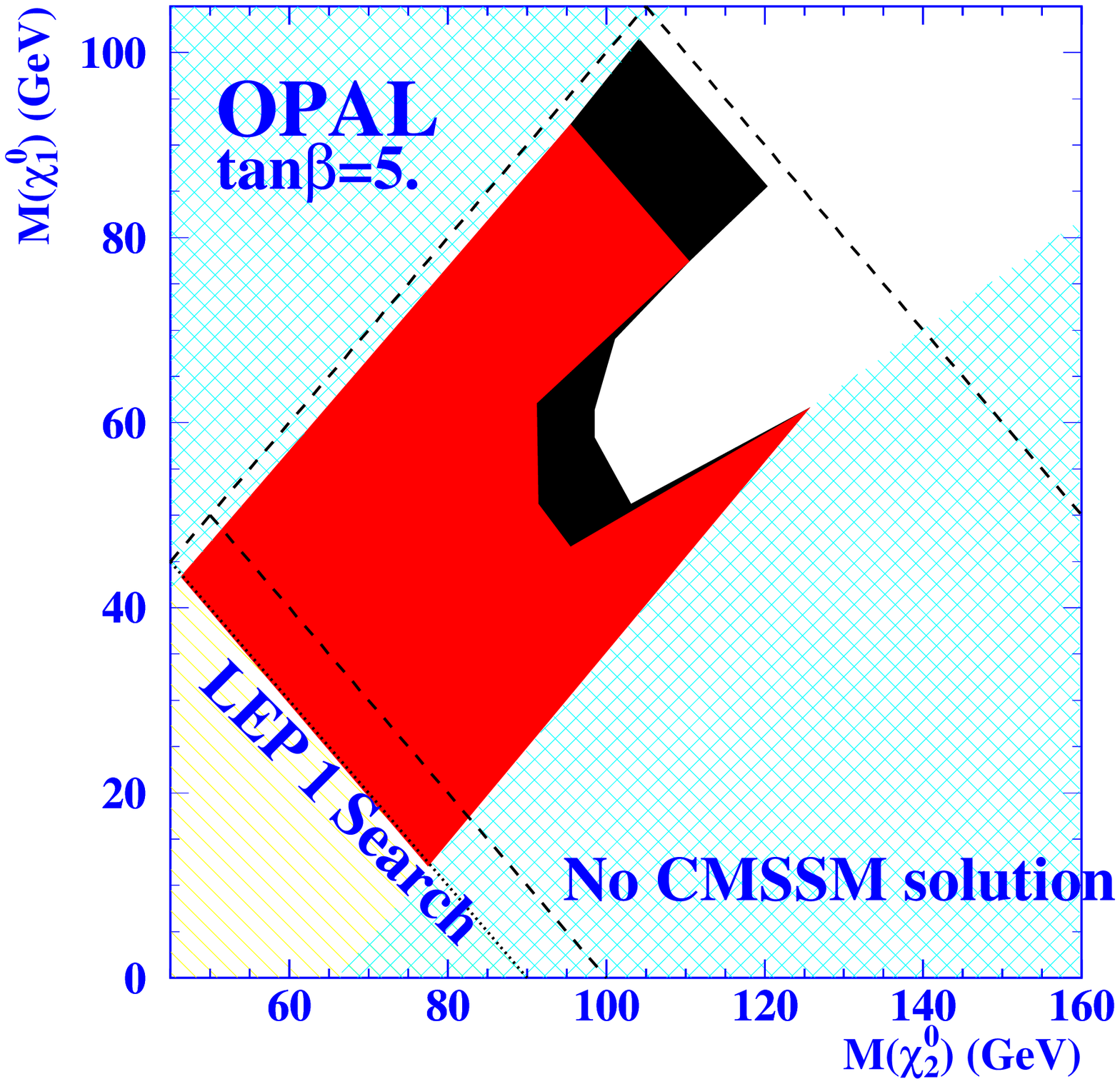}}
\resizebox{0.40\textwidth}{!}{\includegraphics{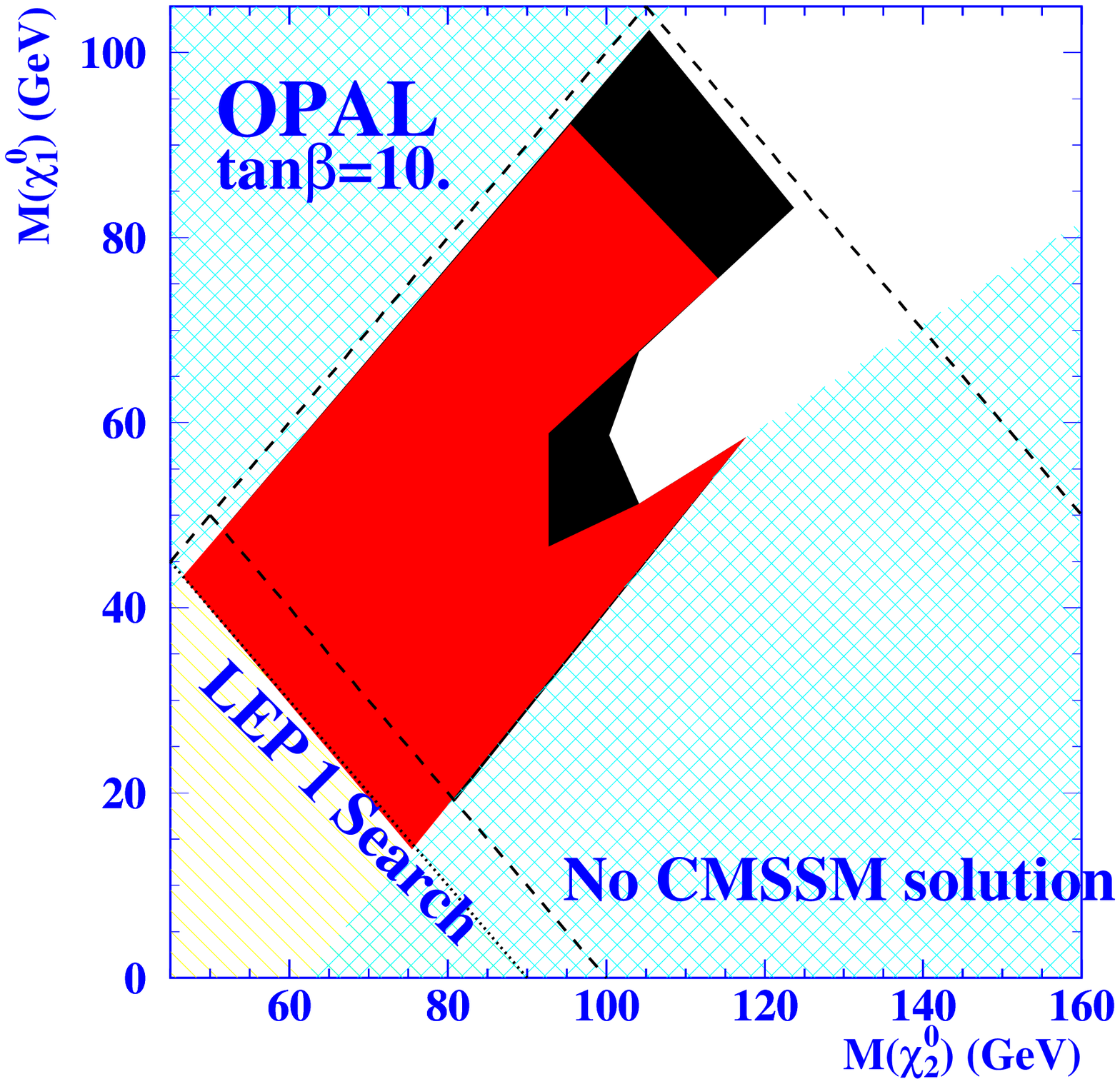}}
\resizebox{0.40\textwidth}{!}{\includegraphics{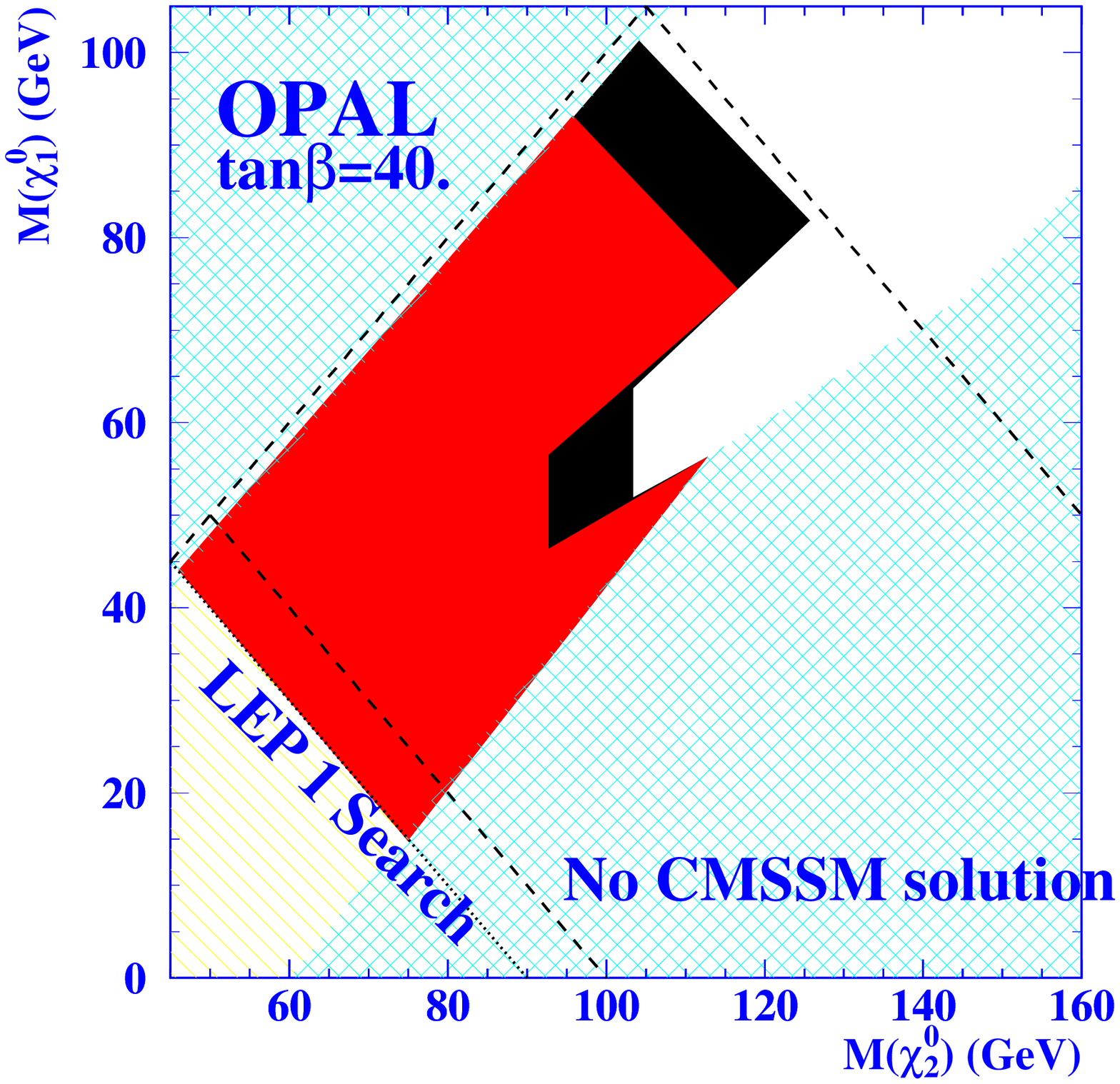}}
\resizebox{0.40\textwidth}{!}{\includegraphics{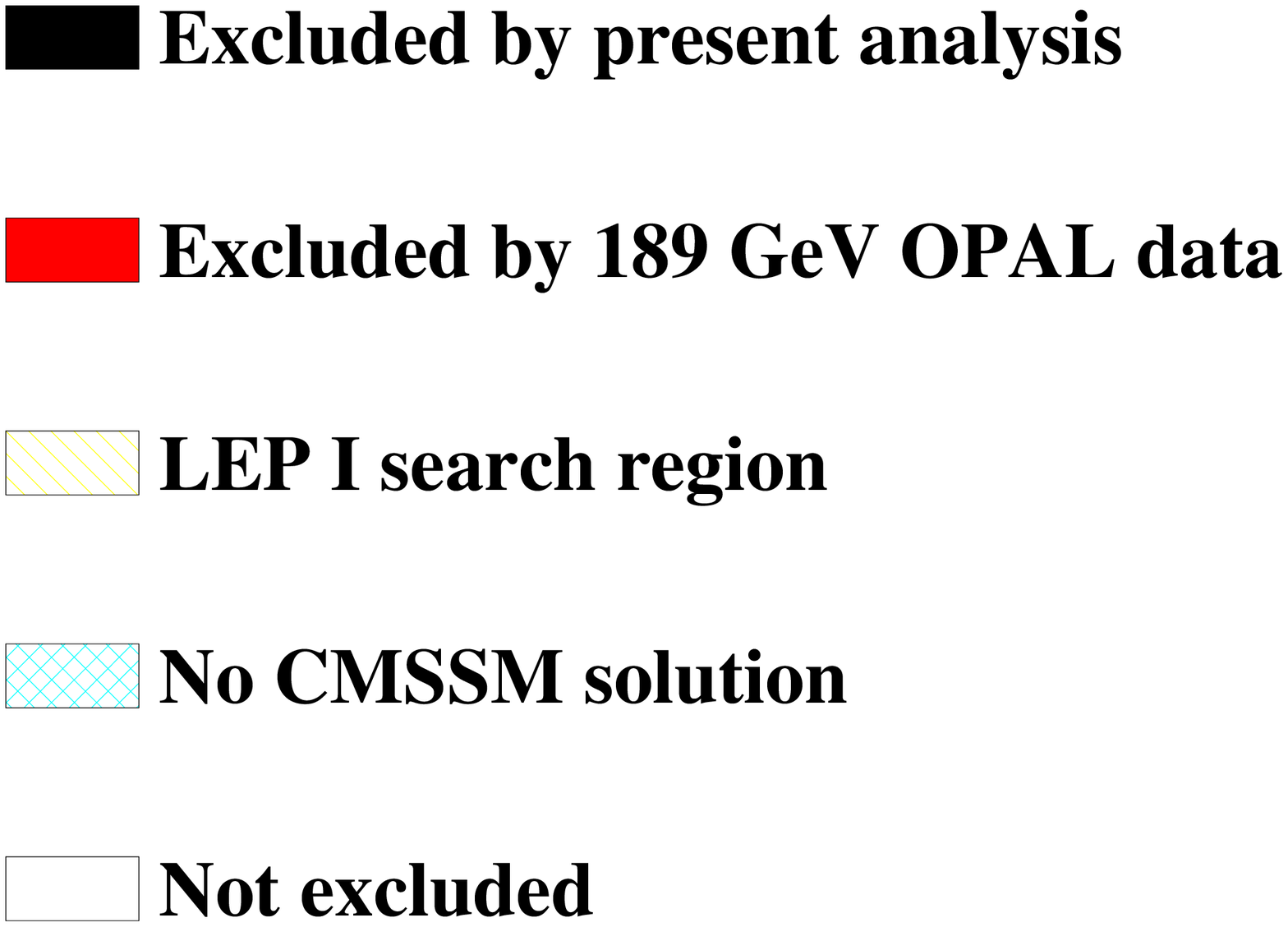}}
}
\caption{The shaded regions of the
  $\tilde\chi^0_2 - \tilde\chi^0_1$ mass plane are excluded by OPAL
  data, with the lighter colour showing the region excluded by 
  previously published searches~\cite{189opal} at centre-of-mass
  energies up to 189~GeV, and the black region being excluded by the
  present search.
  The kinematic limits of the neutralino search
  are indicated by dashed lines.  The dashed line at
  $M_{\tilde\chi^0_2}+M_{\tilde\chi^0_1}=100$~GeV is simply a reminder
  that for the present analysis, no $\tilde\chi^0_2\tilde\chi^0_1$
  Monte Carlo events were generated with lower masses; exclusion shown
  in black below this line is purely due to the chargino searches.
  The singly hatched region labelled 
  ``LEP 1 Search'' is the region kinematically accessible to LEP~1
  searches; it was not considered in this study.  The cross-hatched
  region labelled ``No CMSSM solution'' 
  corresponds to regions of the parameter space having no solution in
  the CMSSM. The regions left white cannot be excluded by the
  data. \label{fig:neutmassexcl}}  
\end{figure}

The final step is to see what masses of charginos and neutralinos are
excluded in the CMSSM.  The kinematically accessible regions of the
$M_{\tilde\chi^\pm_1}, M_{\tilde\chi^0_1}$ and $M_{\tilde\chi^0_2},
M_{\tilde\chi^0_1}$ planes are divided into a 1 by 1~GeV grid.
All of the scanned $(M_2,\mu,\tan\beta)$ points are examined to see
what masses they predict for the lightest chargino and two lightest
neutralinos.  The smallest cross-section predicted in any 1 by
1~GeV grid square is compared with the cross-section limit.  The 
CMSSM exclusions for the chargino mass plane are shown for the five
values of $\tan\beta$ in figure~\ref{fig:charmassexcl} and the
neutralino mass plane exclusions are shown in
figure~\ref{fig:neutmassexcl}.

Assuming that $m_0\ge 500$~GeV and $A_0=0$, the following conclusions
can be made.
For $\Delta M>5$~GeV, charginos with masses less than 101~GeV
are excluded for all $\tan\beta$.  For $\Delta M>5$~GeV,
$\tilde\chi^0_2$ with masses less than 78~GeV and $\tilde\chi^0_1$
with masses less than 40~GeV are excluded for all $\tan\beta$.

\section{Conclusions}
No hint of charginos or neutralinos was observed in the data from
OPAL.  The cross-section limits presented in this paper for charginos
or neutralinos decaying to particular final states may be regarded as
completely model-independent and can be used to set limits on any
model.
Limits are set on the cross-section for chargino pair production by
combining the three chargino final states under the assumption that
all charginos decay to a virtual $\mathrm{W}$, 
which then decays with the usual branching ratios to leptons and hadrons.
Limits are calculated for the specific example of the Constrained 
MSSM in the absence of light sfermions or mixing between right- and
left-handed sfermions. Charginos are excluded almost up to the
kinematic limit set by the maximum LEP beam energy of about 104~GeV.

\clearpage
\section*{Acknowledgements}
We particularly wish to thank the SL Division for the efficient operation
of the LEP accelerator at all energies
 and for their close cooperation with
our experimental group.  In addition to the support staff at our own
institutions we are pleased to acknowledge the  \\
Department of Energy, USA, \\
National Science Foundation, USA, \\
Particle Physics and Astronomy Research Council, UK, \\
Natural Sciences and Engineering Research Council, Canada, \\
Israel Science Foundation, administered by the Israel
Academy of Science and Humanities, \\
Benoziyo Center for High Energy Physics,\\
Japanese Ministry of Education, Culture, Sports, Science and
Technology (MEXT) and a grant under the MEXT International
Science Research Program,\\
Japanese Society for the Promotion of Science (JSPS),\\
German Israeli Bi-national Science Foundation (GIF), \\
Bundesministerium f\"ur Bildung und Forschung, Germany, \\
National Research Council of Canada, \\
Hungarian Foundation for Scientific Research, OTKA T-038240, 
and T-042864,\\
The NWO/NATO Fund for Scientific Research, the Netherlands.\\



\begin{thebibliography}{99}
  \bibitem{susy}
    H.P.~Nilles, Phys.~Rep.\ {\bf 110} (1984) 1;\\
    H.E.~Haber and G.L.~Kane, Phys.~Rep.\ {\bf 117} (1985) 75.

  \bibitem{feng}
    J.L.~Feng and M.J.~Strassler, Phys.~Rev.\ {\bf D51} (1995) 4661.

  \bibitem{ambrosanio}
    S.~Ambrosanio and B.~Mele, Phys.~Rev.\ {\bf D52} (1995) 3900.
  
   \bibitem{acoplanar}The OPAL Collaboration, G. Abbiendi {\em
    et~al.\/}, `Search for Anomalous
    Production of Di-lepton Events with Missing Transverse Momentum in
    $\e^+\e^-$ Collisions at $\sqrt{s}=$ 183--209~GeV',  
    10th July 2003, CERN-EP-2003-040, Accepted by Eur.~Phys~J.~C.
    
 \bibitem{189opal}
    G. Abbiendi {\it et~al.}, Eur.~Phys.~J.\ {\bf C14} (2000) 187.

  \bibitem{182opal}
    G. Abbiendi {\it et~al.}, Eur.~Phys.~J.\ {\bf C8} (1999) 255.
    
  \bibitem{172opal}
    K. Ackerstaff {\it et~al.}, Eur.~Phys.~J. {\bf C2} (1998) 213.
    
  \bibitem{acoplanarold}
    G. Abbiendi {\it et~al.}, Eur.~Phys.~J {\bf C14} (2000) 51.
     
  \bibitem{detector}
    OPAL Collab., K.~Ahmet {\it et~al.}, Nucl.~Instr.\ Meth.\ {\bf A305}
    (1991) 275.
  
  \bibitem{microvertex}
    S.~Anderson {\it et~al.}, Nucl.~Instr.\ Meth.\ {\bf A403} (1998)
    326.
  
  \bibitem{te}
    G.~Aguillion {\it et~al.}, Nucl.~Instr.\ Meth.\ {\bf A417} (1998)
    266.
  
  \bibitem{sw}
    B.E.~Anderson {\it et~al.}, IEEE Trans. on Nucl. Science {\bf 41}
    (1994) 845.
  
  \bibitem{dfgt}
    C.~Dionisi {\it et~al.}, in `Physics at LEP2', eds. G.~Altarelli,
    T.~Sj\"{o}strand and 
    F.~Zwirner, CERN 96-01, vol.2, 337.
    
  \bibitem{jetset}
    T.~Sj\"{o}strand, Comp.~Phys.\ Comm.\ {\bf 82} (1994) 74;\\
    T.~Sj\"{o}strand, Lund University report LU~TP~95-20.
     
  \bibitem{nunugpv}
    G.~Montagna {\it et~al.}, Nucl.~Phys. {\bf B541} (1999) 31.
    
  \bibitem{bhwide}
    S. Jadach, W. P{\l}aczek and B.F.L. Ward, Phys.~Lett {\bf B390}
    (1997) 298.
    
  \bibitem{teegg}
    D. Karlen, Nucl.~Phys. {\bf B289} (1987) 23.
    
  \bibitem{kk2f}
    S.~Jadach, B.F.L.~Ward and Z.~W\c{a}s, Comp.~Phys.\
    Comm.{\bf 130} (2000) {260}; \\ 
    S.~Jadach, B.F.L.~Ward and Z.~W\c{a}s, Phys.~Lett. {\bf B449} (1999)
    97.
    
  \bibitem{radcor}
    F.A.~Berends, R.~Kleiss, Nucl.~Phys. {\bf B186} (1981) 22.
    
  \bibitem{koralw}
    S. Jadach, W. P{\l}aczek, M. Skrzypek, B.F.L. Ward and Z. W\c{a}s,
    Comp.~Phys.\ Comm.{\bf 119} (1999) {272}.
    
  \bibitem{grc4f}
    J.~Fujimoto {\it et~al.}, Comp.~Phys.\ Comm.\ {\bf 100} (1997)
    128.
    
  \bibitem{vermaseren}
    J.A.M.~Vermaseren, Nucl.~Phys.\ {\bf B229} (1983) 347.
    
  \bibitem{bdk}
    F.A.~Berends, P.H.~Daverveldt and R.~Kleiss, Nucl.Phys {\bf B253}
    (1985) 421;\\
    F.A.~Berends, P.H.~Daverveldt and R.~Kleiss, Comput.~Phys.\ Comm.\
    {\bf 40} (1986) 271;\\
    F.A.~Berends, P.H.~Daverveldt and R.~Kleiss, Comput.~Phys.\ Comm.\
    {\bf 40} (1986) 285;\\
    F.A.~Berends, P.H.~Daverveldt and R.~Kleiss, Comput.~Phys.\ Comm.\
    {\bf 40} (1986) 309.
    
  \bibitem{phojet}
    R.~Engel, Z.~Phys.\ {\bf C66} (1995) 203;\\
    R.~Engel and J.~Ranft, Phys.~Rev.\ {\bf D54} (1996) 4246.

  \bibitem{pythia}
    T.~Sj\"ostrand, Comp.~Phys.\ Comm.\ {\bf 82} (1994) 74.\\
    T.~Sj\"ostrand, Lund University report LU~TP~95-20.
    
  \bibitem{herwig}
    G.~Marchesini {\it et~al.}, Comp.~Phys.\ Comm.\ {\bf 67} (1992)
    465.
    
  \bibitem{gopal}
    J.~Allison {\it et~al.}, Nucl.~Instr.\ Meth.\ {\bf A317} (1992)
    47.
    
    
  \bibitem{pc}
    D.~Karlen, Comp.~in.~Physics, {\bf 12} (1998) 380.
    
  \bibitem{futyan}
    D.~Futyan, ``Search for Supersymmetry Using Acoplanar Lepton
    Pair Events at OPAL'', Appendix, PhD. thesis, Univerity of Manchester,
    December 1999.
    {\em http://hepwww.ph.man.ac.uk/theses/DavidFutyan.ps}.
    
  \bibitem{marchant}
    T.E.~Marchant, ``Search for New Physics Using Acoplanar Lepton
    Pair Events in ${\e}^+{\e}^-$ Collisions at $\sqrt{s}=$183-208 GeV'',
    Chapter 5, PhD. thesis, University of Manchester, March 2002.
    {\em http://hepwww.ph.man.ac.uk/theses/TomMarchant.ps}.

  \bibitem{durham}
    S.~Catani {\it et~al.}, Phys.~Lett.\ {\bf B269} (1991) 432.
    
  \bibitem{trackquality}
    K. Ackerstaff {\it et~al.}, Eur.~Phys.~J.\ {\bf C1} (1998) 395.

  \bibitem{susygen}
    N.Ghodbane, S.Katsanevas, P.Morawitz, E.Perez, SUSYGEN 3, hep-ph/9909499.
    
\end{thebibliography}
\end{document}